\documentclass[twocolumn]{aastex631}

\usepackage{graphicx}
\usepackage{txfonts}
\usepackage{hyperref}
\usepackage{cases}
\usepackage{threeparttable}

\usepackage{listings} 
\usepackage{xcolor} 

\hypersetup{
    colorlinks=true,
    citecolor=blue,
    linkcolor=blue,
    urlcolor=blue,
}

\expandafter\def\expandafter\UrlBreaks\expandafter{\UrlBreaks
  \do\a\do\b\do\c\do\d\do\e\do\f\do\g\do\h\do\i\do\j%
  \do\k\do\l\do\m\do\n\do\o\do\p\do\q\do\r\do\s\do\t%
  \do\u\do\v\do\w\do\x\do\y\do\z\do\A\do\B\do\C\do\D%
  \do\E\do\F\do\G\do\H\do\I\do\J\do\K\do\L\do\M\do\N%
  \do\O\do\P\do\Q\do\R\do\S\do\T\do\U\do\V\do\W\do\X%
  \do\Y\do\Z}

\definecolor{WHITE}{RGB}{255, 255, 255} 
\definecolor{codeblackkey}{RGB}{46, 46, 46}
\definecolor{codeblackid}{RGB}{51, 51, 51}
\definecolor{codegraycom}{RGB}{153, 153, 136}
\definecolor{codegrayfrm}{RGB}{153, 153, 153}
\definecolor{codegreennum}{RGB}{0, 153, 153}
\definecolor{coderedstr}{RGB}{221, 17, 68}
\definecolor{codered}{RGB}{192, 52, 29}
\definecolor{coderedhl}{RGB}{251, 229, 255}

\newcommand*{\FormatDigit}[1]{\textcolor{codegreennum}{#1}} 
\newcommand*{\FormatBool}[1]{\textcolor{codegraycom}{#1}} 
\newcommand*{\FormatBase}[1]{\textcolor{black}{#1}} 

\graphicspath{{./}{figures/}}

\shorttitle{A formally motivated retrieval framework applied to the high resolution transmission spectrum of HD~189733~b}
\shortauthors{Blain et al.}
\received{2023 November 06}
\revised{2024 February 13}
\accepted{2024 February 21}

\begin{document}
\title{A formally motivated retrieval framework applied to the high resolution transmission spectrum of HD~189733~b}

\author[0000-0002-1957-0455]{Doriann Blain}
\affiliation{Max-Planck-Institut für Astronomie, Heidelberg, Germany}
\correspondingauthor{blain@mpia-hd.mpg.de}

\author[0000-0002-0516-7956]{Alejandro Sánchez-López}
\affiliation{Leiden Observatory, Leiden University, Leiden, The Netherlands}
\affiliation{Instituto de Astrof{\'i}sica de Andaluc{\'i}a (IAA-CSIC), Glorieta de la Astronom{\'i}a s/n, 
18008 Granada, Spain}

\author[0000-0003-4096-7067]{Paul Mollière}
\affiliation{Max-Planck-Institut für Astronomie, Heidelberg, Germany}



\begin{abstract}

Ground-based high-resolution spectra provide a powerful tool for characterising exoplanet atmospheres. However, they are greatly hampered by the dominating telluric and stellar lines, which need to be removed prior to any analysis. Such removal techniques ("preparing pipelines") deform the spectrum, hence a key point is to account for this process in the forward models used in retrievals. We develop a formal derivation on how to prepare froward models for retrievals, in the case where the telluric and instrumental deformations can be represented as a matrix multiplied element-wise with the data. We also introduce the notion of "Bias Pipeline Metric" (BPM), that can be used to compare the bias potential of preparing pipelines. We use the resulting framework to retrieve simulated observations of 1-D and 3-D exoplanet atmospheres and to re-analyse high-resolution ($\mathcal{R} \approx 80\,400$) near infrared (0.96--1.71 $\mu$m) CARMENES transit data of HD~189733~b. We compare these results with those obtained from a CCF analysis. With our fiducial retrieval, we find a blueshift of the absorption features of $-5.51^{+0.66}_{-0.53}$ km$\cdot$s$^{-1}$. In addition, we retrieve a H$_2$O $\log_{10}$(VMR) of $-2.39^{+0.12}_{-0.16}$ and a temperature of $660^{+6}_{-11}$ K. We are also able to put upper limits for the abundances of CH$_4$, CO, H$_2$S, HCN and NH$_3$, consistent with a sub-solar metallicity atmosphere enriched in H$_2$O. We retrieve a broadened line shape, consistent with rotation- and wind-induced line broadening. Finally, we find a lower limit for the pressure of an opaque cloud consistent with a clear atmosphere, and find no evidence for hazes.

\end{abstract}


%

\section{Introduction}
    \label{sec:introduction}
    
    The characterisation of exoplanet atmospheres provides unique information about their atmospheric composition, dynamics, and aerosol presence, among other properties. Moreover, it may be possible to infer from the chemical composition of a planet how it formed and whether it experienced orbital migration \citep[e.g.,][]{Oberg2011, Madhusudhan2014b, Mordasini2016, Molliere2022}.
    
\begin{figure}[t]
       \centering
       \includegraphics[width=\hsize]{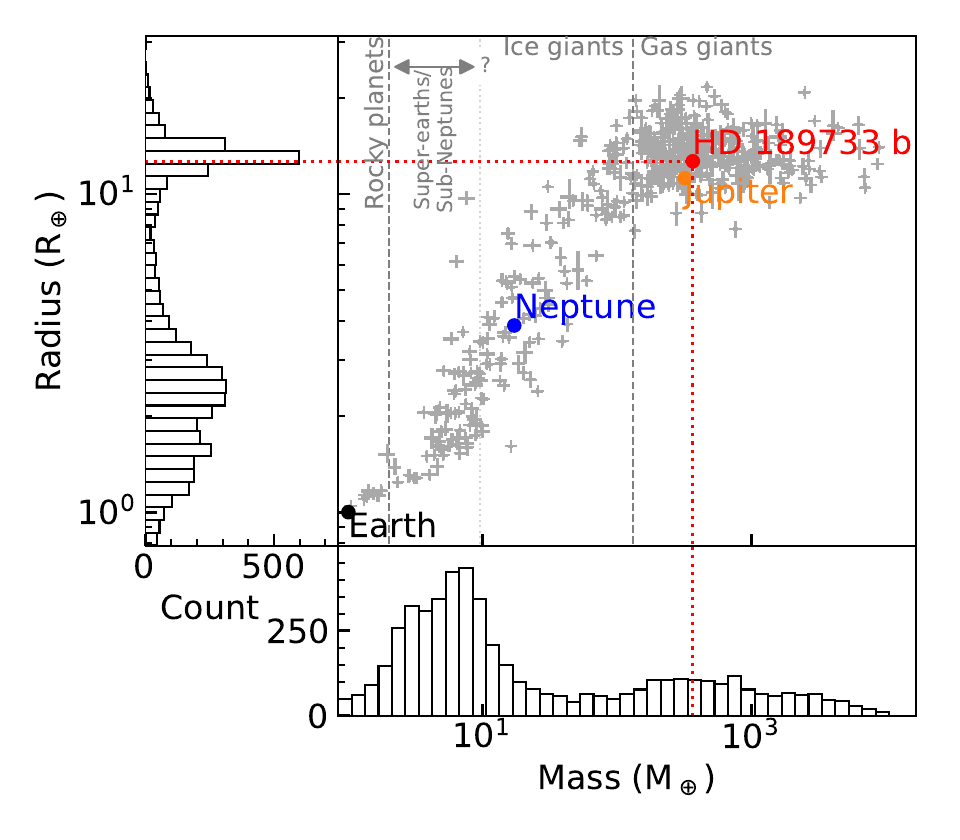}
          \caption{
            Mass-radius distribution of the 5243 confirmed exoplanets to date \citep[28 February 2023,][]{NasaExoplanetArchive}. Not all the confirmed planets have a measured radius and/or mass, thus the total count in the histograms is less than 5243. The scatter plot shows only the 512 planets for which the mass and radius are known within 15$\%$. 
            }
        \label{fig:exoplanet_distribution}
    \end{figure}

    With a mass and radius comparable to that of Jupiter and an equilibrium temperature of 1209 $\pm 11$ K (see \autoref{tab:general_parameters}), \object{HD 189733 b} falls into the "hot Jupiter" category (see \autoref{fig:exoplanet_distribution}). These properties coupled with its bright host star make this planet one with the highest transmission spectroscopy metric \citep[TSM,][]{Kempton2018}, at $\approx$ 770. Consequently, HD~189733~b is one of the most studied exoplanets. Focusing solely on spectral features, previous studies on the last ten years demonstrate the presence of H$_2$O in the atmosphere of HD~189733~b \citep{birkby2013detection, Pont2013, Line2014, danielski20140, McCullough2014, Sing2016, Cabot2018, Tsiaras2018, alonso2019multiple, sanchez2019water, Zhang2020, Boucher2021, Changeat2022, Finnerty2024, Klein2024}. Detections of CH$_4$ \citep{Line2014}, CO \citep{Line2014, Cabot2018, Brogi2019, Finnerty2024}, $^{13}$CO \citep{Finnerty2024}, CO$_2$ \citep{Line2014, Changeat2022}, HCN \citep{Cabot2018}, and Na \citep{Pont2013, Sing2016, Langeveld2022} were also reported. Also, upper limits for FeH \citep{kesseli2020search}, K \citep{Oshagh2020} and NH$_3$ \citep{Finnerty2024} have been established. A spectral slope were also observed at optical wavelengths and attributed to hazes \citep{Pont2013, Lee2014, Sing2016, Pino2018, sanchez2020discriminating}, but could also be interpreted as unocculted star spots \citep{McCullough2014, Oshagh2020}. Hot Jupiters, HD~189733~b included, are likely tidally locked \citep{Lin2004}. Three-dimensional atmospheric models suggest that under these conditions, a super-rotating equatorial jet forms. In the case of HD~189733~b, winds up to $\approx$ 5 km$\cdot$s$^{-1}$ are expected within this jet \citep{Showman2013, Dobbs2013, Lines2018, Flowers2019}.

    Transmission spectroscopy is a commonly used technique to study the atmosphere of transiting exoplanets. It has been successfully performed both with ground- and space-based instruments at low and high resolving power ($\mathcal{R}$), that is, at $\mathcal{R} \approx 10^2$ and $\mathcal{R} \approx 10^5$ respectively (see, e.g., the above-mentioned works). Spectra obtained from space-based instruments have the advantage of being exempt from telluric contamination, although currently, their spectral resolution is limited \citep[$\mathcal{R}$ $\lessapprox$ 3$\,$000,][]{hstdoc, jwstdoc}. From this kind of low-resolution data, it is possible to identify broad-band absorption features (e.g., molecular bands), but the core and wings of narrow spectral lines are not resolved. This can prevent the detection of minor species, and can also lead to ambiguities when the bands of two different species overlap \citep[e.g.][]{Brogi2017, Blain2021, Bezard2022}. Moreover, when hazes or clouds are present in the planet atmosphere, the molecular bands can be dampened or even completely masked, hindering detection \citep[e.g.][]{barstow2016consistent, Sing2016, barstow2020unveiling}. The situation is reversed for spectra obtained from the ground. With the higher resolving powers available, individual lines can be identified unambiguously. In addition, since the core of the lines probes higher altitudes, detection is not as impacted as at lower resolutions when clouds and hazes are present. Another advantage in resolving individual lines is to have information on atmospheric kinematics through the Doppler-shifting of the line positions and the shape (in particular, the width) of the spectral lines. However\footnote{An additional drawback of these data compared to low-resolution data is that, given their current signal to noise ratio, it is challenging to check for partial under- or over-fit (but global acceptable fit) of the data, thus preventing the detection of incorrect modelling or data reduction.}, telluric contamination plays a dominant role in the collected spectra and needs to be accurately removed. This operation is done after the usual raw data reduction step (flats, darks, wavelength calibration), in an operation that we will call "preparing pipeline"\footnote{This name is motivated by the core aim of these operations, which is to prepare the observed data for exoplanet studies.} (also called "pre-processing" or "detrending" in other works).

    \begin{deluxetable*}{@{}lccc@{}}[t]
    \tablecaption{\label{tab:general_parameters} Parameters of HD~189733~b and its star}
        \tablehead{\colhead{Parameter} & \colhead{Value} & & \colhead{References}}
        \startdata
        \textbf{Host star:} \\
        Spectral type										& K2 V													    &					            & 1				\\
        $M_\ast$ (kg)										& 1.630 $\pm$ 0.060 $\times 10^{30}$						& (0.828 M$_\odot$)	            & 2				\\
        $R_\ast$ (Gm)										& 0.543 $\pm$ 0.007									        & (0.780 R$_\odot$)	            & 2				\\
        $T_{\ast,\,\text{eff}}$ (K)							& 5052 $\pm$ 16												&					            & 3				\\
        $g_\ast$ (m$\cdot$s$^{-2}$)                         & 380$^{+1}_{-1}$	        								& (4.49 [cm$\cdot$s$^{-2}$])    & 3				\\
        $[$Fe/H$]$											& -0.030 $\pm$ 0.080										&					            & 4				\\
        $t_{\ast}$ (Gyr)									& 6.80$^{+5.20}_{-4.40}$									&					            & 4				\\
        $V_{\mathrm{sys}}$ (km$\cdot$s$^{-2}$)              & 2.204$^{+0.010}_{-0.011}$                                 &                               & 5             \\
        RA/Dec (J2000, epoch 2015.5)                        & $300.1821223^\circ\,22.7097759^\circ$                     & (20:00:43.71 +22:42:35.19)    & 6             \\
        \textbf{Planet:} \\
        $a_p$ (Gm)											& 4.67643 $\pm$ 0.00036								        & (0.03126 au)			        & 2				\\
        $e$													& 0.027$^{+0.021}_{-0.018}$									&					            & 2				\\
        $i_p$ (degree)										& 85.71$^{+0.00}_{-0.00}$									&					            & 3				\\
        $M_p$ (kg)											& 2.14 $\pm$ 0.15 $\times 10^{27}$							& (1.13 M$_J$)	                & 3				\\
        $R_p$ (km)											& 80$\,$800 $\pm$ 700				                        & (1.13 R$_J$)	                & 3				\\
        $g_p$ (m$\cdot$s$^{-2}$)                            & 21.9 $\pm$ 1.6        								    & (3.34 [cm$\cdot$s$^{-2}$])    & Derived   	\\
        $T_{\text{eq}}$ (K)				                    & 1209 $\pm$ 11	                                            &                               & 5			    \\
        $T_{p,\,\text{int}}$ (K)							& $\approx$ 100											    &					            & 7 (model)		\\
        $P$ (day)							                & 2.2185748039122 $\pm$ 0.0000001765372                     &					            & 6   		    \\
        $T_{14}$ (s)							            & 6463.08504 $\pm$ 7.89926                                  &					            & 6		        \\
        $T_0$ (BJD$_{\text{TDB}}$, day)					    & 2458004.424877 $\pm$ 0.000145			                    &					            & Derived       \\ \hline
        \enddata
    \tablerefs{
    (1)~\citet{Paredes2021}; (2) \citet{Rosenthal2021}; (3) \citet{Stassun2017}; (4) \citet{Bonomo2017}; (5) \citet{Addison2019}; (6) \citet{exofopweb}; (7) \citet{Rogers2010}. Some parameters were acquired from the \citet{NasaExoplanetArchive}.
    }
    \end{deluxetable*}
    
    A well-established technique to analyse high-resolution ground-based data is the cross-correlation (CCF) technique \citep[e.g.][]{snellen2010orbital, brogi2012signature, brogi2018exoplanet, birkby2013detection, birkby2017discovery, Cabot2018, hawker2018evidence, hoeijmakers2018atomic, alonso2019multiple, sanchez2019water, sanchez2020discriminating, sanchez2022searching, kesseli2020search,  kesseli2022atomic, merritt2020non, merritt2021inventory, stangret2020detection, kesseli2021confirmation, landman2021, nugroho2021first,   cont2022atmospheric, cont2022silicon, hoeijmakers2022mantis, prinoth2022titanium}. Cross-correlation has been successfully employed to detect atmospheric species, as well as to identify in some cases spectral line Doppler-shifts -- translated in an offset of rest-velocity -- that has been attributed to high-altitude winds \citep[e.g., for HD~189733~b,][]{Louden2015, Wyttenbach2015, Brogi2016, alonso2019multiple, Flowers2019}. However, this technique gives no or inaccurate information about atmospheric species' abundances or other parameters such as the temperature profile. Indeed, the CCF is sensitive at first order to the lines position and shape, but not to their absolute depth. On the other hand, data obtained from space are often studied via a chi-squared analysis (included or not in a Bayesian framework), directly comparing models with data and giving access to such information. Such direct comparison with a model cannot be done as easily with ground-based data of exoplanets, because the aforementioned preparing pipeline deforms the spectrum (also see \autoref{subsec:preparing_pipeline}), biasing the analysis.
    
    This issue with the preparing pipeline in high-resolution ground-based data analysis appears mostly solved. In recent years, new techniques to perform Bayesian analysis on ground-based data have emerged, with successful retrievals of atmospheric species abundances, temperature, and other properties \citep{Brogi2019, Gandhi2019, Gibson2020, Pelletier2021, Gibson2022}. The methods used have two main steps: first, prepare the data with a preparing pipeline to remove instrumental effects as well as telluric and stellar contamination, then modify the terms in the log-likelihood function used in the Bayesian analysis to take into account the data preparing step. That is, the data are replaced with the prepared data, and the model as well as the uncertainties are modified accordingly. This allows for direct data--model comparisons, and hence, in principle, accurate atmospheric parameter estimations.
    
    However, for both the preparing techniques and the log-likelihood modification, there is currently no agreement on the method to use. The preparing steps are often not described exhaustively, sometimes making reproducibility and comparisons difficult. This need for a consistent, reproducible, and robust preparation technique has already been raised \citep{Cabot2018, Cheverall2023}. Similarly, how the likelihood functions terms are modified can vary significantly between authors -- especially regarding the modification of the forward models. A formal justification for why a forward model should be prepared in a given way is also lacking from the literature, although this is a crucial step. Hence, there is also a need for a consistent method to modify the log-likelihood function terms.
    
    In this work, we use previously analysed high-resolution ground-based observations of a HD~189733~b transit to introduce a formally motivated high-resolution Bayesian retrieval framework in the near infrared. We propose a formal demonstration that this framework is unbiased for a subset of preparing pipelines satisfying given requirements. We also introduce the notion of "Bias Pipeline Metric" of the preparing pipeline, that can be used to compare preparing pipelines among each other. Finally, we present the results of our re-analysis of the data and put new constraints on the planet kinematics, temperature profile, cloud and haze properties, and molecular abundances. Doing so, we demonstrate the viability of our framework on real data.

\section{Observations and data reduction}
    \subsection{Observations}
        \label{subsec:observations}
        \begin{figure}
           \centering
           \includegraphics[width=\hsize]{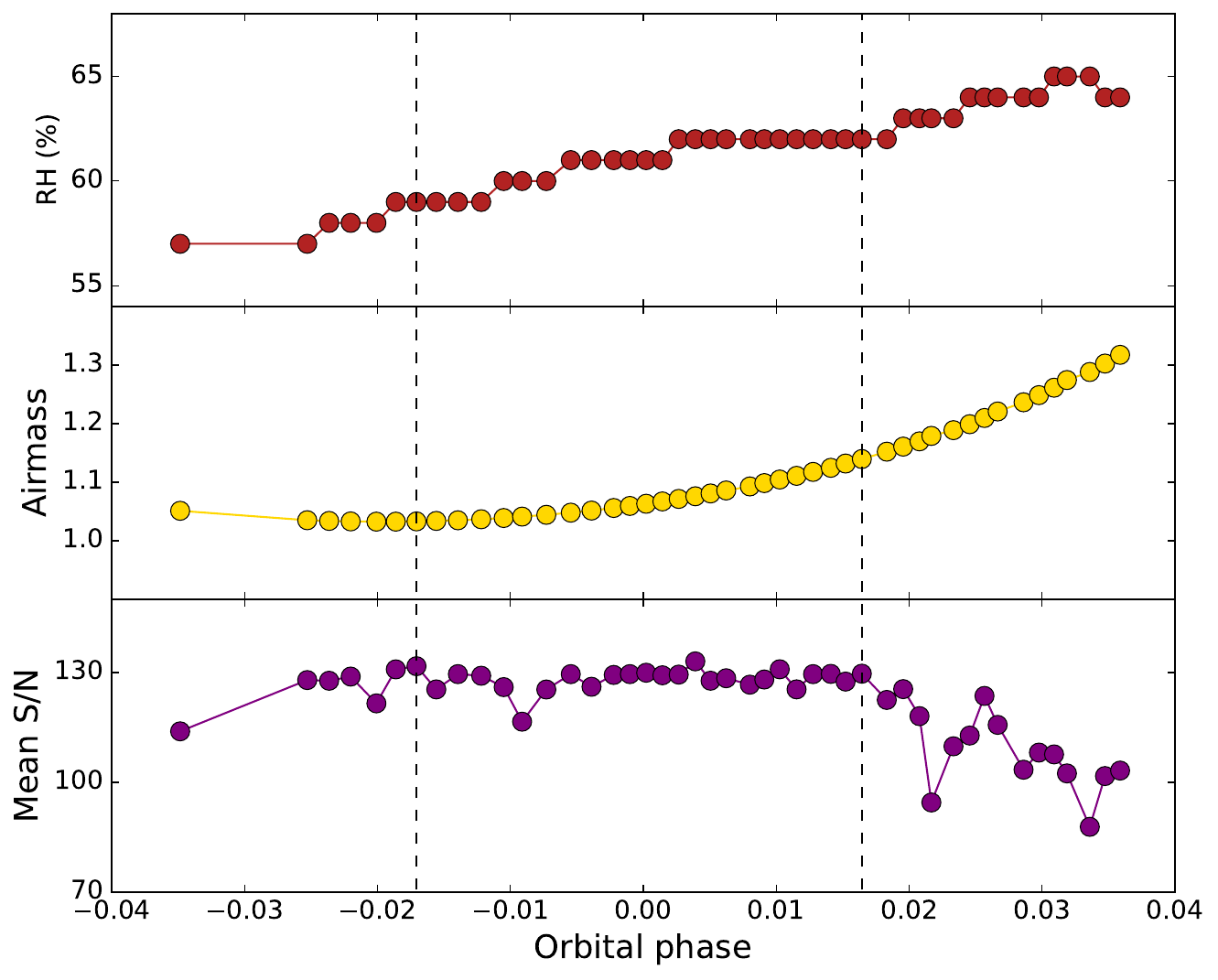}
              \caption{
                Evolution of the relative humidity (RH, top panel), airmass (middle panel), and mean signal-to-noise ratio (S/N, bottom panel) during the observations. The transit event of HD~189733~b occurs between the vertical dashed lines.
                }
            \label{fig:observation_condition}
        \end{figure}
        
        We analysed a transit event of HD~189733~b observed with CARMENES \citep{quirrenbach2016carmenes, quirrenbach2018carmenes} on 7 September 2017, which is publicly available from the Calar Alto Observatory (CAO) archive\footnote{\url{http://caha.sdc.cab.inta-csic.es/calto/}}. This dataset has been previously used to successfully detect water vapor and to infer the hazy nature of the atmosphere of this hot Jupiter \citep{alonso2019multiple, sanchez2019water, sanchez2020discriminating}. In addition, it was used by \citet{Cheverall2023} as a case study for evaluating different preparing methods for CCF analysis. We used the data from the near-infrared (NIR) channel's fiber A (focused on the target star HD\,189733), which covers the spectral range 0.96\,$-$\,1.71\,$\mu$m in 28 echelle orders at a spectral resolution of $\mathcal{R} \approx 80\,400$. The instrument's fiber B was kept on sky during the observations so as to monitor the occurrence of strong sky emission lines, which we found not to impact our analyses in any significant way. Each spectral order is, in turn, built from two 2040\,$\times$\,2040 pixel infrared detectors.
        
        The observations consist of 45 exposures of 198\,s encompassing the planet orbital phases from $-$0.035 to 0.036 (i.e., the primary transit of HD~189733~b). In \autoref{fig:observation_condition} we show the evolution of the relative humidity (RH), airmass, and mean S/N per spectrum during the observations \citep[see additional details in][]{alonso2019multiple}. Even with the low airmasses, the precipitable water vapor content of the atmosphere ranged from 11.7 to 15.9\,mm. This is relatively high compared to the mean value of $\approx 7$ mm at the Calar Alto Observatory and translated into a high number to spectral pixels rejected due to a strong telluric contamination (see details on the telluric correction and masking in Sect.\,\ref{subsec:preparing_pipeline}). Nevertheless, the mean S/N stayed high over the course of the observations, being always above 100 during transit.
    
    \subsection{Data reduction}
        The archive provides the option to download the data already reduced using the standard pipeline {\tt caracal} v2.00 \citep{zechmeister2014flat, caballero2016carmenes}, which provides one-dimensional spectra extracted from the raw observed frames. {\tt caracal} performs a bias correction to remove the unwanted contribution of each order's specific properties, a dark correction to subtract the contribution from thermal dark currents, as well as a flat field subtraction, which removes effects such as artifacts in the images produced by pixel-wise sensitivity differences. In addition, {\tt caracal} performs a wavelength calibration, providing the final processed spectra with wavelengths in vacuum and in the rest frame of the Earth. We took the airmass, julian date, and barycentric velocity ($V_\mathrm{bary}$) corresponding to each exposure from each file's header. The provided uncertainty matrix for each spectral point at each frame is built by {\tt caracal} by estimating the read out noise and photon noise.

        The time stamps given by the {\tt caracal} pipeline are in MJD$_\textrm{UTC}$\footnote{M. Zechmeister and L. Nortmann, private communication.}. We corrected them from the light travel time to the barycentre\footnote{This is done using the \lstinline|astropy| 5.3 (\url{https://www.astropy.org/}) functions \lstinline|astropy.coordinates.Skycoord|, \lstinline|astropy.coordinates.EarthLocation.of_site|, and \lstinline|astropy.time.Time.light_travel_time|. We used the RA and Dec in \autoref{tab:general_parameters} and the site name "CAHA", which corresponds to the CAO (see \autoref{anx:utc_to_tdb_code}). Optional parameters were set to their default values.} to obtain time stamps in BJD$_\textrm{TDB}$, that we used in our analysis. In our case, the difference between the UTC and TDB time stamps is $\textrm{UTC} - \textrm{TDB} = -368$ s on average. Note that this is equivalent to a Doppler velocity shift of $\approx +$ 2 km$\cdot$s$^{-1}$ of the planetary spectral lines at mid transit.

\section{Models}
    \label{sec:models}
    
    \subsection{Transmission spectrum model}
        \label{subsec:spectrum_model}
        
        To model the transmission spectrum of HD~189733~b, we follow the steps below.
        
        \paragraph{Step 1:} we model the planet atmosphere with 100 equally log-spaced pressure levels between $10^{7}$ and $10^{-5}$ Pa. We set the temperature constant with pressure. We use a mass fraction (also called "mass mixing ratio", or MMR) constant with pressure for every species included in our model. We ensure that the sum of the MMRs $\mathcal{X}_i$ of all species $i$ at each level is equal to 1 by doing the following:
        \begin{itemize}
            \item If $\sum_i \mathcal{X}_i > 1$: the MMR of each species is divided by $\sum_i \mathcal{X}_i$.
            \item If $\sum_i \mathcal{X}_i < 1$: H$_2$ and He are added to the atmosphere so that $\mathcal{X}_{\mathrm{H}_2} + \mathcal{X}_{\mathrm{He}} + \sum_i \mathcal{X}_i = 1$, with $\mathcal{X}_{\mathrm{He}}/\mathcal{X}_{\mathrm{H}_2} = 12/37$, which is roughly the ratio found in Jupiter's atmosphere \citep{Zahn1998}.
        \end{itemize}
        The average molar mass (also called "mean molar weight", or MMW) of each atmospheric levels is obtained via $\mathrm{MMW} = (\sum_i \mathrm{X}_i / \mathcal{M}_i)^{-1}$, where $\mathcal{M}_i$ is the molar mass of species $i$, obtained from the  \lstinline|molmass|\footnote{\url{https://github.com/cgohlke/molmass}} library.
        We calculate the planet orbital velocity $v_o$, assuming that the planet's orbit is circular and that its mass is negligible compared to that of its host:
        \begin{eqnarray}
            v_o &=& \sqrt{\frac{G M_*}{a_p}},
        \end{eqnarray}
        where $G$ is the gravitational constant, $M_*$ is the mass of the star, and $a_p$ is the length of the semi-major axis of the planet orbit.
        
        \paragraph{Step 2:} We obtain the planet radial velocity with respect to the observer in the reference frame of the planet's system barycenter $V_p(t)$ following:
        \begin{eqnarray}
            V_p(t) &=& v_o \sin(i_p) \sin(2 \pi \phi(t)),
        \end{eqnarray}
        where $i_p$ is the planet orbital inclination, with $i_p = 90^\circ$ corresponding to the planet-observer axis being contained into the plane of the planet orbit, and $\phi(t) = (t - T_0) / P$ is the planet orbital phases with respect to time $t$, where $T_0$ is the mid-transit time, and $P$ is the planet orbital period. The planet's radial velocity semi-amplitude $v_o \sin(i_p)$ is often denoted $K_p$, and will be retrieved in our setup. A positive radial velocity denotes that the planet is moving away from the observer, while a negative one denotes that the planet is moving toward the observer. Accordingly, $\phi = 0$ when the planet is in front of the star relative to the observer (mid primary transit), while $\phi = 0.5$ when the planet is behind the star (mid secondary transit)\footnote{If $i_p \neq 0^\circ$.}. This value will be used for the Doppler-shifting of the model spectrum in step 5.

        \begin{deluxetable}{l l} 
            \tablecaption{\label{tab:line_list_references} References of the line lists used for our high-resolution spectra}
            \tablehead{\colhead{Species} & \colhead{Reference}}
            \startdata
            CH$_4$      & \citet{hargreaves2020}        \\
            CO          & \citet{rothman2010}           \\
            H$_2$O      & \citet{rothman2010}           \\
            H$_2$S      & \citet{rothman2013}           \\
            HCN         & \citet{harris2006,barber2014} \\
            NH$_3$      & \citet{yurchenko2011}         \\
            \enddata
        \end{deluxetable}
        
        \paragraph{Step 3:} using the petitRADTRANS\footnote{\url{https://gitlab.com/mauricemolli/petitRADTRANS}} (pRT) package \citep{Molliere2019}, we compute the model transit radius of the planet $m_{\theta,0}$ from a set of parameters $\theta$. This transit radius is defined for a set of wavelengths in vacuum and in the rest frame of the model, $\lambda_0$, at $\mathcal{R} = 10^6$. We include the line-by-line opacities of each species $i$ included in step 1, as well as the collision-induced absorptions of H$_2$--H$_2$ and H$_2$--He and the Rayleigh scattering effect of H$_2$ and He. We use the opacities included with petitRADTRANS, listed in \autoref{tab:line_list_references}. To reduce memory usage and improve calculation speed, we downsampled those opacities such as to take 1 out of 4 points along wavelength. Given CARMENES' resolving power, we do not expect this to significantly impact our results. We simulate clouds as an opaque layer topping at a given pressure. Hazes are simulated using an additional opacity following a power-law and defined by its value at 0.35 $\mu$m ($\kappa_0$) and a power-law parameter ($\gamma$).
        
        \paragraph{Step 4:} we scale the transit radii to the star radius $R_\ast$ following:
        \begin{eqnarray}
        		m_{\theta,\mathrm{scale}}(\lambda_0) &=& 1 - \left( \frac{m_{\theta,0}(\lambda_0)}{R_\ast} \right)^2.
        \end{eqnarray}
        
        \paragraph{Step 5:} in order to take into account the Doppler effect caused by the relative velocity between the planet and the observer, we shift each wavelength $\lambda_0$, using:
        \begin{eqnarray}
        		\lambda_\mathrm{shift}(t) &=& \lambda_0 \sqrt{\frac{1 + \frac{V_{\mathrm{obs}}(t)}{c}}{1 - \frac{V_{\mathrm{obs}}(t)}{c}}},
        \end{eqnarray}
        where $\lambda_\mathrm{shift}$(t) contains the shifted wavelengths in the rest frame of the planet at each exposure, $V_{\mathrm{obs}} (t) = V_p(t) + V_\mathrm{sys} + V_\mathrm{bary}(t) + V_{\mathrm{rest}}$, where $V_\mathrm{sys}$ is the radial velocity of the star system with respect to the solar system -- also called systemic velocity --, $V_\mathrm{bary}(t)$ is the radial component of the velocity between the observer and the solar system barycenter projected along the vector pointing from the observer to the star, $V_{\mathrm{rest}}$ is a corrective scalar we retrieve so as to take into account additional sources of dynamics in the planet atmosphere, and $c$ is the speed of light in vacuum. Thereby, we obtain from $m_{\theta,\mathrm{scale}}(\lambda_0)$ a spectral matrix $\mathbf{M}_{\theta,\mathrm{shift}}(t, \lambda_\mathrm{shift})$, that is, one spectrum in the rest frame of the planet for each exposure\footnote{At this point, for a given wavelength index $i$, the values in $\lambda_{\mathrm{shift},i}(t)$ are different along $t$, but the values of $\mathbf{M}_{\theta,\mathrm{shift}}(t, \lambda_{\mathrm{shift},i})$ are identical along $t$ and equal to $m_{\theta,\mathrm{scale}}(\lambda_0)$. No re-binning has been performed yet.}.

        \paragraph{Step 6:} during ingress and egress, the intersection area between the planet's disk and its star's disk changes from 0 at the very beginning ($T_1$) and end ($T_4$) of the transit, to the planet's disk area during the full transit. Since the transit depth is proportional to this area, the planet signal during ingress and egress evolves accordingly. For a CCF analysis, this can be neglected as it is the position of the spectral features and their shape that matters the most for signal detection at first order. For a retrieval analysis however, this is more problematic as the absolute amplitude of the spectral features have an impact on the retrieved values. To take this effect into account, we calculate the wavelength-dependent mutual sky-projected distance between the apparent star centre and the apparent planetary centre ($\delta$) following \citet{Csizmadia2020}:
        \begin{eqnarray}
                \delta(t,\lambda_\mathrm{shift}) &=& \sqrt{b^2 + \left( \left(1 + r_\theta(t,\lambda_\mathrm{shift}) \right)^2 - b^2 \right) \frac{2\phi(t)}{\phi_{14}}},
        \end{eqnarray}
        where $b = a_p R_\ast \cos(i_p)$ is the impact parameter assuming a circular orbit, $r_\theta(t, \lambda_\mathrm{shift}) = \sqrt{1 - \mathbf{M}_{\theta,\mathrm{shift}}(t, \lambda_\mathrm{shift})}$ is simply the normalised transit radii, and $\phi_{14} = T_{14} / P$ is the phase arc described by the planet over the course of the total transit, with $T_{14}$ the planet's total transit time and $P$ the planet's orbital period. Then, we assume that at each wavelength the planet is an opaque, perfectly dark sphere and neglect the limb-darkening effect of the star. We calculate the corrected scaled transit radii $\mathbf{M}_{\theta,\mathrm{transit}}(t,\lambda_\mathrm{shift})$ following \citet{Mandel2002}:
        \begin{subnumcases} {\mathbf{M}_{\theta,\mathrm{transit}} =}
            $1,$ & if $\delta > 1 + r_\theta,$ \\
            1 - \left( r_\theta^2 c_0 + c_1 - \sqrt{\Delta} \right) / \pi, & if $r_{-} < \delta \leq r_{+},$ \\
            \mathbf{M}_{\theta,\mathrm{shift}}, & if $\delta \leq 1 - r_\theta$, \\
            $0,$ & if $\delta \leq r_\theta - 1,$
        \end{subnumcases}
        where the dependencies on $\lambda_\mathrm{shift}$ and $t$ are implied, and where $c_0(t,\lambda_\mathrm{shift}) = \arccos((\delta^2 - \mathbf{M}_{\theta,\mathrm{shift}})/2r_\theta\delta)$, $c_1(t,\lambda_\mathrm{shift}) = \arccos((\delta^2 + \mathbf{M}_{\theta,\mathrm{shift}})/2\delta)$, $\Delta(t,\lambda_\mathrm{shift}) = (4\delta^2 - (\mathbf{M}_{\theta,\mathrm{shift}} + \delta^2)^2)/4$, $r_{-} = \left|1 -  r_\theta\right|$, and $r_{+} = 1 + r_\theta$. Because of its effect on the lines amplitude, this step is crucial to retrievals, in contrast with CCF analysis. 
        
        Note that we assume that the planet's atmosphere is spatially (i.e., along latitudes and longitudes) uniform. As previously mentioned, models \citep[e.g.][]{Flowers2019} show that we should expect HD 189733 b's atmosphere to be, in contrast, spatially asymmetric, in particular considering wind velocity (see \autoref{sec:introduction}). A consequence is that the overall Doppler shift effect on the atmospheric spectral lines during ingress and egress ($\left|1 -  r_\theta\right| < \delta \leq 1 + r_\theta,$) should be different from this effect during the full transit ($\delta \leq 1 - r_\theta$). On the other hand, the spectral lines amplitudes are dampened as the planet eclipses partially its star's disk. This Doppler shift effect discrepancy is thus the strongest when the spectral lines amplitudes are the weakest. Moreover, in our case a majority of the analysed data are taken during the full transit (see \autoref{subsec:exposure_selection}). Hence, we do not expect this effect to strongly impact our results.
        
        \paragraph{Step 7:} in order to simulate the effect of the instrument line spread function (LSF), we convolve $\mathbf{M}_{\theta,\mathrm{transit}}(t, \lambda_\mathrm{shift})$ at each exposure by a Gaussian filter of standard deviation $\sigma_{\mathrm{conv}}$ given by:
        \begin{eqnarray}
        		\sigma_{\mathrm{conv}} &=& \frac{\lambda}{\mathcal{R}_{\mathrm{C}}} \frac{1}{\langle \Delta \lambda \rangle} \frac{1}{2 \sqrt{ 2 \ln{(2)}}},
        \end{eqnarray}
        where $\langle \Delta \lambda \rangle$ is the average step between the wavelengths $\lambda$, $\mathcal{R}_{\mathrm{C}}$ is the resolving power of the CARMENES NIR channel (expected to be $80\,400$), and $\ln$ denotes the natural logarithm. The rightmost term is used to convert the full width at half maximum (FWHM) of the LSF into the Gaussian filter's standard deviation. A Gaussian-kernel filter\footnote{The \lstinline|scipy| 1.8.0 (\url{https://scipy.org/}) function \lstinline|scipy.ndimage.gaussian_filter1d|,  with the optional arguments set to their default value} $\mathcal{G}$ is applied on the data. This gives us the shifted, convolved spectrum $\mathbf{M}_{\theta,\mathrm{conv}}(t, \lambda_\mathrm{shift})$ via:
        \begin{eqnarray}
        		\mathbf{M}_{\theta,\mathrm{conv}}(t, \lambda_\mathrm{shift}) &=& \mathcal{G}(\mathbf{M}_{\theta,\mathrm{shift}}(t, \lambda_\mathrm{shift}), \sigma_{\mathrm{conv}}).
        \end{eqnarray}
        
\begin{figure}
            \centering
            \includegraphics[width=\hsize]{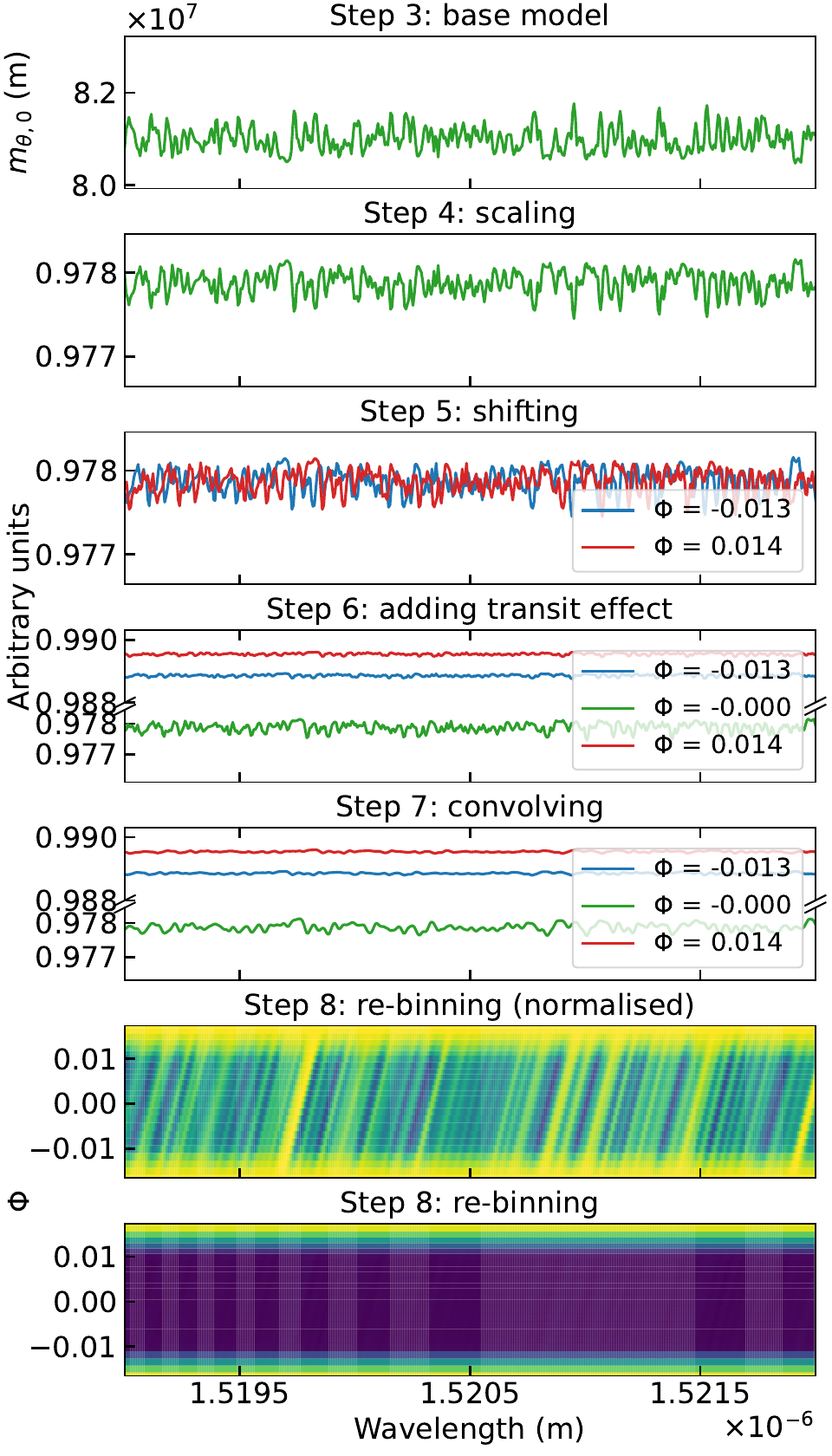}
            \caption{
                Illustration of the model construction process for a region of order 46 of the studied transit. Steps 1 and 2 do not involve spectra and are not represented. The rows displaying steps 6 and 7 have a broken y-axis in order to better show the spectra despite the transit effect. In the penultimate row, the spectra have been normalised over wavelengths for illustrative purposes. The bottom panel represents $\mathbf{M}_\theta(t, \lambda)$ without modification.
            }
            \label{fig:model_steps}
        \end{figure}

        \paragraph{Step 8:} $\mathbf{M}_{\theta,\mathrm{conv}}(t, \lambda_\mathrm{shift})$ is re-binned to the CARMENES wavelengths $\lambda$ to obtain the final spectral model $\mathbf{M}_{\theta}(t, \lambda)$. This is done using the \lstinline|rebin_spectrum| function of petitRADTRANS. In short, \lstinline|rebin_spectrum| constructs bin boundaries from the observational wavelength array which always lie in the middle between two neighboring wavelength points. At the edges of the spectrum the bin is assumed to be symmetric about the wavelength value of the observations. The model is then binned by calculating its integral over the bin width, and dividing the integral by that width afterwards. For the integral a linear wavelength dependence between the model points is assumed, so a trapezoidal rule is used for integration. Re-binning is preferred over interpolation because the former preserves the spectrum integral (so it is closer to the instrument's physics, where detector pixels collect the light on a order), at a negligible computational cost compared to that of our complete forward model calculation. We qualify this model as 1-D because the parameters depend only on the pressure. In \autoref{fig:model_steps} we show an illustration of these steps.
        
    \subsection{Simulated data}
        \label{subsec:simulated_data}

        The simulated data first follow the exact same steps than those listed in~\autoref{subsec:spectrum_model}. We replace $\theta$, the retrieved parameters, with $\Theta$, a "true" set of parameters. We do not include the effect of stellar lines in this model. In order to simulate the effect of the Earth atmosphere and of the instrument, we follow these additional steps:
        
        \paragraph{Step 6 bis:} this step is inserted between step 6 and step 7 of \autoref{subsec:spectrum_model}. We obtain the wavelength-dependant transmittance of the Earth's atmosphere $\mathbf{T}_{\oplus, 0}$ at wavelengths $\lambda_\oplus$, at a resolving power of $10^6$, and for an airmass $\mu = 1$, using the sky model calculator SKYCALC\footnote{\url{https://www.eso.org/observing/etc/bin/gen/form?INS.MODE=swspectr+INS.NAME=SKYCALC}}. The airmass at the time of each exposure $\mu(t)$ is obtained from the CARACAL pipeline. We obtain the wavelength- and time-dependent transmittance of the Earth atmosphere, $\mathbf{T}_{\oplus}(t, \lambda_\oplus)$, via:

        \begin{figure}
           \centering
           \includegraphics[width=\hsize]{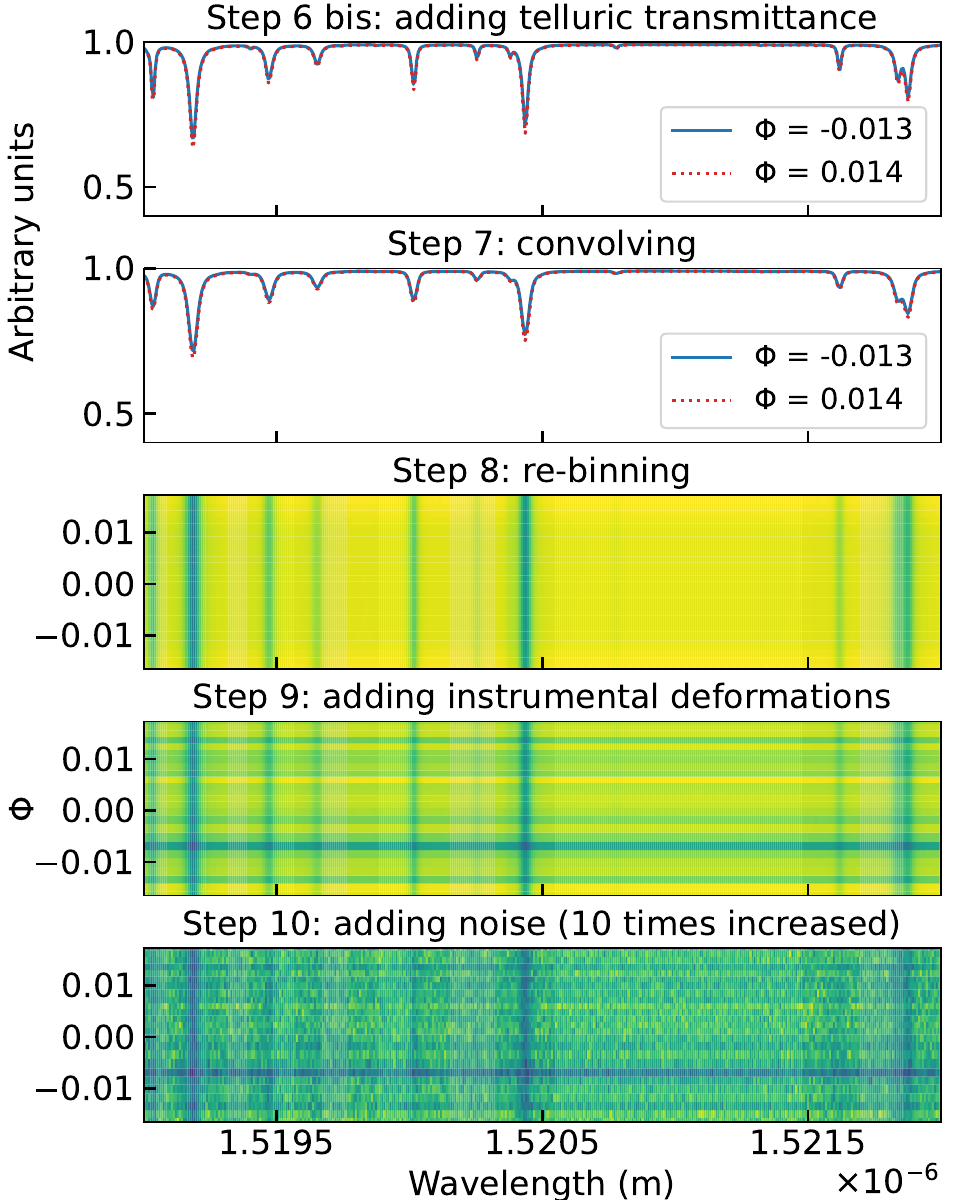}
            \caption{
                Illustration of the simulated data construction process for a region of order 46 of the studied transit. Steps 3 to 5 are identical to \autoref{fig:model_steps}. The slight differences between the telluric line shapes with orbital phase are mainly due to the change in airmass. For illustrative purposes, the noise in step 9 has been increased by a factor 10 compared to the amount expected for CARMENES.
            }
            \label{fig:simulated_data_steps}
        \end{figure}
        
        \begin{eqnarray}
            \label{eq:telluric_transmittance}
        		\mathbf{T}_{\oplus}(t, \lambda_\oplus) &=& \exp\left(\mu(t) \ln(\mathbf{T}_{\oplus, 0}(\lambda_\oplus))\right).
        \end{eqnarray}
        In contrast with the steps described in \autoref{subsec:spectrum_model}, here we use an intermediate wavelength grid $\lambda_\mathrm{int}$ at the same resolving power than $\lambda_\mathrm{shift}(t)$, and such that the grid is included in the intersection of $\lambda_\mathrm{shift}(t)$ along $t$. This intermediate grid is used to ensure that the transmittances of the Earth's atmosphere vary smoothly with time after the final re-binning performed in step 8. For each exposure, both $\mathbf{M}_{\theta,\mathrm{shift}}(t, \lambda_\mathrm{shift})$ and $\mathbf{T}_{\oplus}(t, \lambda_\oplus)$ are re-binned to the corresponding $\lambda_\mathrm{int}$ using the same function as in step 8. The re-binned $\mathbf{T}_{\oplus}(t, \lambda_\mathrm{int})$ is included into the re-binned spectrum $\mathbf{M}_{\theta,\mathrm{int}}(t, \lambda_\mathrm{int})$ via:
        \begin{eqnarray}
        		\mathbf{M}_{\Theta,\mathrm{int},\mathbf{T}}(t, \lambda_\mathrm{int}) &=& \mathbf{M}_{\Theta,\mathrm{int}}(t, \lambda_\mathrm{int}) \circ \mathbf{T}_{\oplus}(t, \lambda_\mathrm{int}),
        \end{eqnarray}
        with "$\circ$" denoting the element-wise product, also called "Hadamard product" \citep{Million2007}.
        Then, steps 7 and 8 of \autoref{subsec:spectrum_model} are applied to $\mathbf{M}_{\Theta,\mathrm{int},\mathbf{T}}(t, \lambda_\mathrm{int})$ without any modification, to obtain $\mathbf{M}_{\Theta,\mathbf{T}}(t, \lambda)$.
        
        \paragraph{Step 9:} the variation of observed flux level, pseudo-continuum and blaze function, that we regroup under the term "instrumental deformations", are represented with a time- and wavelength-dependent matrix $\mathbf{X}(t, \lambda)$. To model $\mathbf{X}$ we used the fit given by the first step of our preparing pipeline (see \autoref{subsec:preparing_pipeline}) on our data. We obtain the noiseless simulated data via:
        \begin{eqnarray}
        		\mathbf{F}_{\mathrm{sim},0}(t, \lambda) &=& \mathbf{M}_{\Theta,\mathbf{T}}(t, \lambda) \circ \mathbf{X}(t, \lambda),
        \end{eqnarray}

        \begin{figure}
           \centering
           \includegraphics[width=\hsize]{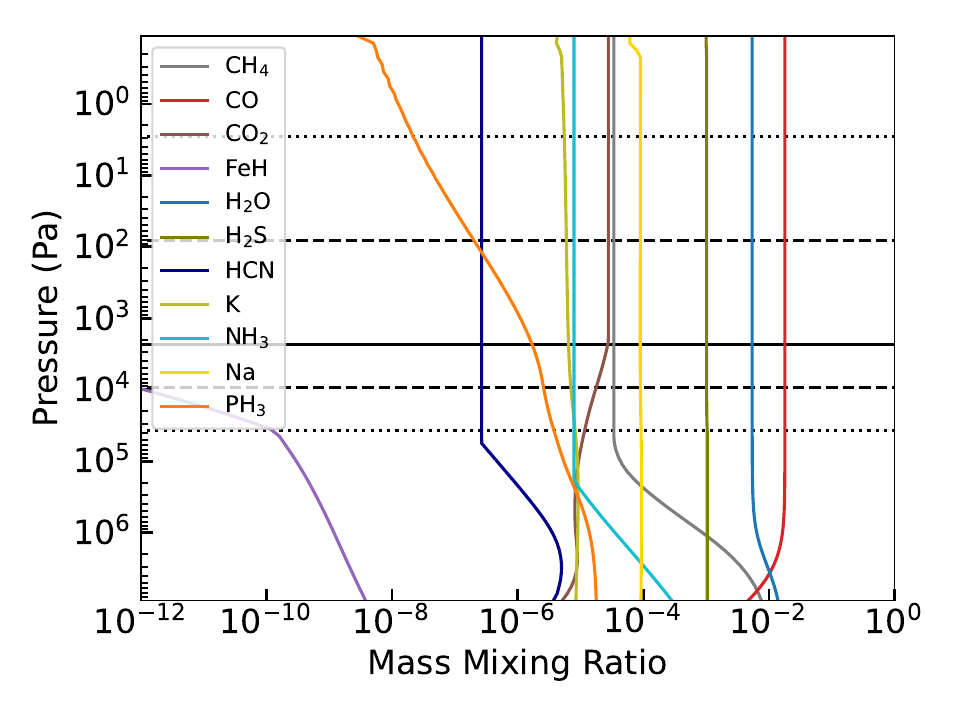}
              \caption{
                Mass mixing ratios obtained from our self-consistent model, for 3 times the solar metallicity. Only the absorbing species MMRs are represented, except TiO and VO, which had MMRs below $10^{-12}$. The horizontal dotted and dashed black lines correspond respectively to the pressure range concentrating $95\%$ and $68\%$ of the transmission contribution of our petitRADTRANS base model (see \autoref{subsec:validation}) over the CARMENES wavelength range. The horizontal solid black line corresponds to the pressure of the maximum contribution.
            }
            \label{fig:exorem_mmr}
        \end{figure}
        
        \paragraph{Step 10:} the noisy simulated data are finally obtained from:
        \begin{eqnarray}
        	\label{eq:simulated_data}
        		\mathbf{F}_{\mathrm{sim}}(t, \lambda) &=& \mathbf{F}_{\mathrm{sim}, 0}(t, \lambda) + \mathbf{N}(t, \lambda),
        \end{eqnarray}
        where $\mathbf{N}(t, \lambda)$ is the wavelength- and time-dependent Gaussian noise of the data\footnote{This is obtained using the \lstinline|numpy| 1.24.3 (\url{https://numpy.org/}) function \lstinline|numpy.random.default_rng(seed=12345).normal()| with parameters $\mathrm{loc} = 0$, $\mathrm{scale} = \mathbf{U}_\mathbf{N}(t,\lambda)$ and $\mathrm{size}$ is a tuple representing the shape of $\mathbf{U}_\mathbf{N}(t,\lambda)$. In this work we always use the same seed value.}. The scale of the noise $\mathbf{U}_\mathbf{N}(t, \lambda)$, i.e. the data uncertainties, is given by the CARACAL pipeline. In \autoref{fig:simulated_data_steps} we show an illustration of these steps.

        \begin{figure}
           \centering
           \includegraphics[width=\hsize]{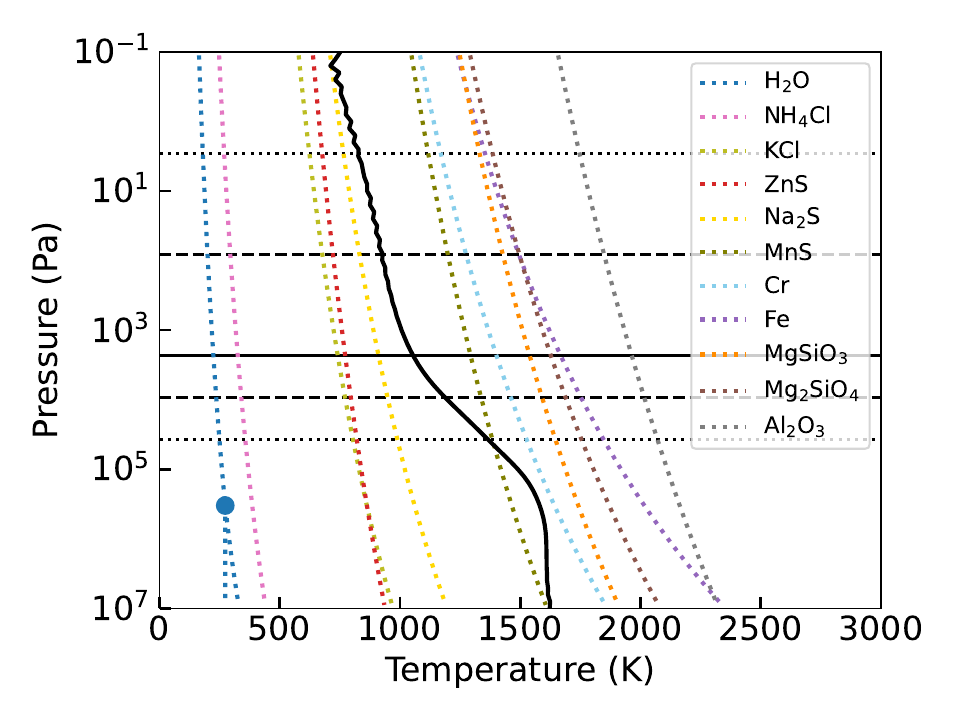}
              \caption{
                Temperature profile and condensation curves obtained from our self-consistent model, for 3 times the solar metallicity. 
                The horizontal dotted and dashed black lines correspond respectively to the pressure range concentrating $95\%$ and $68\%$ of the transmission contribution of our petitRADTRANS base model (see \autoref{subsec:validation}) over the CARMENES wavelength range. The horizontal solid black line corresponds to the pressure of the maximum contribution.
            }
            \label{fig:exorem_tpr}
        \end{figure}
        
    \subsection{Self-consistent model}
        \label{subsec:self-consistent_model}
        
        In order to check the physical consistency of our retrieved results, we use a self-consistent model generated by the Exo-REM software\footnote{{\sloppy \url{https://gitlab.obspm.fr/Exoplanet-Atmospheres-LESIA/exorem}}}. Exo-REM calculates 1D radiative-equilibrium models for H$_2$-dominated (atmospheric metallicity $Z \lessapprox 1000$ times solar) cold to hot (equilibrium temperature $\lessapprox$ 2000 K) planetary atmospheres. The software is fully described by \citet{Baudino2015, Baudino2017, Charnay2018, Blain2021}. To summarise, fluxes in Exo-REM are calculated using the two-stream approximation, assuming hemispheric closure. The radiative-convective equilibrium is solved assuming that the net flux (radiative plus convective) is conservative. The conservation of flux over the pressure grid is solved iteratively using a constrained linear inversion method.

        \begin{figure*}
            \centering
           \includegraphics[width=\linewidth]{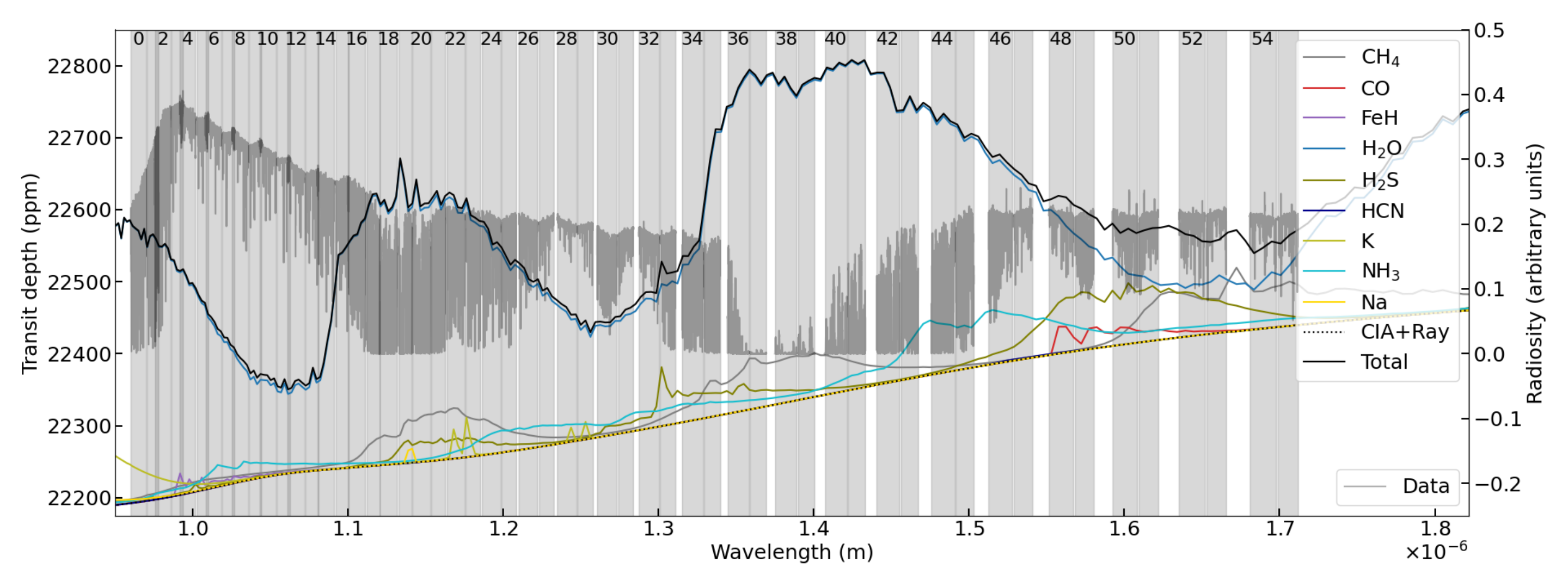}
            \caption{HD 189733 b low-resolution ($\mathcal{R} \approx 500$) transit depth and species contributions simulated with Exo-REM. The gray areas represent the wavelength range of the CARMENES orders. On top are indicated the index number of the even-numbered orders. The dotted line labelled "CIA+Ray" represents the combined contributions of the H$_2$--H$_2$, H$_2$--He and H$_2$O--H$_2$O collision-induced absorptions and the effect of Rayleigh scattering. The contributions of CO$_2$, PH$_3$, TiO and VO were negligible in this specral region and thus are not represented. On the background, in black, are represented the time-averaged data in arbitrary units.}
            \label{fig:species_contribution}
        \end{figure*}
        
        Our HD~189733~b self-consistent model was parameterised following the same procedure described in \citet{Blain2021}, except that we did not include the radiative effect of clouds in this simulation. The parameters used are displayed in \autoref{tab:general_parameters}. Since HD~189733~b has a mass and radius similar to that of Jupiter, and since its star's metallicity is similar to that of the Sun, we chose to use an atmospheric metallicity of 3 times the solar metallicity, which is approximately the value measured in the upper atmosphere of Jupiter \citep{Atreya2020}. The MMRs and temperature profile obtained are displayed respectively in \autoref{fig:exorem_mmr} and \autoref{fig:exorem_tpr}. The simulated temperature profile is shown crossing the condensation curve of MnS at $3.7\times10^4$ Pa, where the contribution to the transmission spectrum is relatively low ($\lessapprox$ 2.5$\%$ of the total contribution). The corresponding low-resolution ($\mathcal{R} \approx 500$) transmission spectrum and the contribution of some absorber species over the wavelength range of CARMENES are represented in \autoref{fig:species_contribution}. 
        
        This latter figure shows which species we can expect to detect according to this model. In addition to H$_2$O, significant contributions of CH$_4$, CO, H$_2$S and NH$_3$ are expected assuming chemical disequilibrium \citep[see][]{Blain2021} with a self-consistent eddy diffusion coefficient \citep[see][]{Charnay2018}. We expect the contributions of FeH, K and Na to be too low and too localised to be detectable. The other absorbers included in this model, namely CO$_2$, HCN, PH$_3$, TiO and VO, have negligible contributions over the CARMENES NIR spectral region. However, HCN has a band around 1.5 $\mu$m and was detected in the atmosphere of this planet \citep{Cabot2018} with a peak CCF detection corresponding to a volume mixing ratio of $10^{-6}$, which is 40 times what our Exo-REM model predicts. There is therefore a possibility that the Exo-REM chemical model does not describe this species' chemistry well, due to unknown or underestimated processes. In conclusion, we decided to include CH$_4$, CO, H$_2$O, H$_2$S, HCN, and NH$_3$ opacities in our retrieval model.

    \subsection{3-D transmission spectrum model}
        \label{subsec:3d_transmission_spectrum_model}
        
        We generated three-dimensional transmission spectra using pRT-Orange, which will be described in an upcoming paper (Mollière et al., in prep.).
        In short, pRT-Orange divides the planet atmosphere into (orange fruit) segments that are connected at the planet's pole-to-pole axis. The number, longitudinal location and width of the segments can be specified freely. Within each segment the atmosphere is described by a one-dimensional vertical structure which can be obtained using the full set of options available in classic (1-D) pRT model. pRT-Orange's current setup thus assumes that the atmospheric properties do not change as a function of latitude. Changing the segment properties and location as a function of latitude is currently under investigation. The setup of pRT-Orange thus allows to straightforwardly capture gradients in atmospheric properties that vary longitudinally, such as in the day-to-night or morning-evening differences.
        
        The three-dimensional transmission spectrum is obtained by slicing the planet atmosphere into disks parallel to the equatorial plane. In each disk, the two-dimensional transmission problem is solved. This is very similar to pRT's one-dimensional treatment, with the added complication that atmospheric properties change as a function of longitude.
        The slices' locations are chosen by transforming the height above the equatorial plane into a coordinate in which constant distances mean constant area fractions of the planet's terminator annulus, and then discretizing the coordinate using Gaussian quadrature. Integrating the annulus area, weighted by the atmospheric transmission obtained from the two-dimensional disks, then results in the effective planet area as a function of wavelength. We benchmarked our setup with gCMCRT \citep{lee2022}, finding excellent agreement.

        For our 3-D spectra of HD~189733~b we post-processsed the GCM results reported in \citet{drummond2018}\footnote{\url{https://ore.exeter.ac.uk/repository/handle/10871/34691}}. In particular, we use their cloud-free model at equilibrium chemistry, at solar composition (metallicity $Z = 1$). For simplicity, we separated the planet in 36 segments spaced equidistantly in longitude and assumed that their temperature-pressure structures are well represented by the equatorial structure from \citet{drummond2018}, at the same longitude. We used their full (also latitudinally varying) velocity structure and the planetary rotation rate to derive line-of-sight velocities at all discretised locations of the planet, and Doppler-shifted the opacities accordingly. We used the segments' pressure-temperature profiles and equilibrium chemistry to determine the atmospheric composition in each segment. To coincide better with the self-consistent Exo-REM models, we set a metallicity of $Z = 3$ -- instead of the $Z = 1$ used for the GCM model -- for the equilibrium chemistry calculations. We assume that this increase in metallicity would not significantly affect the GCM results or P-T profiles.

\section{Methods}
    \subsection{Preparing pipeline}
        \label{subsec:preparing_pipeline}
        
        In order to retrieve valuable information from the data, it is necessary to have a proper parameterisation of each atmospheric property in our forward model. While all properties in \autoref{subsec:spectrum_model} can easily be parameterised, it is not the case for $\mathbf{T}_\oplus(t, \lambda)$, $\mathbf{X}(t, \lambda)$ and the stellar lines. They would require a lot of parameters to retrieve, and Earth atmospheric modelling is computationally expensive. We will represent these "non-retrievable"\footnote{It would not be practical to retrieve those parameters, but their retrieval is not impossible per se.} parameters with a so-called "deformation matrix" $\mathbf{D}(t, \lambda) \equiv \mathbf{X}(t, \lambda) \circ \mathbf{T}(t, \lambda)$. Here, $\mathbf{T}(t, \lambda)$ represents the telluric transmittance and stellar lines component of the noiseless data, which would be obtained via $\mathbf{T} = \mathbf{M}_{\Theta,\mathbf{T}} / \mathbf{M}_\Theta$ (see \autoref{subsec:spectrum_model} and \autoref{subsec:simulated_data}). By applying a preparing pipeline $P_\mathbf{R}$, it is possible, in principle, to remove most of $\mathbf{D}(t, \lambda)$ from data. Since the CARMENES data are coming from different orders, we apply the preparing pipeline on each of these orders separately. For a set of spectra $\mathbf{F}(t, \lambda)$, the pipeline $P_\mathbf{R}(\mathbf{F})$ steps are described below:
        
        \paragraph{Step 1:} We clean the effect of $\mathbf{X}(t, \lambda)$ by dividing each exposure of $\mathbf{F}(t, \lambda)$ by a corresponding second-order polynomial fit of $\mathbf{F}(t, \lambda)$ over wavelength\footnote{This is done using the \lstinline|numpy| 1.24.3 (\url{https://numpy.org/}) function \lstinline|numpy.polynomial.Polynomial.fit| with parameters $x = \lambda_i$, $y = \mathbf{F}_{ij}$, $\mathrm{deg} = 2$ and $w = \mathbf{1}$, where $i$ and $j$ denote respectively the order and the time. The other parameters of the function are set to their default value. Note that from the function documentation, $w$ is supposed to be $1 / \mathbf{U}_{\mathbf{N},ij}$. However, we obtained a better result by not using this.}. If we call the fit $\overline{\mathbf{X}}(t, \lambda)$, we obtain the $\mathbf{X}$-corrected ("normalised") spectrum $\mathbf{F}_{\overline{\mathbf{X}}}(t, \lambda)$ with:
        \begin{eqnarray}
        	\label{eq:pipeline_step1}
        		\mathbf{F}_{\overline{\mathbf{X}}}(t, \lambda) &=& \mathbf{F}(t, \lambda) \oslash \overline{\mathbf{X}}(t, \lambda),
        \end{eqnarray}
        where "$\oslash$" denotes the element-wise division, also called "Hadamard division" \citep{Wetzstein2012}. Following the propagation errors, the uncertainties of $\mathbf{F}_{\overline{\mathbf{X}}}(t, \lambda)$ are $\mathbf{U}_{\mathbf{N},\overline{\mathbf{X}}}(t,\lambda) = \mathbf{U}_{\mathbf{N}}(t,\lambda) \oslash \left|\overline{\mathbf{X}}(t, \lambda)\right| \circ \sqrt{n_\lambda(t)}$. The variance correction factor of the fit $n_\lambda(t)$ is given by $n_\lambda(t) = (N_\lambda'(t) - d) / N_\lambda$, where $N_\lambda'(t)$ is the number of non-masked points along wavelength at time $t$, $N_\lambda$ is the number of wavelength points, and $d = 3$ is the number of degrees of freedom of a second-order polynomial fit. This correction is required to accurately estimate the effect of the fit on the uncertainties. This effect is generally regarded as negligible for a large number of fitted points, but it must be highlighted that here $N$ is the number of points used for each individual fit, not the total number of points in the data. In this first step the fit is not performed over our $\approx 10^6$ total data points, but one time for each exposure, over the wavelengths. This corresponds to a maximum of 2040 data points fitted each time, and less when some of the points are masked. This number is still large in most orders, but in the second step (see below) the fit will be preformed one time for each wavelength, over the exposures. This corresponds to a maximum of 26 (see \autoref{subsec:exposure_selection}) points fitted each time. Hence, not accounting for $n_\lambda(t)$ in our case can lead to a significant overestimation of the uncertainties, and, after a retrieval, to a model that misleadingly appears to overfit the data.
        
        \paragraph{Step 2:} We fit the effect of $\mathbf{T}(t, \lambda)$ with a second-order polynomial fit\footnote{This is done using the \lstinline|numpy| 1.24.3 function \lstinline|numpy.polynomial.Polynomial.fit| with parameters $x = \mu$, $y = \ln{\left(\mathbf{F}_{\overline{\mathbf{X}},ik}\right)}$, $\mathrm{deg} = 2$ and $w = \mathbf{1}$. The other parameters of the function are set to their default value.} of $\ln(\mathbf{F}_{\overline{\mathbf{X}}}(t, \lambda))$ over airmass: $\ln{\left(\mathbf{T}_\oplus(t, \lambda)\right)}$ is indeed linearly dependent on the airmass (see \autoref{eq:telluric_transmittance}). Note that this correction also works for stellar lines, that are independent on airmass: the fitting polynomial will simply have its airmass-dependent term close to 0. The second order is used to fit for eventual other slow temporal variations, like precipitable water vapor (PWV) variations when the weather is stable. Next, we divide $\mathbf{F}_{\overline{\mathbf{X}}}(t, \lambda)$ with the exponential of the fit. If we call the fit $\overline{\mathbf{T}}(t, \lambda)$, we obtain the prepared spectrum $\mathbf{F}_{\overline{\mathbf{T}}}(t, \lambda)$ with:
        \begin{eqnarray}
        	\label{eq:pipeline_step2}
        		\mathbf{F}_{\overline{\mathbf{T}}}(t, \lambda) &=& \mathbf{F}_{\overline{\mathbf{X}}}(t, \lambda) \oslash \exp{\left(\overline{\mathbf{T}}(t, \lambda)\right)}.
        \end{eqnarray}
        In addition, we mask $\mathbf{F}_{\overline{\mathbf{T}}}(t, \lambda)$ where $ \exp{\left(\overline{\mathbf{T}}(t, \lambda)\right)} < 0.8$. This is done to remove from the analysis data that would be too affected by the telluric lines. The value of the threshold was chosen as a good balance between masking tellurics and not masking too many useful data, although we did not perform extensive testing of the effect of this parameter.
        
        From this pipeline, we also obtain what we will call a preparing matrix $\mathbf{R}_\mathbf{F}(t, \lambda) \equiv 1 \oslash \left(\overline{\mathbf{X}}(t, \lambda) \circ \exp{\left(\overline{\mathbf{T}}_\oplus(t, \lambda)\right)}\right)$. Similarly to the previous step, we obtain the variance correction factor of the fit $n_t(\lambda) = (N_t'(\lambda) - d) / N_t$, where $N_t'(\lambda)$ is the number of non-masked points along time at wavelength $\lambda$ and $N_t$ is the number of time points. All of the preparing pipeline steps can be represented as a function $P_\mathbf{R}$:
        \begin{eqnarray}
        	\label{eq:pipeline}
        		P_\mathbf{R}(\mathbf{F}) &=& \mathbf{F}(t, \lambda) \circ \mathbf{R}_\mathbf{F}(t, \lambda) \equiv \mathbf{F}_{\overline{\mathbf{T}}}(t, \lambda).
        \end{eqnarray}
        Note that the preparing matrix $\mathbf{R}$ will change depending on the input of $P_\mathbf{R}$ (e.g., $P_\mathbf{R}(\mathbf{M}_\theta) = \mathbf{M}_\theta \circ \mathbf{R}_{\mathbf{M}_\theta}(t, \lambda)$, with $\mathbf{R}_{\mathbf{M}_\theta} \neq \mathbf{R}_\mathbf{F}$ since the fits of $\mathbf{M}_\theta$ from step 1 and step 2 are different from the fits of $\mathbf{F}$), and hence it needs to be recalculated for each set of parameters during the retrieval. The uncertainties on $P_\mathbf{R}(\mathbf{F})$, following the propagation of errors, are:
        \begin{eqnarray}
        	\label{eq:pipeline_uncertainties}
        		\mathbf{U}_\mathbf{R}(t, \lambda) &=& \mathbf{U}_\mathbf{N}(t, \lambda) \circ \left|\mathbf{R}_\mathbf{F}(t, \lambda)\right| \circ \sqrt{n(t, \lambda)},
        \end{eqnarray}
        where $n(t, \lambda) = n_\lambda(t) \circ n_t(\lambda)$.

        \begin{figure}
           \centering
           \includegraphics[width=\hsize]{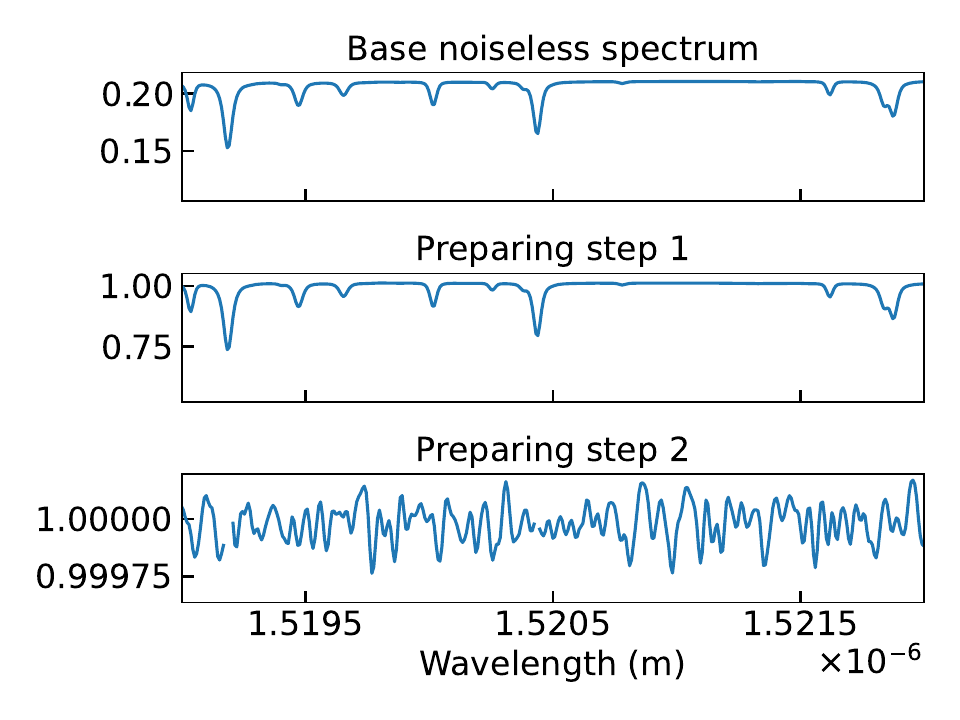}
              \caption{
                Illustration of the preparing effect on noiseless simulated data corresponding to order 46. Only the spectrum corresponding to orbital phase -0.0073 is shown. The spectrum used is the same as in step 9 of \autoref{fig:simulated_data_steps}. The spectra are given in arbitrary units.
            }
            \label{fig:preparing_pipeline_1d}
        \end{figure}
        
\begin{figure}
           \centering
           \includegraphics[width=\hsize]{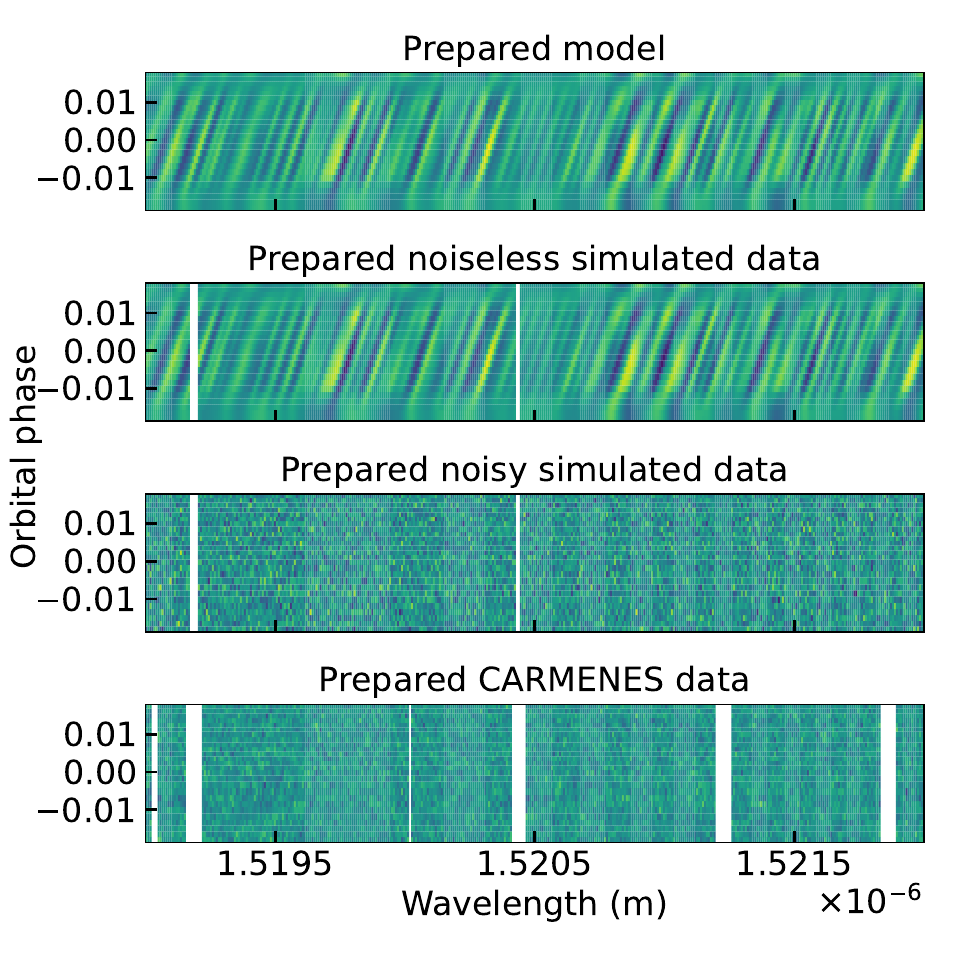}
              \caption{
                Illustration of the preparing pipeline effect on the order 46 of the studied data. First row: prepared model. The model is the same as in step 8 of \autoref{fig:model_steps}. While deformed, the spectral lines are clearly visible. The transit effect is also still visible. Second row: prepared noiseless simulated data. The simulated data used are the same as in step 9 of \autoref{fig:simulated_data_steps}. Third row: same as second row but with added noise. The lines are no longer visible to the naked eye. Bottom row: prepared CARMENES data. The white vertical stripes are masked telluric or stellar lines.
            }
            \label{fig:preparing_pipeline}
        \end{figure}

        In \autoref{fig:preparing_pipeline_1d} and \autoref{fig:preparing_pipeline} we show illustrations of these steps, respectively for one and all the exposures of one order. For convenience we will refer to this pipeline as "Polyfit". Note that the latter is similar to the preparing pipeline developed by e.g. \citet{Brogi2019}. We also implemented the SysRem \citep{Tamuz2005} preparing pipeline, this is detailed in \autoref{anx:sysrem_pipeline_effect}. For this pipeline or principal component analysis-based pipelines (PCA), it has been shown that the equivalent of \autoref{eq:pipeline}, applied to the forward model, can be optimised to be much more computationally efficient \citep{Gibson2022, Klein2024}. We have not investigated if an equivalent optimisation could be implemented for "Polyfit".

    \subsection{Cross-correlation setup}
        \label{subsec:ccf_setup}
        In order to test our methods, we used the cross-correlation technique to investigate the presence of the previously detected H$_2$O signal in this dataset. To that end, we used our base model of the planet's transit radius, computed as described in \autoref{subsec:spectrum_model}, as template to perform the cross-correlation analysis. This model has already been put through the same preparing pipeline as the data, following the procedures discussed in \autoref{subsec:preparing_pipeline}. This is done in order to take into account the distortion of a potential real signal produced by the analysis methods performed to remove the telluric, stellar, and instrumental contributions \citep[see, e.g.,][]{Brogi2019}. We stress that no artificial telluric or stellar lines were added to the model in this step.
        
\begin{figure}
           \centering
           \includegraphics[width=\hsize]{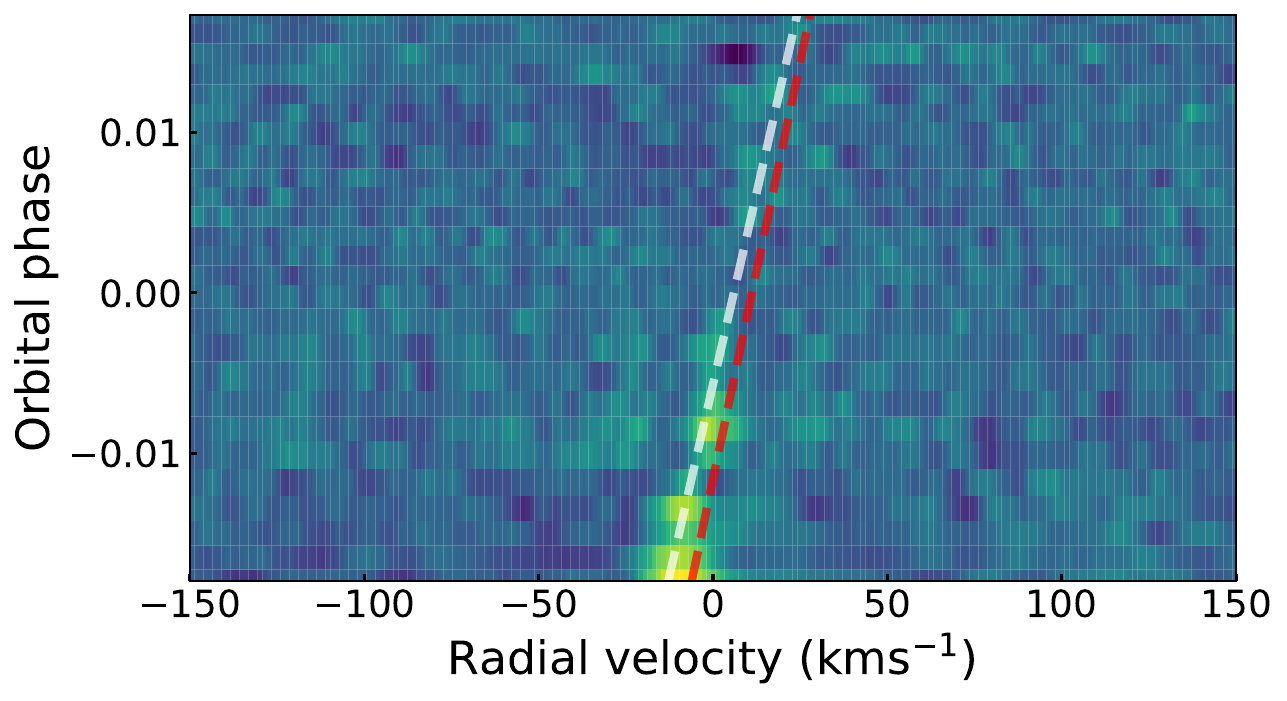}
              \caption{
                In-transit cross-correlation matrix of the prepared data with our base model in the Earth's rest frame and for the order selection of \citet{alonso2019multiple}. The horizontal axis shows the different Doppler shifts applied to the model and the vertical axis shows the exposures at different planet orbital phases. The white line represents the expected exoplanet velocities with respect to the Earth during the observations, assuming the maximum significance K$_{\text{P}}$ of the CCF analyses (166\,km$\cdot$s$^{-1}$) and the observed $-$5.2 km$\cdot$s$^{-1}$ blueshift. The red line shows the expected planet velocities with respect to the Earth considering the expected planet K$_{\text{P}}$ ($\sim$153\,km$\cdot$s$^{-1}$) and no additional sources of dynamics in the atmosphere. 
            }
            \label{fig:cc_erf}
        \end{figure}

        Consecutively, we cross-correlated the prepared data matrix $P_\mathbf{R}(\mathbf{F}) \equiv \mathbf{F}_\mathbf{R}$ with our prepared model $P_\mathbf{R}(\mathbf{M}_{\theta}) \equiv \mathbf{M}_{\theta,\mathbf{R}}$ over a wide range of exoplanet radial velocities, from $-$150\,km$\cdot$s$^{-1}$ to 150\,km$\cdot$s$^{-1}$ in steps of 1.3\,km$\cdot$s$^{-1}$ (i.e., the mean velocity step between pixels in the CARMENES NIR channel). This analysis was conducted independently for each NIR order of the instrument to inspect the individual cross-correlation matrices in a search for strong residuals. We used a conservative order selection following \citet{alonso2019multiple} and performed the cross-correlation following
        \begin{eqnarray}
            \label{eq:ccf}
        	CC(t, \varv) = \frac{\sum_i\, \left( \mathbf{F}_{\mathbf{R},i} - \langle \mathbf{F}_{\mathbf{R}} \rangle \right) \, \left( \mathbf{M}_{\theta,\mathbf{R},i}(\varv) - \langle \mathbf{M}_{\theta,\mathbf{R}} \rangle \right) } {\sqrt{\sum_i\, \left( \mathbf{F}_{\mathbf{R},i} - \langle \mathbf{F}_{\mathbf{R}} \rangle \right) ^2\sum_i\, \left( \mathbf{M}_{\theta,\mathbf{R},i}(\varv) - \langle \mathbf{M}_{\theta,\mathbf{R}} \rangle \right) ^2}},
         \end{eqnarray}
         where $i$ denotes the wavelength element, $\varv$ is a specific lag (i.e., Doppler shift applied to the model) and where we have dropped the dependencies with time and wavelength for clarity. $\langle \mathbf{F}_{\mathbf{R}} \rangle$ and $\langle \mathbf{M}_{\theta,\mathbf{R}} \rangle$ represent the mean value of the corresponding vectors.  With this, we obtain a cross-correlation matrix in the Earth's rest frame, as shown in Fig.\,\ref{fig:cc_erf}, where we expect a potential planet signal to follow the expected exoplanet velocities with respect to the Earth (V$_{\text{obs}}(t)$), which roughly ranged from $\sim$\,$-$22\,km$\cdot$s$^{-1}$ to $\sim$\,45\,km$\cdot$s$^{-1}$ during this event.

         The final cross-correlation function (CCF) is then obtained by co-adding in time the cross-correlation matrix. If a signal from HD~189733~b's atmosphere is indeed present in the data, we would expect the co-added CCF to present a significant peak if we shift the matrix to the rest-frame of the exoplanet. In order to add robustness to our analysis and as it is usual in similar works, in this step the planet's $K_p$ is assumed to be unknown, hence allowing us to explore residuals or noise structures to aid our assessment of a potential signal's significance (see \autoref{subsec:ccf_results}). For the latter, we calculated the S/N of the CCFs at each $K_p$ dividing the peak CCF value by the CCF's standard deviation, excluding a $\pm$\,15\,km$\cdot$s$^{-1}$ region around the maximum.
            
    \subsection{High-resolution retrieval setup}
        \label{subsec:hr_framework}
        \subsubsection{Taking into account the preparing pipeline}
            \label{subsubsec:account_p_r}
            Due to the use of the preparing pipeline on the data, great care must be taken to correctly setup the retrieval in order to avoid biases. Let us have a model $\mathbf{M}_\theta(t, \lambda)$ with retrieved parameters $\theta$ and data $\mathbf{F}(t, \lambda)$ with uncertainties $\mathbf{U}_\mathbf{N}(t, \lambda)$. 
            
            For the following demonstration we make no assumption regarding the specific operations performed by the preparing pipeline (i.e. it does not need to follow the steps described in \autoref{subsec:preparing_pipeline}), but we assume that it can still be described as in \autoref{eq:pipeline}: i.e. as the data Hadamard-multiplied by any matrix $\mathbf{R}$. We also assume that the uncertainties of the prepared data $\mathbf{U}_\mathbf{R}(t, \lambda)$ have been accurately estimated. If we have a model perfectly describing the planet spectra, if $\Theta$ is the true set of parameters (as in \autoref{subsec:simulated_data}), and if $\mathbf{D}$ and $\mathbf{N}$ are respectively the true deformation matrix and the true noise matrix, then the observed data can be written as:
            \begin{eqnarray}
                \label{eq:data_rewritten}
                    \mathbf{F} \equiv \mathbf{M}_{\Theta} \circ \mathbf{D} + \mathbf{N}, 
            \end{eqnarray}
            and we can write from \autoref{eq:pipeline}: 
            \begin{eqnarray}
                \label{eq:reduced_data_rewrited}
                    P_\mathbf{R}(\mathbf{M}_{\Theta} \circ \mathbf{D} + \mathbf{N}) 	&=& (\mathbf{M}_{\Theta} \circ \mathbf{D} + \mathbf{N}) \circ \mathbf{R}_\mathbf{F},
            \end{eqnarray}
            where the time- and wavelength-dependencies of all terms  are implied. Not knowing $\Theta$ nor $\mathbf{N}$ but knowing $\mathbf{D}$, we would need to compare the prepared data $P_\mathbf{R}(\mathbf{M}_{\Theta} \circ \mathbf{D} + \mathbf{N})$ with:
            \begin{eqnarray}
                \label{eq:ideal_retrieval}
                    \mathbf{M}_{\mathrm{ret}, \theta} &=& \mathbf{M}_{\theta} \circ \mathbf{D} \circ \mathbf{R}_\mathbf{F}.
            \end{eqnarray}
            As a proof, we can calculate the reduced chi squared, defined as:
            \begin{eqnarray}
                \label{eq:chi2}
                    \chi^2_\nu(f, m, \sigma) &=& \frac{1}{N} \sum_i \left( \frac{f_i - m_i}{\sigma_i} \right)^2,
            \end{eqnarray}
            where $f$, $m$ and $\sigma$ are vectors representing respectively data, a model and the uncertainties on $f$, $N$ is the number of elements in the vectors, and $i$ denotes the vector element. If \autoref{eq:data_rewritten} is true and if we use $\theta = \Theta$, we obtain $\chi^2_\nu(P_\mathbf{R}(\mathbf{F}) , \mathbf{M}_{\mathrm{ret}, \Theta}, \mathbf{U}_\mathbf{R}) \approx 1$, which is the expected value for a model well-describing the data.
            
            However, for real observations $\mathbf{D}(t, \lambda)$ is unknown. We can rewrite \autoref{eq:reduced_data_rewrited} to obtain:
            \begin{eqnarray}
                \label{eq:deformation_matrix_rewrite}
                    \mathbf{D} 	&=& \frac{P_\mathbf{R}(\mathbf{M}_{\Theta} \circ \mathbf{D} + \mathbf{N}) - \mathbf{N} \circ \mathbf{R}_\mathbf{F}}{\mathbf{M}_{\Theta} \circ \mathbf{R}_\mathbf{F}},
            \end{eqnarray}
            where the fraction bar here and from now represents a Hadamard division. Injecting this equation into \autoref{eq:ideal_retrieval}, we obtain:
            \begin{eqnarray}
                \label{eq:ideal_retrieval_rewriten}
                    \mathbf{M}_{\mathrm{ret}, \theta}    &=& \frac{\mathbf{M}_{\theta}}{\mathbf{M}_{\Theta}} \circ \left( P_\mathbf{R}(\mathbf{M}_{\Theta} \circ \mathbf{D} + \mathbf{N}) - \mathbf{N} \circ \mathbf{R}_\mathbf{F} \right).
            \end{eqnarray}
            
            As stated in \autoref{subsec:preparing_pipeline}, the goal of $P_\mathbf{R}$ is to remove $\mathbf{D}$ from the data as much as possible. Hence, an ideal preparing pipeline would be such that $\mathbf{R}_\mathbf{F} = \mathbf{1} \oslash \mathbf{D}$. We now assume that the preparing matrix $\mathbf{R}_\mathbf{F}$ of a non-ideal pipeline on data $\mathbf{F}$ can be written as:
            \begin{eqnarray}
                \label{eq:realisitc_pipeline}
                    \mathbf{R}_\mathbf{F} &=& \mathbf{A}(\mathbf{M}_{\Theta}) \oslash \mathbf{D} + \mathbf{B}(\mathbf{M}_{\Theta}, \mathbf{D}, \mathbf{N}),
            \end{eqnarray}
            where $\mathbf{A}$ and $\mathbf{B}$ are time- and wavelength-dependent matrices. The elements between parenthesis are here used only to highlight the matrices dependencies. The matrix $\mathbf{A}$\footnote{To illustrate what $\mathbf{A}$ may represent, we can use the following simplistic example. If $P_\mathbf{R}$ is the mean function, $\mathbf{M}_{\Theta}$ is constant, $\mathbf{D}$ is constant, and $\mathbf{N} = \mathbf{0}$, then $\mathbf{R}_\mathbf{F} = \mathbf{1} \oslash (\mathbf{M}_{\Theta} \circ \mathbf{D})$, with $\mathbf{A} = \mathbf{1} \oslash \mathbf{M}_{\Theta}$ (and $\mathbf{B} = \mathbf{0}$).} can depend only on $\mathbf{M}_{\Theta}$. The matrix $\mathbf{B}$ can depend on $\mathbf{M}_{\Theta}$, $\mathbf{D}$ and $\mathbf{N}$. These two matrices represent together the "imperfections" of a non-ideal preparing pipeline, that is, how it "catches" some of the signal, and how it imperfectly removes $\mathbf{D}$. A preparing pipeline effective at removing $\mathbf{D}$ would be such that $| \mathbf{B} | \ll | \mathbf{A} \oslash \mathbf{D} |$. Indeed, if this condition is not respected, then the effect of $\mathbf{D}$ on the prepared data would still be significant, and the correction would be significantly biased by $\mathbf{N}$. If we assume that this condition is respected, we can write from \autoref{eq:reduced_data_rewrited}:
            \begin{eqnarray}
            	\label{eq:realisitc_pipeline_effect}
            		P_\mathbf{R}(\mathbf{M}_{\Theta} \circ \mathbf{D} + \mathbf{N}) 	&=& \left( \mathbf{M}_{\Theta} \circ \mathbf{D} + \mathbf{N} \right) \circ \left( \mathbf{A} \oslash \mathbf{D} + \mathbf{B}(\mathbf{D}, \mathbf{N}) \right) \nonumber \\
            									        &\approx& \mathbf{M}_{\Theta} \circ \mathbf{A} + \mathbf{N} \circ \mathbf{R}_\mathbf{F},
            \end{eqnarray}
            where the dependencies on $\mathbf{M}_{\Theta}$ of $\mathbf{A}$ and $\mathbf{B}$ are implied.
            It follows that, if there is no deformation matrix nor noise, i.e. if we replace $\mathbf{D}$ by a matrix of ones and $\mathbf{N}$ by a matrix of zeros, using \autoref{eq:realisitc_pipeline_effect} and the fact that $\mathbf{A}$ depends only on $\mathbf{M}_\Theta$:
            \begin{eqnarray}
                \label{eq:realisitc_pipeline_effect_m1}
                    P_\mathbf{R}(\mathbf{M}_{\Theta}) &=& \mathbf{M}_{\Theta} \circ \left( \mathbf{A} + \mathbf{B}(\mathbf{1}, \mathbf{0}) \right) \nonumber\\
                                    &\approx& \mathbf{M}_{\Theta} \circ \mathbf{A} \nonumber\\
                                    &\approx& P_\mathbf{R}(\mathbf{M}_{\Theta} \circ \mathbf{D} + \mathbf{N}) - \mathbf{N} \circ \mathbf{R}_\mathbf{F},
            \end{eqnarray}
            where the dependencies on $\mathbf{M}_{\Theta}$ of $\mathbf{A}$ and $\mathbf{B}$ are implied. Injecting \autoref{eq:realisitc_pipeline_effect_m1} into \autoref{eq:ideal_retrieval_rewriten}, we obtain:
            \begin{eqnarray}
            	\label{eq:retrieval_true_parameters}
            		\mathbf{M}_{\mathrm{ret}, \theta}(t, \lambda) &\approx \frac{\mathbf{M}_{\theta}}{\mathbf{M}_{\Theta}} \circ P_\mathbf{R}(\mathbf{M}_{\Theta}).
            \end{eqnarray}
            In a real case, we obviously do not know the true set of parameters $\Theta$, but we can make the assumption that the retrieval best fit will correspond to a good approximation of $\Theta$ -- which is true if the model accurately represents the physics of the data. Thus, we can write:
            \begin{eqnarray}
                \label{eq:actual_retrieval}
                    \mathbf{M}_{\mathrm{ret}, \theta} (t, \lambda) &\approx& P_\mathbf{R}(\mathbf{M}_{\theta}) \\ 
                        &\approx& \mathbf{M}_{\theta} \circ \mathbf{R}_{\mathbf{M}_{\theta}}. \nonumber
            \end{eqnarray}
            
            To summarize, if a preparing pipeline can be written as in \autoref{eq:pipeline} and has small residuals coming from $\mathbf{D}$ and $\mathbf{N}$ -- which is arguably a desirable property of preparing pipelines --, in other words if it respects the following condition, using the notation of \autoref{eq:reduced_data_rewrited}:
            \begin{eqnarray}
                \label{eq:bpm}
                    \mathbf{1} - \frac{P_\mathbf{R}(\mathbf{M}_{\Theta}) + \mathbf{N} \circ \mathbf{R}_\mathbf{F}}{P_\mathbf{R}(\mathbf{F})} &\approx& \mathbf{0},
            \end{eqnarray}
            then the optimal retrieval model is given by \autoref{eq:actual_retrieval}, that is, applying the preparing pipeline directly on the forward models. This condition can easily be tested on synthetic data, for which $\Theta$ is known, using any realistic $\mathbf{D}$ and $\mathbf{N}$. We will call the values given by the left term of this equation "Bias Pipeline Metric" (BPM)\footnote{In the case of SysRem, the above demonstration does not apply and the BPM cannot be calculated because its effect cannot be written as in \autoref{eq:pipeline}.}. The BPM can be seen as analogous to $\chi^2_\nu$ for preparing pipelines, especially if we compute its absolute mean. A $|\langle\mathrm{BPM}\rangle|$ closer to 0 is more desirable, so the BPM can be a tool to compare preparing pipelines following \autoref{eq:pipeline} with each other.
        
        \subsubsection{Log-likelihood function and retrieval algorithm}
        \label{subsubsec:log_l}
        In order to retrieve the parameters $\theta$ (such as the temperature profile, species abundances, etc.) from the data, we use the multimodal nested sampling algorithm \lstinline|MultiNest| \citep{Feroz2009} and its Python wrapper \lstinline|PyMultiNest| \citep{Buchner2014}. This algorithm explores the parameter space with coordinates $\theta$ defined by priors, and uses a log-likelihood function $\ln(\mathcal{L})$ to compare the model to the data and estimate which set of parameter (the posteriors) best fit the data according to Bayesian statistics. For high-resolution retrievals, \cite{Brogi2019, Gandhi2019, Gibson2020, Gibson2022} make the CCF appear in the classical log-likelihood function. However, the versions of this function with and without the CCF term are strictly equivalent from a mathematical standpoint \citep[see, e.g.][]{Gibson2020, Gibson2022}, and none have a significant advantage over the other from a computational cost standpoint. The CCF is thus not necessary for high-resolution retrievals. This is well known, but often not sufficiently highlighted by the authors who make the choice to use the CCF version of the log-likelihood function. Assuming Gaussian noise, we follow \cite{Gibson2020} and use one of their intermediate steps while dropping constant terms. \cite{Gibson2020} also uses two scaling parameters $\alpha$ and $\beta$, respectively for the model and the uncertainties. We assume that our model correctly parameterises the atmosphere's scale height, so we set $\alpha$ to 1. To fit for $\beta$, we need to assume that our model is necessarily correct, and that the uncertainties may be uniformly incorrect over orders, exposures and wavelengths. Given the relative simplicity of our model, we cannot reasonably assume that it is necessarily correct. Moreover, we observed that the CARMENES uncertainties may be slightly overestimated, and not uniformly so (see \autoref{subsec:uncertainties_check}). Hence, we also chose to fix $\beta$ to 1 (we show results retrieving $\beta$ in \autoref{anx:marginalizing_the_uncertainties}). From the demonstration in \autoref{subsubsec:account_p_r}, we use:
        \begin{eqnarray}
            \label{eq:log_likelihood}
                \ln(\mathcal{L}) &=& - \frac{1}{2} \sum \left( \frac{P_\mathbf{R}(\mathbf{F}) - P_\mathbf{R}(\mathbf{M}_\theta)}{\mathbf{U}_\mathbf{R}} \right)^2,
        \end{eqnarray}
        where the sum is over every non-masked element of the matrices (orders, exposures, wavelengths)\footnote{For comparison with a selection of other works, and neglecting constant terms and factors, $P_\mathbf{R}(\mathbf{M}_\theta)$ in \autoref{eq:log_likelihood} is replaced by $P_\mathbf{R}(\mathbf{M}_\theta \circ \mathbf{F}) \oslash P_\mathbf{R}(\mathbf{F})$ in \citet{Brogi2019}, by $P_\mathbf{R}(\mathbf{M}_\theta \circ \mathbf{F}) - P_\mathbf{R}(\mathbf{F})$ in \citet{Brogi2023} (L. Pino, private communication), and by $P_\mathbf{R}(\mathbf{M}_\theta \oslash \mathbf{R}_{\mathbf{F}})$ in \citet{Pelletier2021}.}. Note that $P_\mathbf{R}(\mathbf{F})$ and $\mathbf{U}_\mathbf{R}$ need to be calculated only once, while $P_\mathbf{R}(\mathbf{M}_\theta)$ is calculated at every log-likelihood evaluation, with every steps of $P_\mathbf{R}$ performed on $\mathbf{M}_\theta$. From \autoref{subsubsec:account_p_r}, if the model $\mathbf{M}_\theta$ is accurate, the parameters are not degenerate, the uncertainties $\mathbf{U}_\mathbf{R}$ correctly evaluated and if \autoref{eq:bpm} is respected, then using \autoref{eq:log_likelihood} should in principle guarantee an unbiased retrieval.

        Our \lstinline|PyMultiNest| retrievals use 100 live points, an evidence tolerance of 0.5 -- which is the recommended value -- and a sampling efficiency of 0.8, which is recommended for parameter estimation\footnote{\url{https://github.com/farhanferoz/MultiNest/blob/master/README.md}. A sampling efficiency of 0.3 is recommended for evidence evaluation, however we did not find a significant difference between the log-evidence computed with these two parameter values.}. We used the non-constant efficiency mode of \lstinline|MultiNest|, which is significantly slower than the constant efficiency mode but more accurate in constraint and evidence estimation \citep[see e.g.][]{Chubb2022}.

    \subsection{Exposure selection}
        \label{subsec:exposure_selection}
        The planet total transit duration is given in \autoref{tab:general_parameters}. Our observations span nearly 12$\,$000 s, meaning that a bit less than half of our exposures contain no transmission spectra from the planet. We can thus optimise the speed of our retrievals significantly by selecting only the relevant exposures. If we knew $T_0$ and $T_{14}$ perfectly well, we could just select the exposures corresponding to the planet's total transit. While we know these two parameters to a precision of $\approx 10$ s, the CARMENES time stamps are only given "for guidance"\footnote{M. Zechmeister, private communication}. Hence, we chose to retrieve the offset of $T_0$ -- that we will abbreviate into $T_0$ for convenience from now -- using a prior spanning $\pm 300$ s in our retrievals to provide us with a higher precision. We selected the exposures such that all possible total transits for all possible $T_0$ values are encompassed in the selection. This corresponds to 26 exposures, from the CARMENES times $2\,458\,004.385018$ to $2\,458\,004.463298$ BJD$_\textrm{TDB}$ (day).
    
    \subsection{Order selection}
        \label{subsec:order_selection}
        Not all of the collected CARMENES data contain valuable information. Using some orders might lead to biased results due the combination of the low mean Earth's transmittance in them, imperfections in our modelling of physical phenomena and in our preparing pipeline, as well as unaccounted noise sources. The latter includes unpredictable changes in the instrumental response, in the telluric lines due random changes in the observing conditions, correlated noise, etc. This is true for a CCF analysis, but even more so for a retrieval analysis: these imperfections can significantly bias the retrieved values. In order to perform our retrievals in good conditions, we have to perform an order selection with the goal of maximising the extraction of useful information and to minimise biases. 
        
        A conservative approach would be to simply use all the available data. However, doing so for our selected exposures gives a maximum significance CCF peak (following \autoref{subsec:ccf_setup}) at $K_p = 97$ km$\cdot$s$^{-1}$ and $V_{\textrm{rest}} = -2.6$ km$\cdot$s$^{-1}$. For comparison we expect $K_p \approx 153$ km$\cdot$s$^{-1}$ and $V_{\textrm{rest}} \approx -4$ km$\cdot$s$^{-1}$, as reported by, for example, by \citet{sanchez2019water}. In contrast, we retrieved sensible values with our Polyfit retrieval (see \autoref{anx:retrieval_with_all_orders}), suggesting that retrievals with our framework are more stable than CCF analysis to this kind of perturbation.
        
        We also tried several variations of approaches based on using the CCF S/N to detect a synthetic signal injected into the data, order by order. The orders selected by these approaches often corresponded to the one with the most important telluric contamination, which is the opposite of the desired behaviour. We also developed an algorithmic order selection (see \autoref{anx:order_selection_algorithm}), but it seems prone to biases and we decided to not use it. We note that more detailed works on this issue \citep{Cheverall2023, Debras2023} reached a similar conclusion.

        While we could have performed our analysis with retrievals on all orders, we ultimately chose to use the \citet{alonso2019multiple} and \citet{sanchez2019water} selection to be able to compare our results with our own CCF analysis as well as to these previous works. Following their approach, we removed orders 18 to 21 and 36 to 41, due to the very strong telluric water absorptions present at these wavelengths. This results in a CCF peak at $K_p = 166$ km$\cdot$s$^{-1}$ and $V_{\textrm{rest}} = -5.2$ km$\cdot$s$^{-1}$ (see \autoref{subsec:ccf_results}), which is consistent with their results, given the error bars. We thus considered this selection as satisfactory and used it from here.
    
    \subsection{Uncertainties check}
        \label{subsec:uncertainties_check}
        From the setup described in \autoref{subsec:hr_framework} and \autoref{eq:log_likelihood}, there are three ways our retrieval could fail at retrieving the planet atmospheric parameters:
        \begin{enumerate}
            \item Using an improper model, either because of a parameterisation leading to biases, incorrect physics, or inaccurate line lists.
            \item Having residuals in the prepared data due to an imperfect preparing pipeline.
            \item Using an inaccurate estimation of the uncertainties.
        \end{enumerate}
        The line lists and physical models we are using are state-of-the-art, although they may be improvable, which is beyond the scope of this paper. We discussed our methodology to limit and test parameterisations leading to biases in \autoref{subsec:exposure_selection} and \autoref{subsec:validation}. Because our pipeline is perfectible, we expect residuals to be left in the prepared data. We discussed our methodology to limit residuals in \autoref{subsec:order_selection}. We are thus left to check the accuracy of our data uncertainties.

        To estimate if the uncertainties we are using are accurate, we calculate the order- and time-dependent standard deviations along wavelength of $P_\mathbf{R}(\mathbf{F})$ and compare them with the order- and time-dependent average of $\mathbf{U}_\mathbf{R}$ along wavelength. In the case of Gaussian noise, these two values should roughly be equal, any deviation may be a hint for an under- or overestimation of the uncertainties. We found that, on average and considering only the selected exposures and our order selection, $\mathbf{U}_\mathbf{R}$ was overestimated by a factor $k_\sigma \approx$ 1.15\footnote{$k_\sigma$ may be seen as the inverse of the $\beta$ scaling parameter (see \autoref{subsubsec:log_l}).} compared to the standard deviations of $P_\mathbf{R}(\mathbf{F})$. We do not observe such discrepancy when doing the same comparison with simulated noisy observations: for the retrieval described in \autoref{anx:retrieval_on_simulated_data_with_simulated_noise}, we obtained $k_\sigma = 0.996$. The origin of this apparent overestimation is probably linked to the CARACAL pipeline. We did not investigate this further, chose to be conservative, and used the CARMENES uncertainties ($\mathbf{U}_\mathbf{N}$) without modification in our retrievals. As a consequence, the constraints we will establish may be less precise than what they could optimally be. However, when evaluating the $\chi^2_\nu$ of the data against our retrieved models, we will correct the uncertainties by this factor. This is nevertheless an imperfect way of doing so, as $k_\sigma$ may vary with order and exposure.
        
    \subsection{High-resolution retrieval setup validation}
        \label{subsec:validation}
        \subsubsection{BPM}
            \label{subsubsec:validation_bpm}
            \begin{deluxetable}{l l c}
                \tablecaption{\label{tab:prt_base_model_mmr}Mass mixing ratios used for our simulated data and equivalent volume mixing ratios (VMR).}  
                \tablehead{\colhead{Species} & \colhead{MMR} & \colhead{$\log_{10}$(VMR)}}
                \startdata
                CH$_4$		& $3.4\times10^{-5}$	& $-5.3$ \\
                CO			& $1.8\times10^{-2}$	& $-2.8$ \\
                H$_2$O		& $5.4\times10^{-3}$	& $-3.2$ \\
                H$_2$S		& $1.0\times10^{-3}$	& $-4.2$\\
                HCN		    & $2.7\times10^{-7}$	& $-7.6$\\
                NH$_3$		& $7.9\times10^{-6}$	& $-6.0$\\
                \enddata
            \end{deluxetable}
    
            \begin{figure}
                \centering
                \includegraphics[width=\hsize]{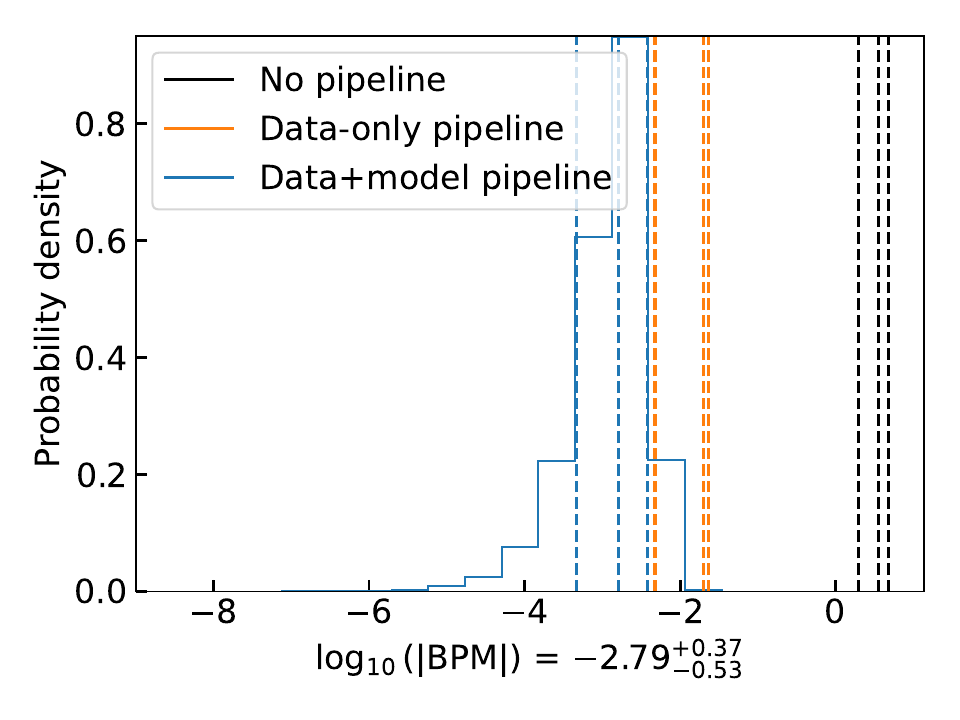}
                  \caption{
                    Probability density distribution of the decimal logarithm of the absolute value of the BPM of different setups with our order selection. Black: BPM without pipeline ($\mathbf{R}_\mathbf{F} = 1$). Orange: BPM when applying the pipeline only to the data. Blue: standard BPM (\autoref{eq:bpm}). The vertical lines represents the 0.16, 0.50 and 0.84 quantiles of the distributions.
                }
                \label{fig:BPM}
            \end{figure}
            
            Our nominal petitRADTRANS model is built following the steps described in \autoref{subsec:spectrum_model} and the parameters listed in \autoref{tab:general_parameters} and \autoref{tab:prt_base_model_mmr}. We used the aforementioned exposure selection and the order selection described in \autoref{subsec:order_selection}. Our simulated noisy data are built in the same way, following in addition the steps described in \autoref{subsec:simulated_data}, and using $\mathbf{U}_\mathbf{N}(t,\lambda)$ to build the noise matrix. We first test if our preparing pipeline respects the condition described by \autoref{eq:bpm}. Using our noisy simulated data, we obtain an absolute mean BPM of $2.75 \times 10^{-5}$ over our $2\,319\,330$ selected data points for that specific noise realisation. The maximum of the BPM absolute value is $3.42 \times 10^{-2}$. The distribution of our BPM values is represented in \autoref{fig:BPM}. For comparison, when applying no pipeline, i.e. when we replace $\mathbf{R}_\mathbf{F}$ in \autoref{eq:bpm} by $\mathbf{1}$, we obtain an absolute mean BPM of $122$ (with an absolute median of $3.51$). When applying the pipeline only to the data (i.e. $P_\mathbf{R}(\mathbf{M}_\theta)$ replaced by $\mathbf{M}_\theta$ in \autoref{eq:bpm}), we obtain an absolute mean BPM of $1.60 \times 10^{-2}$. 
            
            While this results are encouraging -- the mean BPM of our setup is several orders of magnitude lower than the BPM without pipeline, and since the mean BPM is arguably close to 0 --, it is difficult to draw a conclusion on the validity of "Polyfit" solely from the BPM. Indeed, this indicator is useful for comparisons, but we lack values from other trusted preparing pipelines following \autoref{eq:pipeline}, which is not the case of e.g. SysRem. We develop this more in \autoref{subanx:bpm}.

        \subsubsection{Simulated retrievals}
            \label{subsubsec:validation_simulated_retrievals}
            
\begin{figure*}
               \centering
                \includegraphics[width=\hsize]{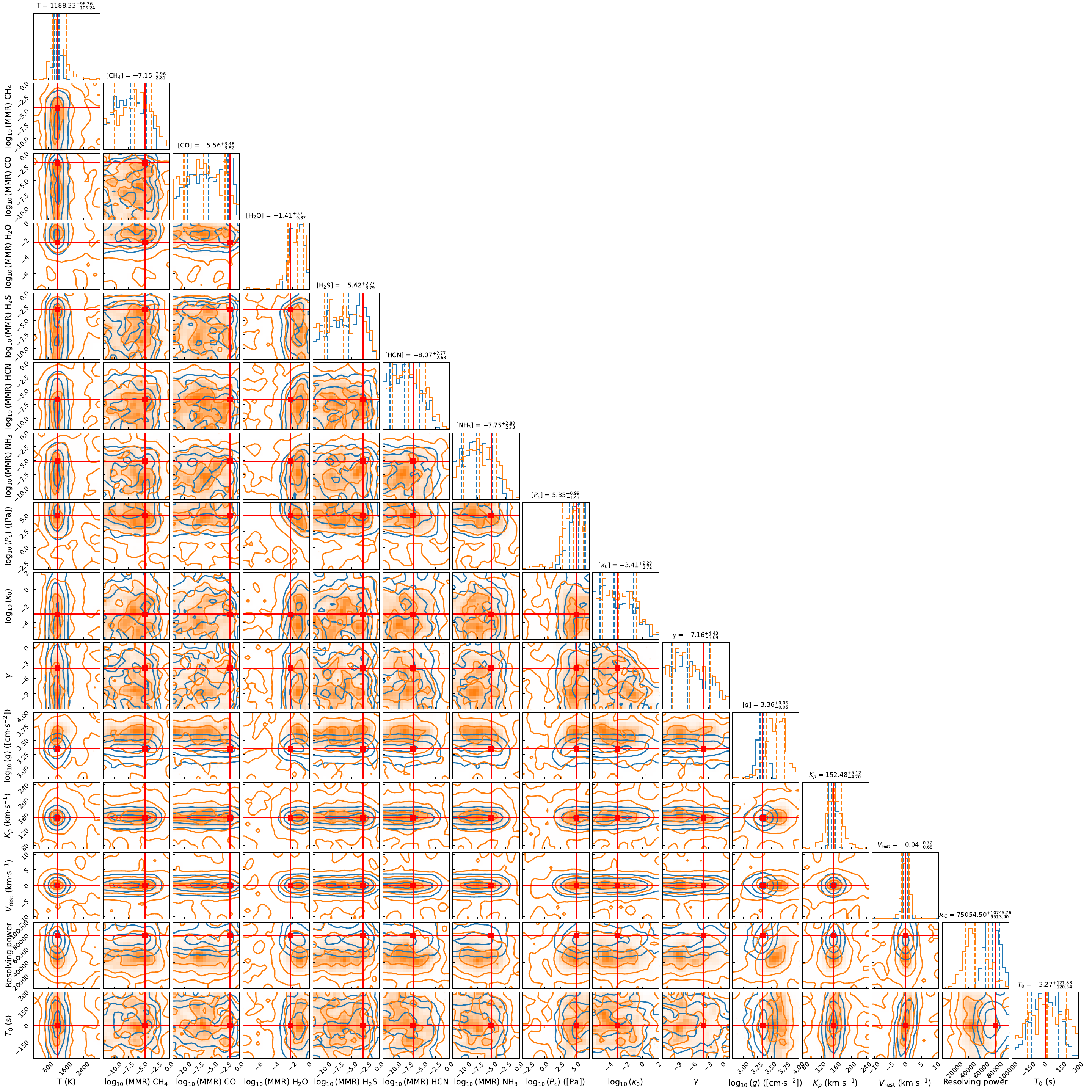}
                \caption{Posterior probability distributions of our simulations. Blue: pRT 1-D simulation using Polyfit. Orange: pRT-Orange 3-D simulation using Polyfit. 
                The solid red lines corresponds to the model input values for the 1-D model. The dashed vertical blue lines represents the 0.16, 0.50 and 0.84 quantiles (i.e. median and 1-$\sigma$ error bar) of the distributions for the 1-D simulation. On top of each column is indicated the median and the 1-$\sigma$ error bar of the retrieved parameters for the 1-D simulation. The contours in the 2D histograms corresponds to the 1, 2 and 3-$\sigma$ contours.}
                \label{fig:expected_retrieval}
            \end{figure*}

            \paragraph{Setup} To confirm the BPM results, we perform a retrieval with our framework on noiseless simulated data. That is, we take $\mathbf{N}(t, \lambda)$ in \autoref{eq:simulated_data} as a matrix of zeros, but we keep the uncertainties as $\mathbf{U}_\mathbf{N}(t,\lambda)$. The reason for setting the noise to $\mathbf{0}$ is that we are interested in estimating the retrieval setup bias, not the bias introduced by a random realisation of the noise. Typically, we would need to run many retrievals with different $\mathbf{N}(t, \lambda)$ realisations so as to isolate and neglect the effect of noise. However, using our noiseless simulated data provides us directly with the desired results. This was also highlighted by \citet{Feng2018} (see their Section 5.2), who shown that the combination of a large number of noisy simulated data retrievals converges to the posteriors obtained from the retrieval of noiseless simulated data. In order to obtain a rough estimation of biases introduced by 3-D effects, we also retrieve the pRT-Orange 3-D model described in \autoref{subsec:3d_transmission_spectrum_model} with the same setup as for our simulated 1-D model. The setup and results are compiled in \autoref{tab:simulation_retrieval}, and resulting posterior probability distributions are shown in \autoref{fig:expected_retrieval}.

            \begin{deluxetable}{l l l l}
            	\tablecaption{\label{tab:simulation_retrieval}Setup and results of our noiseless 1-D simulated data retrieval with Polyfit}	
            	\centering	 
            	\tablehead{\colhead{Parameter} & \colhead{Prior} & \colhead{Posterior} & \colhead{Input}}
                \startdata
            	$T$ (K)											& $\mathcal{U}(100,\,4000)$					 & $1188^{+106}_{-96}$				& 1209			\\
            	$[$CH$_4$]										& $\mathcal{U}(-12,\,0)$					 & $-7.15^{+2.81}_{-2.96}$			& -4.47			\\
            	$[$CO]											& $\mathcal{U}(-12,\,0)$					 & $-5.56^{+3.48}_{-3.82}$			& -1.74			\\
            	$[$H$_2$O]										& $\mathcal{U}(-12,\,0)$					 & $-1.41^{+0.71}_{-0.87}$			& -2.27			\\
            	$[$H$_2$S]										& $\mathcal{U}(-12,\,0)$					 & $-5.62^{+2.77}_{-3.79}$			& -3.00			\\
            	$[$HCN]											& $\mathcal{U}(-12,\,0)$					 & $-8.07^{+2.77}_{-2.63}$			& -6.57			\\
            	$[$NH$_3$]										& $\mathcal{U}(-12,\,0)$					 & $-7.75^{+2.80}_{-2.73}$			& -5.10			\\
            	$\log_{10}(P_c)$ [Pa]							& $\mathcal{U}(-5,\,7)$						 & $5.35^{+0.99}_{-1.43}$			& 5.00			\\
            	$\log_{10}(\kappa_0)$							& $\mathcal{U}(-6,\,2)$						 & $-3.41^{+2.29}_{-1.72}$			& -3.00			\\
            	$\gamma$										& $\mathcal{U}(-12,\,1)$					 & $-7.16^{+4.43}_{-3.09}$			& -4.00			\\
            	$\log_{10}(g)$ [cm$\cdot$s$^{-2}$]				& $\mathcal{U}(2.5,\,4.0)$					 & $3.36^{+0.06}_{-0.06}$			& 3.35			\\
            	$K_p$ (km$\cdot$s$^{-1}$)						& $\mathcal{U}(70,\,250)$					 & $152.48^{+5.12}_{-4.70}$			& 152.53		\\
            	$V_{\mathrm{rest}}$ (km$\cdot$s$^{-1}$)			& $\mathcal{U}(-20,\,20)$					 & $-0.04^{+0.72}_{-0.68}$			& 0.00			\\
            	$\mathcal{R}_{\mathrm{C}}$						& $\mathcal{U}(10^3,\, 10^5)$				 & $75\,100^{+10\,700}_{-9\,500}$	& $80\,400$		\\
            	$T_0$ (s)										& $\mathcal{U}(-300,\,300)$					 & $-3^{+122}_{-121}$				& 0				\\
                \enddata
            	\tablecomments{
            	$\mathcal{U}(x_1, x_2)$ denotes a uniform prior: a prior with a constant positive probability density between its boundaries $x_1$ and $x_2$, and a probability density $= 0$ outside of its boundaries. The values for $T_0$ are given relative to its value in \autoref{tab:general_parameters}.
            }
            \end{deluxetable}

            \paragraph{1-D results} For the retrievals on the 1-D simulated data, the probability distributions for most of the retrieved parameters peaks at values very close to the true model values. It can be noted that the H$_2$O abundance posterior's median is roughly 1-sigma away from the true value. This is mainly caused by the effect of other retrieved parameters combined with our preparing pipeline's imperfection: the simulated data are deformed by $\mathbf{D}$, while the forward models are not. Hence, a prepared forward model with the true parameters will not be exactly equal to the prepared simulated data. In \autoref{anx:retrieval_on_noisless_simulated_data_with_fewer_parameters}, we retrieve the same simulated data with less free parameters. In that case, the H$_2$O abundance posterior peak is not significantly shifted.

\begin{figure}
                \centering
                \includegraphics[width=\hsize]{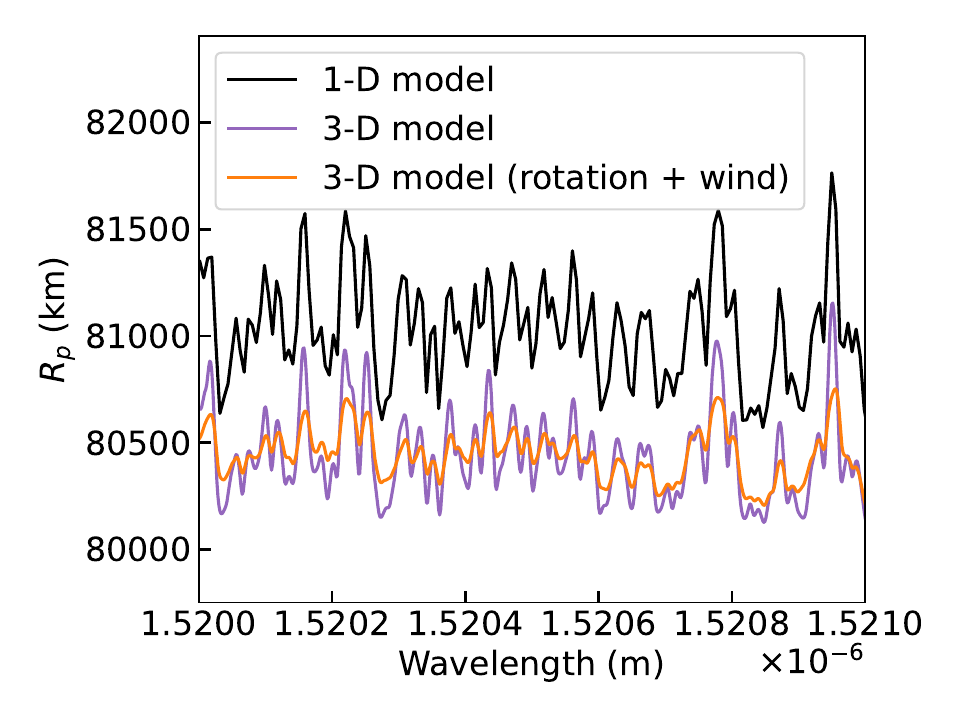}
                  \caption{
                    Comparison between our different pRT models. Black: 1-D model described in \autoref{subsec:spectrum_model}. Purple: 3-D pRT-Orange model without rotation and winds. The 3-D spectrum is different from the 1-D spectrum because of the former has 3-D geometry with varying temperature profiles and uses equilibrium chemistry. Orange: 3-D pRT-Orange model including rotation and winds. While we observe some shifts in the lines when only including rotation, the GCM's strongly super-rotating equatorial jet separates these into two small lines left and right from the central line position, where the blueshifted one is slightly stronger as it corresponds to the hotter morning terminator. A dominating central peak remains in the middle, corresponding to the higher latitudes, which are above the equatorial jet and therefore dominated by the planet's rotation.
                }
                \label{fig:3d_model_comparison}
            \end{figure}

\begin{figure}
                \centering
                \includegraphics[width=\hsize]{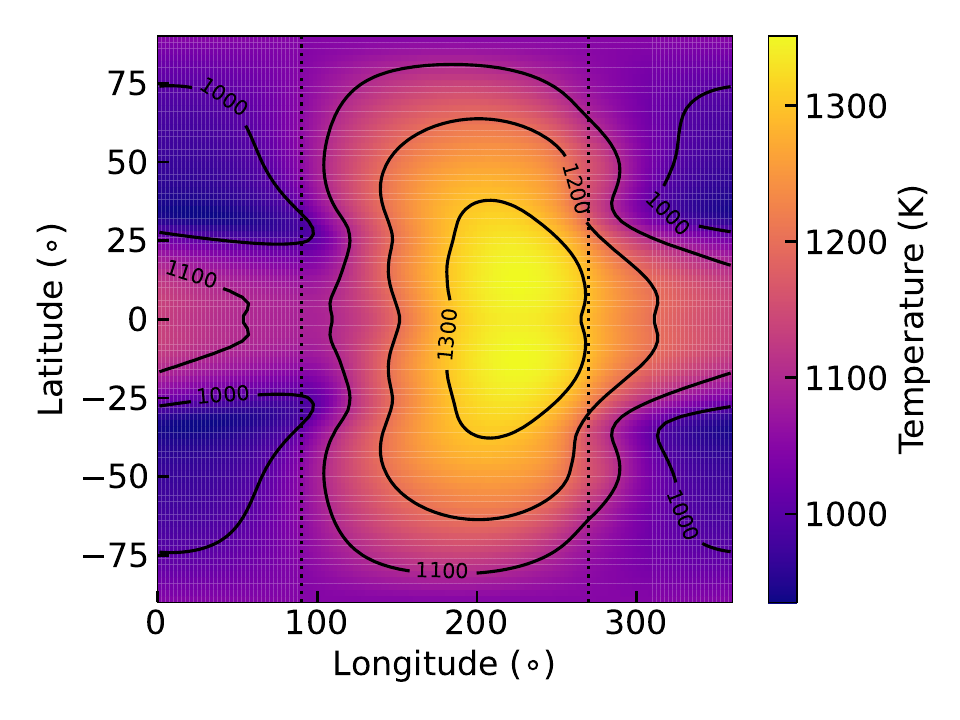}
                  \caption{
                    Temperatures of the 3-D model used \citep{drummond2018} at a pressure of $10^4$ Pa, corresponding roughly to the pressure of the maximum average contribution for this CARMENES dataset. The dotted lines corresponds to the leading (or morning) terminator and to the trailing (or evening) terminator. In the pRT-Orange model used, the longitude-dependent temperatures are set, at all latitudes, to the corresponding equatorial (latitude = $0^\circ$) temperatures of this model.
                }
                \label{fig:3d_model_temperature}
            \end{figure}

            \paragraph{3-D results} For the retrievals on the 3-D simulated data, we can make several remarks:
            \begin{itemize}
                \item The resolving power is biased towards lower values, while the surface gravity is biased towards higher ones. This is caused by the planet rotation and winds, which deform the line shapes. This effect is displayed in \autoref{fig:3d_model_comparison}. We note that in this configuration, we did not retrieve a lower temperature nor a higher H$_2$O MMR as it was highlighted by \citet{MacDonald2020}. In their case it was caused by an asymmetry in temperature and composition at the terminators, with most of the effect coming from the difference in composition. Our 3-D model partially reproduces this asymmetry (see \autoref{fig:3d_model_temperature}), but the difference in H$_2$O MMR may be too small for the effect to be significant.
                \item The posteriors are overall wider, probably because the 1-D model is not able to fit as accurately the spectral features deformed by the 3-D effect.
                \item The simulated winds and planet rotation are not translated in any significant shift in $V_{\mathrm{rest}}$. This is due to the combined effect of rotation and winds cancelling-out in this model. It thus cannot reproduce the rest velocity blueshifts previously observed.
                \item Otherwise, the other posteriors look similar to the one obtained with the 1-D simulated model.
            \end{itemize}

            \paragraph{Posterior analysis} Some comments can be made on the shape of some posteriors, which are not all Gaussian-like. For the CH$_4$, CO, H$_2$S, HCN, and NH$_3$ MMRs, this is due to a combination of low sensitivity of the spectral shape to these parameters (the selected wavelength coverage cover no or weak lines of these species) and the sensitivity of the model to the atmospheric mean molar mass. Beyond a given abundance, some species lines may start to imprint distinguishable features on the spectra, while the effect was inconclusive at lower abundances. Even if a species lines have a negligible effect on the spectrum, increasing this species MMR will in turn increase the mean molar mass, which has an impact on the spectrum. Hence, we obtain uniform posterior probability distributions with a sharp cut-off near the true species MMR value. This is very similar for the cloud deck pressure: a cloud deck pressure below $\approx$ $10^4$ Pa will have a negligible effect on the spectral shape, but stops being negligible above that value. The explanation is again the same for $\kappa_0$ and $\gamma$, but these two parameters are highly correlated: a high haze opacity is possible only with a steep spectral slope in order to avoid a strong effect on the spectral shape.
            
            Note that we did not retrieve the planet radius, which is often a free parameter in low resolution transmission spectra analysis. In transmission spectroscopy, the planet radius has two main effects: offset the transit radius and change the scale height (by affecting the rate of change of $g$ in the atmosphere), so the lines amplitude. However, our preparing pipeline "normalise" the spectrum, so the offsetting effect of the planet radius is negated. The amplitude effect remains, but large variations of radius are necessary to affect this property significantly.

            We note that \citet{Debras2023} also performed a posterior analysis using simulated HD 189733 b SPIRou transmission spectra, which covers a similar spectral range at a similar resolving power. They however used different techniques than ours to build the simulated data and prepare them, as well as a different log-likelihood function. Their results notably show posterior peaks more than 1-$\sigma$ away from the truth in some cases. This includes the H$_2$O abundance posterior, which seems biased toward lower values. Due to \citet{Debras2023} choice to inject their synthetic model into real data, it is unclear if these biases arise from the technique used or are noise-induced.

            \paragraph{Validation summary} To summarise, our setup is in principle essentially unbiased assuming that our model accurately describe the data. If this assumption holds we should be able to constrain $\mathcal{R}_{\mathrm{C}}$, $K_p$, $V_r$, $g$, $T$, as well as the H$_2$O MMR. We don't expect to be able to have a significant detection of the other tested molecules as well as on the cloud and haze parameters. However, our 1-D model probably misrepresents some 3-D effects. Using our pRT-Orange simulated 3-D retrieval results, we can expect a bias of $\approx -2 \sigma$ for $\mathcal{R}_C$. Due to the relative simplicity of pRT-Orange however, we expect to have missed or underestimated some 3-D-induced biases (e.g., the effect of clouds and their patchiness, or the impact of the latitude-dependent atmospheric structure).

\section{Results}
    In this section we present the results of our CCF analysis and retrieval framework for the CARMENES data described in \autoref{subsec:observations}.
    \subsection{Cross-correlation}
        \label{subsec:ccf_results}
\begin{figure}
           \centering
           \includegraphics[width=\hsize]{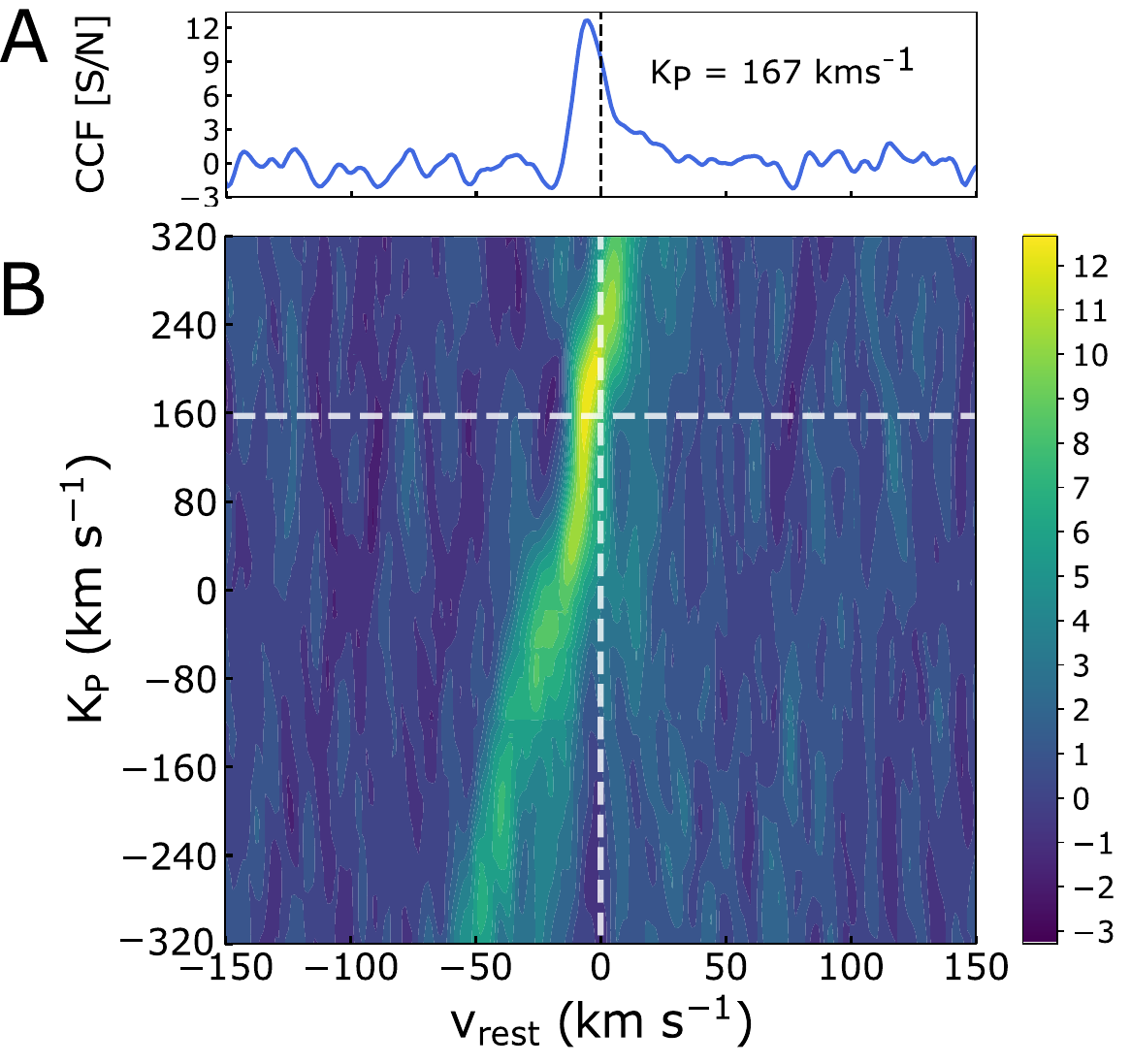}
              \caption{
                Results of the cross-correlation analysis for the order selection of \citet{alonso2019multiple} using our base model. A: Slice through the maximum significance CCF (166 km$\cdot$s$^{-1}$) showing the detected CCF peak. B: Cross-correlation map (expressed as S/N) of potential signals for H$_2$O with respect to the exoplanet rest-frame velocity (horizontal axis) and $K_p$ (vertical axis). The horizontal dashed line marks the expected $K_p$ of the planet ($\approx$153 km$\cdot$s$^{-1}$). The vertical dashed line marks the rest-frame velocity of HD~189733~b when no additional sources of dynamics are considered (i.e., 0 km$\cdot$s$^{-1}$).
            }
            \label{fig:ccf_results}
        \end{figure}

        The resulting S/N map as a function of K$_{\text{P}}$ and V$_{\mathrm{rest}}$ (the latter obtained as V$_{\mathrm{obs}}$ for each K$_{\text{P}}$ value) is shown in \autoref{fig:ccf_results}. A slice through the K$_{\text{P}}$ of the maximum significance 1-D-CCF peak is also shown. The map reveals a region of high cross-correlation values, peaking at a S/N of $\approx$\,12.4 for a K$_{\text{P}}$ of $167 \pm 34$ km$\cdot$s$^{-1}$ and a $V_{\mathrm{rest}} = -5.2 \pm 2.6$ km$\cdot$s$^{-1}$. The uncertainties were determined by a reduction of one in the S/N of the obtained 1-D CCF.

        This result is consistent with the analyses of this dataset presented in \citet{alonso2019multiple} and \citet{sanchez2019water}, considering the significantly different telluric corrections and exposure selection. That is, the expected $K_p$ of HD~189733~b, computed from the literature ($152.1 \pm 2.9$ km$\cdot$s$^{-1}$) is within the uncertainty intervals of our highest-significance signal and the CCF peak position is consistent with the previously reported blueshift, suggesting winds flowing from the planet's dayside to its nightside. We note that the S/N map shows a trace of high cross-correlation that spans to low $K_p$ values. This is caused by two regions of high values of cross correlation appearing at orbital phases between approximately $-$0.018 and $-$0.005 (see \autoref{fig:cc_erf}). 

        Although these regions align well with the expected exoplanet trail in the first half of the transit, we cannot discard the possibility of these values being affected by telluric residuals, as the exoplanet velocities with respect to the Earth at these times are low. As a safety test, we run our CCF analysis excluding all spectra at orbital phases below $-$0.01 and found the exoplanet signal persists, at a S/N of $\sim$8, a $K_p$ of $\sim 148$\,km$\cdot$s$^{-1}$ and a $V_{\mathrm{rest}}$ about $-6.5$\,km$\cdot$s$^{-1}$, with less telluric residuals in the $K_p - V_{\mathrm{rest}}$ map. The lower significance we obtained in this case was due to avoiding the region where the exoplanet signal with respect to the Earth is the lowest, so this measurement is in principle less affected by overlapping telluric H$_2$O lines. However, we also note that the highest S/N exposures and lower target airmasses also occurred at these phases of the early transit, so excluding these spectra from the analysis also potentially removed some exoplanet contributions.
        
    \subsection{Retrievals}
        \label{subsec:retrievals}
        
\begin{figure*}
            \centering
            \includegraphics[width=\hsize]{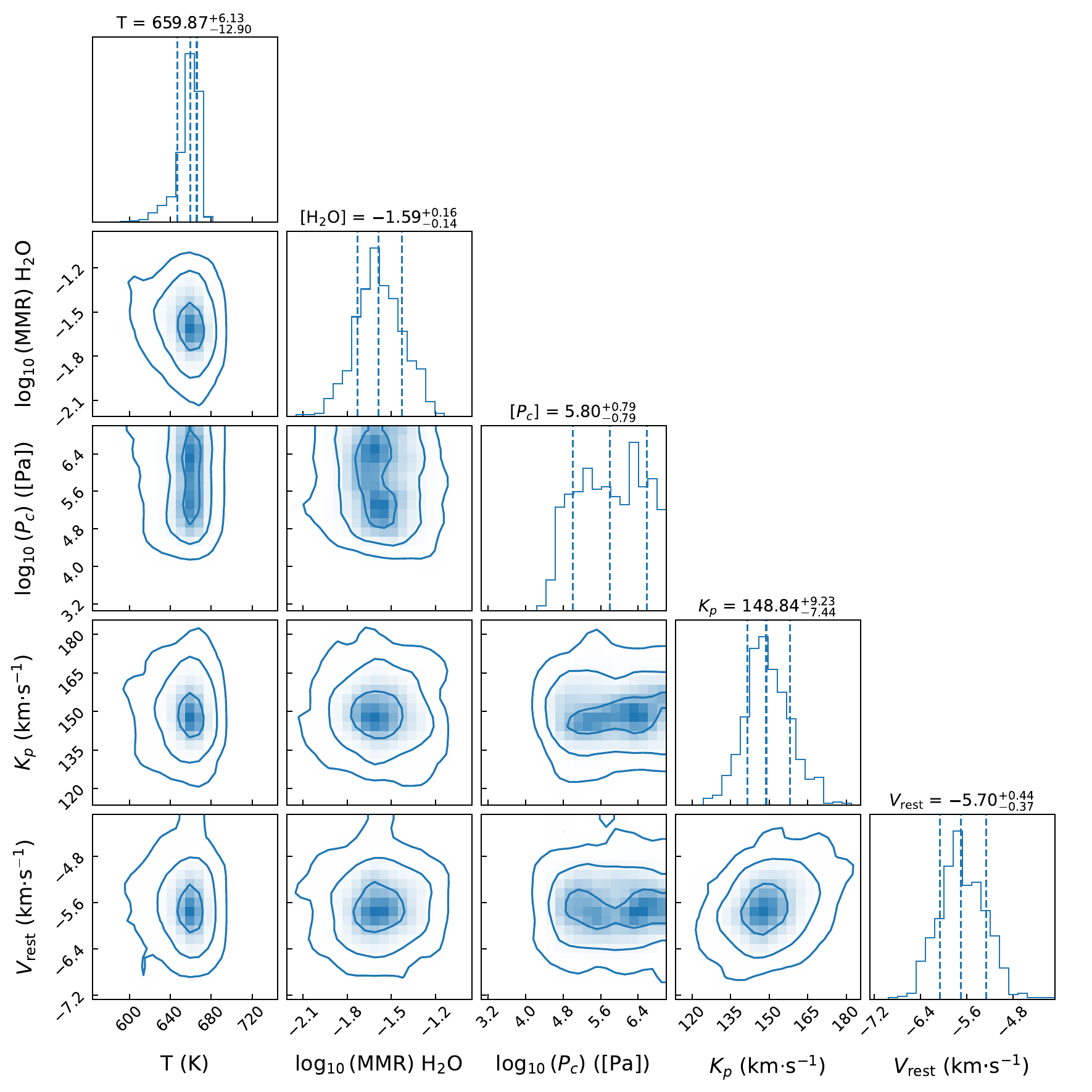}
              \caption{
                Posterior probability distributions for P-01, using 100 live points. 
                The dashed vertical blue lines represents the 0.16, 0.50 and 0.84 quantiles (i.e. median and 1-$\sigma$ error bar) of the distributions obtained with Polyfit. On top of each column is indicated the median and the 1-$\sigma$ error bar of the retrieved parameters using Polyfit. The contours in the 2D histograms corresponds to the 1, 2 and 3-$\sigma$ contours.
            }
            \label{fig:corner_r_01}
        \end{figure*}

        We analyse the data using the procedures and base model described in \autoref{subsec:spectrum_model}, \autoref{subsec:preparing_pipeline}, and \autoref{subsec:hr_framework}. We start with retrieval setups similar to that of \citet{Boucher2021}, that we label "P-01", retrieving only $K_p$, $V_{\mathrm{rest}}$, $T$, $P_c$ and the H$_2$O MMR. The corresponding corner plot is displayed in \autoref{fig:corner_r_01}. With Polyfit we are able to constrain all the parameters, and to put a lower limit on the pressure of an opaque cloud layer.

\begin{figure*}
            \centering
            \vspace{-25pt}
            \includegraphics[width=\hsize]{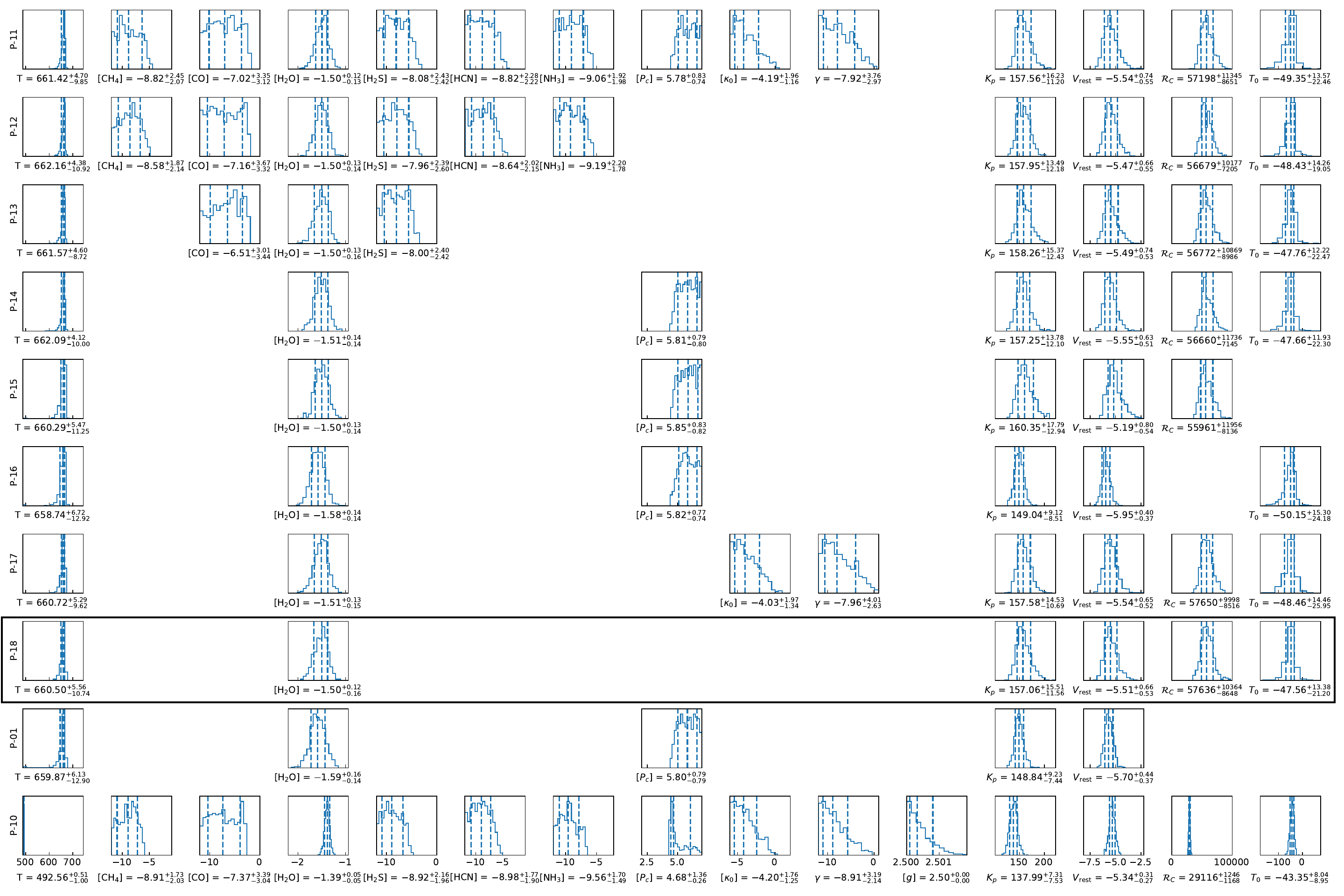}
              \caption{
                Posterior probability distributions for all of the setups in \autoref{tab:retrievals_comparison}. The dashed vertical blue lines represents the 0.16, 0.50 and 0.84 quantiles (i.e. median and 1-$\sigma$ error bar) of the distributions obtained. Below each panel is indicated the median and the 1-$\sigma$ error bar of the retrieved parameters using the setup. The units are the same as in \autoref{fig:expected_retrieval}. The black rectangle highlights our fiducial setup.
            }
            \label{fig:retrievals_posteriors}
            \vspace{-40pt}
        \end{figure*}

        We know from our retrieval on simulated data (\autoref{subsec:validation}) that we should be able to put constraints on more parameters than those retrieved with our "01" setup. We thus performed additional retrievals including these parameters, that we label "1x". In order to test different hypotheses and to estimate the significance of retrieving some parameters, we compare the difference in log-evidence ($\Delta\ln(\mathcal{Z})$) with respect to a model with a cloud layer and where all species MMR are set to $10^{-12}$, that we label "00". The results are listed in \autoref{tab:retrievals_comparison} and the corresponding posteriors are displayed in \autoref{fig:retrievals_posteriors}. A summary of our results can be found in \autoref{tab:summary_retrieved_values}\footnote{As an indication, the run times for the P-01 setup was $\approx 2.5$ h using 64 processes on a Intel Xeon Platinum 8360Y CPU (2.40 GHz). The run time is dependent among other things on the number of free parameters and, more importantly, on the number of live points used. Not taking into account edge of prior convergence, our longest run time was for P-11 ($\approx 6.2$ h).}.

        The P-10 setup had $g$ converging toward the lower edge of its prior. This means that the width of the $g$ posterior converged to 0, which could affect the posterior of other parameters. The log-evidences and results reported for this setup should not be considered as accurate. 

        In general, the log-likelihood does not change significantly with the number of retrieved parameters, but the log-evidence roughly decreases, meaning that these removed parameters -- which we set if they are not retrieved such that they do not significantly impact the spectrum (see note in \autoref{tab:retrievals_comparison}) -- are not necessary to describe the data. The exception are the resolving power and mid transit time, which increase slightly the log-evidence, although not significantly so ($\Delta\ln(\mathcal{Z}) \approx 1$, i.e. $\approx 2 \sigma$). The P-18 setup is the most favoured one with a $\Delta\ln(\mathcal{Z})$ of 206 ($\approx 20 \sigma$) compared to the featureless model P-00. For comparison, retrievals on simulated data in the same conditions, using a H$_2$O MMR of $3.2 \times 10^{-2}$, no hazes and no clouds, we expected a $\Delta\ln(\mathcal{Z})$ of 77 ($\approx 13 \sigma$). Because it is the most favoured setup, we will consider results from P-18 as our fiducial ones. However, since this favouritism is not significant ($\Delta\ln(\mathcal{Z}) \approx 7$, i.e. $\approx 4\sigma$ with P-11), we do not reject the other setups, and we will discuss their results in complement of P-18.

        \begin{deluxetable*}{l l l l l}
            \tablecaption{\label{tab:retrievals_comparison}Log-evidence, log-likelihood and reduced $\chi^2$ of our retrievals}  
            \centering	 
            \tablehead{\colhead{Retrieved parameters setups} & \colhead{$\Delta\ln(\mathcal{Z})$\tablenotemark{a}} & \colhead{Best-fit $\ln(\mathcal{L})$} & \colhead{Best-fit $\chi_\nu^2$\tablenotemark{b}} & \colhead{$k_\sigma$}}
            \startdata				
            (P-00): $T$, $P_c$, $K_p$, $V_\mathrm{rest}$, $\mathcal{R}_C$, $T_0$															& $0.00$	& $-842\,400.10$	& $1.010$	& $1.146$ \\
            (P-01): $T$, H$_2$O, $P_c$, $K_p$, $V_\mathrm{rest}$																			& $202.31$	& $-842\,180.77$	& $1.009$	& $1.146$ \\
            (P-10): $T$, all species\tablenotemark{c}, $P_c$, $\kappa_0$, $\gamma$, $g$, $K_p$, $V_\mathrm{rest}$, $\mathcal{R}_C$, $T_0$	& $1184.00$\tablenotemark{d}	& $-841\,173.72$	& $1.008$	& $1.146$ \\ 
            (P-11): $T$, all species\tablenotemark{c}, $P_c$, $\kappa_0$, $\gamma$, $K_p$, $V_\mathrm{rest}$, $\mathcal{R}_C$, $T_0$		& $199.48$	& $-842\,174.70$	& $1.009$	& $1.146$ \\
            (P-12): $T$, all species\tablenotemark{c}, $K_p$, $V_\mathrm{rest}$,  $\mathcal{R}_C$, $T_0$									& $203.01$	& $-842\,175.02$	& $1.009$	& $1.146$ \\
            (P-13): $T$, CO, H$_2$O, H$_2$S, $K_p$, $V_\mathrm{rest}$, $\mathcal{R}_C$, $T_0$												& $205.69$	& $-842\,175.17$	& $1.009$	& $1.146$ \\
            (P-14): $T$, H$_2$O, $P_c$, $K_p$, $V_\mathrm{rest}$, $\mathcal{R}_C$, $T_0$													& $204.70$	& $-842\,174.76$	& $1.009$	& $1.146$ \\
            (P-15): $T$, H$_2$O, $P_c$, $K_p$, $V_\mathrm{rest}$, $\mathcal{R}_C$															& $203.65$	& $-842\,178.61$	& $1.009$	& $1.146$ \\
            (P-16): $T$, H$_2$O, $P_c$, $K_p$, $V_\mathrm{rest}$, $T_0$																		& $203.56$	& $-842\,177.16$	& $1.009$	& $1.146$ \\
            (P-17): $T$, H$_2$O, $\kappa_0$, $\gamma$, $K_p$, $V_\mathrm{rest}$, $\mathcal{R}_C$, $T_0$										& $205.06$	& $-842\,175.13$	& $1.009$	& $1.146$ \\
            (P-18): $T$, H$_2$O, $K_p$, $V_\mathrm{rest}$, $\mathcal{R}_C$, $T_0$															& $206.37$	& $-842\,174.94$	& $1.009$	& $1.146$ \\
            \enddata
            \tablecomments{
                If a parameter does not appear in a setup: if it is a species, a MMR of $10^{-12}$ is used; for the haze parameters, the values $\kappa_0 = 10^{-6}$ and $\gamma = -12$ are used; for $P_c$, we use $10^7$ Pa; otherwise the value in \autoref{tab:general_parameters} is used. The uncertainty on $\ln(\mathcal{Z})$ reported by MultiNest is less than $0.01$ in all cases.
            }
            \tablenotetext{a}{Compared to a setup without any species (MMR = $10^{-12}$).}
            \tablenotetext{b}{Calculated from $-2 k_\sigma^2 \ln(\mathcal{L}) / N$, where here $N$ is the number of non-masked data points.}
            \tablenotetext{c}{CH$_4$, CO, H$_2$O, H$_2$S, NH$_3$.}
            \tablenotetext{d}{Edge of prior convergence.}
		\end{deluxetable*}

\begin{figure*}
           \centering
           \includegraphics[width=\linewidth]{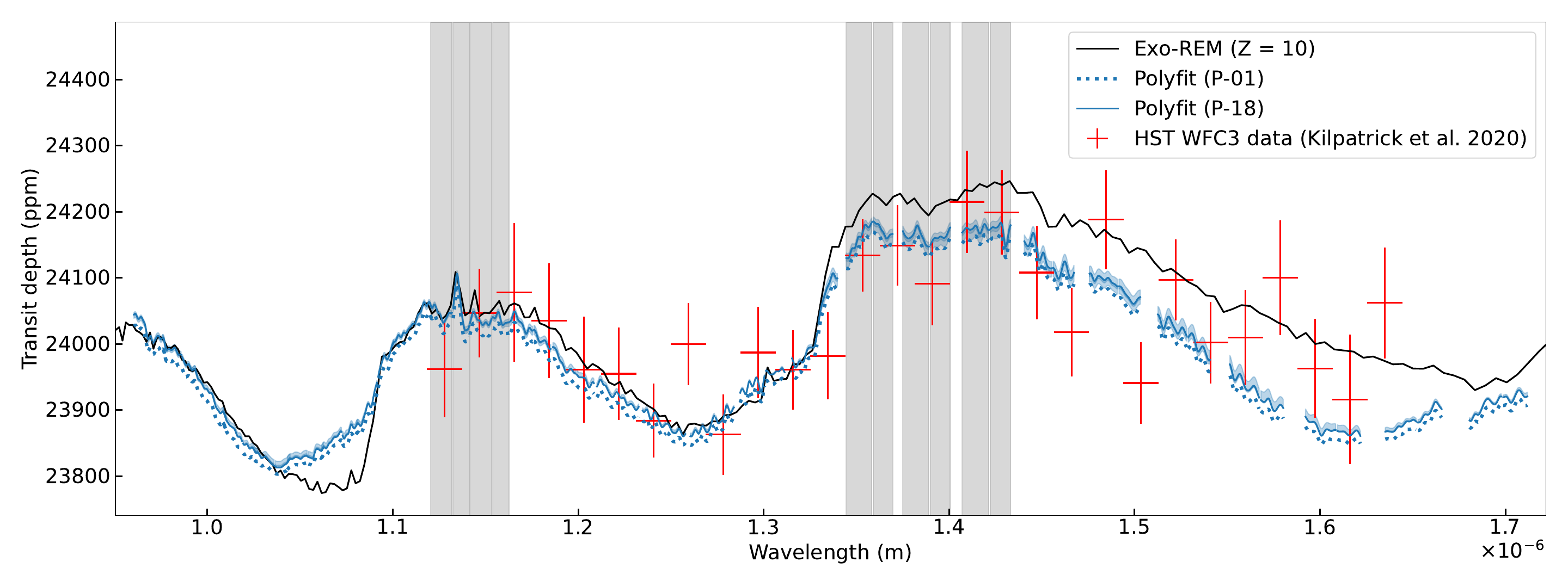}
            \caption{HD 189733 b low-resolution ($\mathcal{R} \approx 500$) transit depth for a selection of models. Black: Exo-REM model with a 10 times solar metallicity. Dotted blue: best-fit model from the P-01 setup. 
            Solid blue: best-fit model from the P-18 setup. 
            Red: HST WFC3 data from \citet{Kilpatrick2020}. The red vertical bars represent the 1-$\sigma$ uncertainties on the HST data, and the red horizontal bars represent the width of the data wavelength bins. The dark and light colored areas represent respectively the envelope of the 1 and 3-$\sigma$ models, for P-18 (blue). The grey areas represents the wavelength range of the orders discarded from our selection. The planet radius of the Exo-REM model (at $10^5$ Pa), and the P-models (at $10^3$ Pa), has been increased by respectively $1200$ km and $3600$ km compared to the value in \autoref{tab:general_parameters}.}
            \label{fig:models_comparison}
        \end{figure*}

        Most of our results with Polyfit are compatible with $K_p \approx 155$ km$\cdot$s$^{-1}$, $V_\mathrm{rest} \approx -5$ km$\cdot$s$^{-1}$, a H$_2$O MMR of $\approx 0.03$, and a temperature of $\approx 650$ K. We find upper limits for the abundance of the other species included in our model, and for the top pressure of an opaque cloud layer. We also find no evidence for hazes. In all cases, the $\chi^2_\nu$ is slightly above one, with a value of 1.009 excluding P-10. This value does not seem consistent with a strong overfitting or underfitting of the data. However, as discussed in \autoref{subsec:uncertainties_check}, the $\chi^2_\nu$ is hard to accurately estimate in our case, so its value should be considered with caution. In \autoref{fig:models_comparison}, we display a qualitative comparison of a selection of our models, with their resolution lowered, with the Hubble Space Telescope (HST) WFC3 data from \citet{Kilpatrick2020}.

        \begin{deluxetable}{l l l l}
            \tablecaption{\label{tab:summary_retrieved_values}Summary of our retrieved values.}  
            \centering   
            \tablehead{\colhead{Parameter} & \colhead{Posterior} & \colhead{Setup}}
            \startdata
            $\mathcal{R}_{\mathrm{C}}$ ($10^3$)				& $57.6^{+10.4}_{-8.6}$                     & 								     & P-18		\\
            $K_p$ (km$\cdot$s$^{-1}$)						& $157.1^{+15.5}_{-11.6}$				    &								     & P-18		\\
            $V_{\mathrm{rest}}$ (km$\cdot$s$^{-1}$)			& $-5.51^{+0.66}_{-0.53}$					&								     & P-18		\\
            $\log_{10}(g)$ [cm$\cdot$s$^{-2}$]			    & --					                    & 		                             & --		\\
            $T$ (K)											& $661^{+6}_{-11}$		                    &								     & P-18		\\
            $[$CH$_4$]						                & $\lessapprox -5.79$    					&	$\lessapprox -6.62$ VMR	         & P-12		\\
            $[$CO]							                & $\lessapprox -2.64$      					&	$\lessapprox -3.71$ VMR  	     & P-12		\\
            $[$H$_2$O]						                & $-1.50^{+0.12}_{-0.16}$   				&	$-2.38^{+0.12}_{-0.16}$ VMR      & P-18		\\
            $[$H$_2$S]						                & $\lessapprox -4.50$   					&	$\lessapprox -5.65$ VMR		     & P-12		\\
            $[$HCN]							                & $\lessapprox -5.64$    					&	$\lessapprox -6.69$ VMR		     & P-12		\\
            $[$NH$_3$]						                & $\lessapprox -6.35$    					&	$\lessapprox -7.20$ VMR		     & P-12		\\
            $\log_{10}(P_c)$ [Pa]							& $\gtrapprox 4.7$							&								     & P-14		\\
            $\log_{10}(\kappa)$ 							& --					                    &   	                       	     & --		\\
            $\gamma$ 							            & --					                    &                          		     & --		\\
            \enddata
            \tablecomments{
            The indicated error bars corresponds to $\pm 1 \sigma$. When a limit is indicated, it corresponds to the $95 \%$ upper or lower limit. A dash indicates that we were not able to put constraints on the parameter. The VMR values are obtained from an atmosphere composed of the median H$_2$O MMR value and the $95 \%$ MMR upper limits of the other species, and obeying the same rules as in step 1 of \autoref{subsec:spectrum_model}.
            }
        \end{deluxetable}

\section{Discussion}        
    \label{sec:discussion}
    \subsection{Retrieved kinematics parameters}
        \subsubsection{Orbital velocity semi-amplitude}
            \label{subsubsec:retrieved_kp}
            All of our Polyfit results, except those from P-10, are compatible at 1 $\sigma$ with the values of $K_p$ found by \citet{alonso2019multiple} and \citet{sanchez2019water} (respectively $K_p = 160^{+45}_{-33}$ km$\cdot$s$^{-1}$ and $K_p = 147^{+33}_{-28}$ km$\cdot$s$^{-1}$), who analysed the exact same dataset. Our retrieved value of $K_p$ with the P-18 setup ($157.1^{+15.5}_{-11.6}$ km$\cdot$s$^{-1}$) is also fully consistent with our computed value from the literature ($K_p = 152.1 \pm 2.9$ km$\cdot$s$^{-1}$, from \autoref{tab:general_parameters}), as well as the value retrieved by e.g. \citet{Brogi2016},  \citet{Damiano2019}, \citet{Boucher2021}, \citet{Klein2024} and \citet{Finnerty2024} (respectively $179^{+22}_{-21}$ km$\cdot$s$^{-1}$, $167^{+32}_{-21}$ km$\cdot$s$^{-1}$, $151 \pm 10$ km$\cdot$s$^{-1}$, $151.8^{+15.8}_{-12.6}$ km$\cdot$s$^{-1}$, and $160 \pm 8$ km$\cdot$s$^{-1}$ with their isothermal temperature profile model).
            
            Our retrieved $K_p$ for HD~189733~b is fully consistent with our CCF results, considering the uncertainties of both measurements. However, we note that the retrieved value is more constrained, with less than half the error bars, which seems to indicate our framework is able to better study the semi-amplitude of the orbital velocity than the regular CCF analyses.

        \subsubsection{Rest velocity shift}
            \label{subsubsec:rest_velocity_shift}
            
\begin{figure}
               \centering
               \includegraphics[width=\hsize]{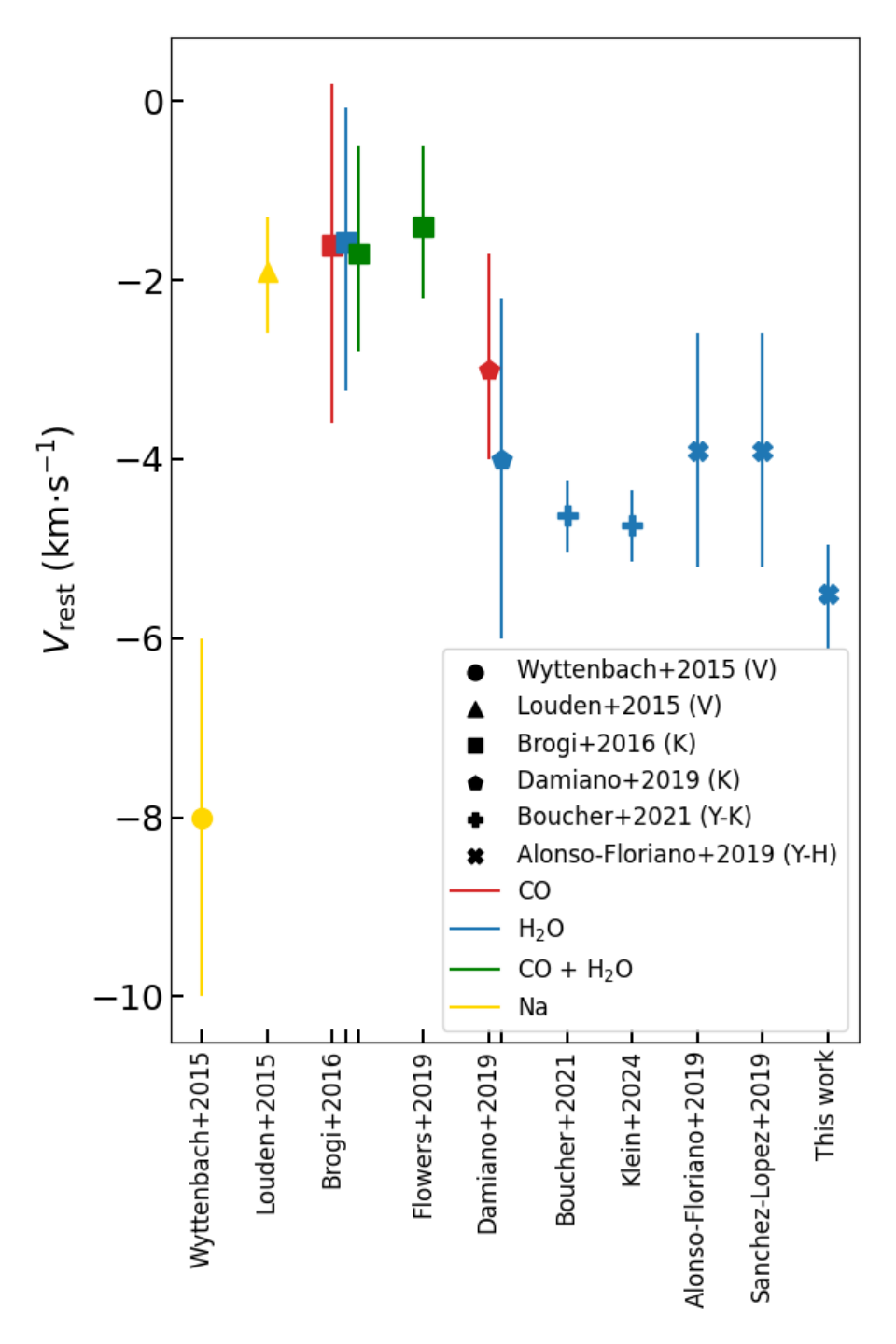}
                  \caption{
                    Summary of the estimated rest velocity of HD 189733 b in the literature. The marker shape indicates the source of the data. In parenthesis, next to the data source, is displayed the spectral band coverage of the data. The color indicates which absorber was used to study the data. The label "CO + H$_2$O" indicates that these two species opacities were present in the model used.
                }
                \label{fig:v_rest_comparison}
            \end{figure}

            Like many previous studies (see previous paragraph for example), we report a significantly negative $V_\mathrm{rest}$ ($-5.51^{+0.66}_{-0.53}$ km$\cdot$s$^{-1}$), that is, an overall blueshift of HD~189733~b spectral absorption in our selected exposures. We obtain a similar value with our CCF analysis, but with four times larger error bars compared to our retrieval analysis. This value is consistent with the results from \citet{alonso2019multiple}, \citet{Damiano2019}, \citet{sanchez2019water}, \citet{Boucher2021} and \citet{Klein2024} (respectively $-3.9\pm1.3$ km$\cdot$s$^{-1}$ for the first two, $-4.0^{+2.0}_{-1.8}$ km$\cdot$s$^{-1}$ for the third, $-4.62^{+0.41}_{-0.39}$ km$\cdot$s$^{-1}$ for the fourth and $-4.73^{+0.61}_{-0.56}$ km$\cdot$s$^{-1}$ for the latter). This is however significantly more than the values reported by \citet{Louden2015}, \citet{Brogi2016}, \citet{Flowers2019} (respectively $-1.9^{+0.7}_{-0.6}$ km$\cdot$s$^{-1}$ averaged over their full transit, $-1.7^{+1.1}_{-1.2}$ km$\cdot$s$^{-1}$, and $-1.4^{+0.8}_{-0.9}$ km$\cdot$s$^{-1}$), and significantly less than the value retrieved by \citet{Wyttenbach2015} (8 $\pm$ 2 km$\cdot$s$^{-1}$). We summarise these results in \autoref{fig:v_rest_comparison}. Note that the results obtained by \citet{alonso2019multiple} and \citet{sanchez2019water}, using the same data as ours, were obtained without applying our time stamp correction (see \autoref{subsec:observations}).

            The cause of this rest velocity shift has been attributed to a combination of the winds in a super-rotating equatorial jet, the rotation of the planet, and, most importantly, day to night winds. The overall effect on HD~189733~b has been modelled by \citet{Flowers2019}. They showed that the temperature difference between the day side and the night side of the planet creates winds redistributing heat from the former to the latter. In transmission spectra, this means that winds are overall oriented towards the observer, thus blueshifting the spectral lines. In addition, the equatorial jet moves the hottest point of the planet to the east of the sub-stellar point. This creates an asymmetry between the terminators, with the eastern, or trailing one, being hotter thus more inflated compared to the leading one. During the transit, the trailing terminator, with the jet oriented towards the observer, thus occupies a larger area on the stellar disk than the leading terminator, hence the properties of the former tend to dominate the overall spectrum. However, according to \citet{Flowers2019} models, these effects seem insufficient to explain our retrieved value. Moreover, the retrieval on our 3-D model did not result in any significant shift in velocity. The blueshift we are observing seems thus difficult to explain with these models. We note that \citet{Louden2015} retrieved a trailing terminator velocity of $-5.3^{+1.0}_{-1.4}$ km$\cdot$s$^{-1}$. We can interpret our retrieved value as the trailing terminator dominating our observations more than expected, as also suggested by \citet{Boucher2021}, but further modelling work is required to verify this hypothesis.
    
            As noted above, it seems that there are discrepancies in the measurements of $V_\mathrm{rest}$ in the literature. This has been previously reported, comparing the value found by \citet{Wyttenbach2015} with those from other works. Two different interpretations were proposed: \citet{Louden2015} showed that \citet{Wyttenbach2015} high blueshift could have originated from the Rossiter-McLaughlin effect leading to the detection of a spurious signal, while \citet{Brogi2016} suggested that the difference could arise from the probing of a different atmospheric regime. We can expand on this discussion by highlighting the differences and similarities between those works. From \autoref{fig:v_rest_comparison}, it seems that there is no clear correlation between the estimated $V_\mathrm{rest}$ and the studied species or spectral band. Strong wind speed variation with altitude or strong chemical composition variation with longitude (e.g. H$_2$O depletion on one terminator), does not seem to be expected from 3-D models \citep[e.g.][]{Flowers2019}. The explanation that the different shift observed come from probing species experiencing different kinematic regimes seem thus unlikely. If these differences are real, and do not arise from from data reduction or data analysis artifacts, an alternative explanation could be meteorological variations, but a much more detailed study would be necessary to confirm this.
            
            A point to highlight is that $V_\mathrm{rest}$ is correlated with $T_0$. The consequence is that, if the latter is not well evaluated, it may lead to an inaccurate estimation of $V_\mathrm{rest}$, which might explain the discrepancy between some of the studies. Note also that $V_\mathrm{rest}$ and $T_0$ are not degenerate: while their shifting effect on the lines is identical, $T_0$ has an additional indirect effect on the lines' amplitude within partial transit exposures, because it sets the time of ingress and egress.

        \subsubsection{Resolving power}
            While not strictly speaking a kinematic parameter, the resolving power is a useful proxy to estimate line broadening due to atmospheric dynamics during the transit. As mentioned in \autoref{subsec:validation}, due to the combined Doppler effect of the planet rotation and winds, the spectral lines of the transit spectrum appear broadened. This can be roughly modelled with a lower resolving power compared to what is expected from the instrument. With our 3-D model retrieval we expected to retrieve $\mathcal{R}_C = 52\,000 \pm 17\,000$. This is fully consistent with what we retrieve with our P-18 setup: $\mathcal{R}_C = 58\,000^{+10\,000}_{-9\,000}$, and roughly half the CARMENES resolving power. \citet{Brogi2016} observed instead that their retrieval favoured larger resolving power values than those of the instrument they used, but it was attributed to the properties of the CCF analysis they performed, which our retrieval does not have.

        \subsubsection{Mid-transit time}
            The mid-transit time has, to our knowledge, never been retrieved in studies similar to ours, most likely because this parameter is usually very well known from transit light curve analyses. However, its correlation and non-degeneracy with $V_\mathrm{rest}$ (see \autoref{subsubsec:rest_velocity_shift}), which can be used to estimate wind speeds in General Circulation Model (GCM) simulations, makes it an important parameter to retrieve. Moreover, $K_p$ is usually also a well-known parameter, but is ubiquitously retrieved. Therefore, there appears to be no strong reason to not also explore the effects from retrieving $T_0$. It can be used both as a sanity check for the data time stamps, for the modelled ingress/egress weighting function, as well as to more accurately measure $V_\mathrm{rest}$.

            With our P-18 setup, we retrieve $T_0 = -47^{+13}_{-21}$ s, which is fully consistent with the expected value, taking error bars into account, and represents significantly less than the average duration of an exposure. Therefore, we interpret this result as an additional confirmation that we are retrieving the planet's signal and that our data time stamps are accurate.
            
    \subsection{Retrieved atmospheric parameters}
        \label{subsec:retrieval_discussion}

        \subsubsection{Temperature}
            \label{subsubsec:isothermal_temperature}
            The temperature that we retrieve -- $661^{+6}_{-11}$ K with our P-18 setup -- is significantly lower than the equilibrium temperature of the planet, the temperature that we could expect according to our 1-D self-consistent model, and the temperatures at the terminator in our 3-D model. As mentioned in \autoref{subsec:validation}, an overall colder retrieved temperature can be explained by asymmetric terminators \citep{MacDonald2020}. Moreover, our result is fully consistent with the results from other studies of transmission spectra, for example, \citet{Boucher2021} and \citet{Klein2024} (respectively $T = 698^{+80}_{-172}$ K and $698^{+147}_{-155}$ K in their isothermal analysis) or \citet{Tsiaras2018} ($T = 621.49 \pm 139.05$ K). This value is also consistent with the best fit value found by \citet{Brogi2016} ($T \approx 500$ K), and with the value found by \citet{McCullough2014} ($T \approx 700$ K), although no error bars were provided in these works. Our value is also roughly 2 $\sigma$ away from the value found by \citet{Madhusudhan2014} ($T = 787^{+511}_{-47}$ K). 

            \begin{deluxetable}{l l l l}
                \tablecaption{\label{tab:exorem_z10_vmr}$\log_{10}$ volume mixing ratios at $10^4$ Pa of a HD~189733~b Exo-REM model for different metallicities and carbon-to-oxygen ratios.}  
                \centering   
                \tablehead{&\colhead{$Z = 1$} & \colhead{$Z = 3$} & \colhead{$Z = 10$} \\ \colhead{Species} & \colhead{C/O $= 0.55$} & \colhead{C/O $= 0.1$} & \colhead{C/O $= 0.55$}}
                \startdata
                CH$_4$		& $-5.1$		& $-6.7$		& $-5.7$	\\
                CO			& $-3.3$		& $-3.6$		& $-2.3$	\\
                H$_2$O		& $-3.6$		& $-2.7$		& $-2.6$	\\
                H$_2$S		& $-4.6$		& $-4.1$		& $-3.6$	\\
                HCN		    & $-7.6$		& $-8.5$		& $-7.6$	\\
                NH$_3$		& $-6.1$		& $-6.1$		& $-5.8$	\\
                \enddata
                \tablecomments{
                    The VMRs with a C/O ratio of $0.1$ are obtained by fixing the oxygen elemental abundance (O/H $= Z$) and lowering the carbon elemental abundance compared to its solar abundance (C/H $= 0.18\,Z$). Note that the same C/O ratio can be obtained by fixing the the carbon elemental abundance and increasing the oxygen elemental abundance, but produces different VMRs (typically higher abundances of CO and lower abundances of H$_2$O).
                }
            \end{deluxetable}
            
            However, the temperature that we retrieve with our simulated 3-D data is $T = 1225^{+409}_{-253}$, which is not consistent with our findings in the real data. It can be expected that our 3-D model does not perfectly represent  HD~189733~b, thus explaining an underestimated bias in temperature. Our retrieved temperature is also not consistent with that obtained by \citet{Zhang2020}, who retrieved $T = 1089^{+110}_{-120} K$. The explanation could be their use of equilibrium chemistry and the prevalence of more species in their spectral ranges (0.3--30 $\mu$m): the specific combination of abundances that are required to fit their data might be forbidden by low temperatures. In contrast, our data are dominated by mainly H$_2$O features, and the temperature and H$_2$O MMR can vary independently. For these reasons, it is likely that the value of $T$ retrieved by \citet{Zhang2020} is more accurate than ours.

        \subsubsection{H\texorpdfstring{$_2$}~O abundance} 
            
            Our retrieved H$_2$O $\log_{10}($MMR$)$ -- that is, $-1.50^{+0.12}_{-0.16}$ with our P-18 setup, corresponding to a $\log_{10}$ volume mixing ratio (VMR) of $-2.39^{+0.12}_{-0.16}$ -- is compatible with an atmospheric metallicity of $\approx 10$ times solar and a solar ($0.55$) carbon-to-oxygen (C/O) ratio (see \autoref{tab:exorem_z10_vmr}), which is similar to that of Saturn \citep{Atreya2020}. It is also compatible with an atmospheric metallicity of $\approx 3$ times solar and a C/O ratio of 0.1.
        
            Comparing with other works using retrievals on high-resolution data, our retrieved H$_2$O VMR is consistent with \citet{Klein2024} ($-2.95^{+0.75}_{-0.53}$), and \citet{Finnerty2024} ($-2.9 \pm 0.4$). In contrast, it is not consistent with \citet{Boucher2021}, who obtained a H$_2$O $\log_{10}($VMR$)$ of $-4.4 \pm 0.4$. However, their model preparation technique is based on model injection in a reconstruction of the observed data, which differs significantly from our \autoref{eq:actual_retrieval}. It is possible that this technique introduces a bias in their estimation of the H$_2$O abundance, but a proper benchmarking is necessary to confirm this.

            For other works using high-resolution data, but not providing error bars, our retrieved H$_2$O VMR is also $\gtrapprox 5 \sigma$ from the best fit values found by respectively \citet{danielski20140} (H$_2$O $\log_{10}($VMR$)$ $\approx -4$ to $-3.3$), and \citet{McCullough2014} (H$_2$O $\log_{10}($VMR$)$ $\approx -3.3$), although \citet{danielski20140} only tested models with $T \gtrapprox 1000$ K. The H$_2$O $\log_{10}($VMR$)$ estimated by \citet{alonso2019multiple} and \citet{sanchez2019water} is -5, while it is $\approx$ -3 for \citet{Brogi2016} and \citet{Flowers2019}\footnote{Their best fit is obtained with local equilibrium chemistry for a solar metallicity atmosphere.}, but these values come from a limited parameter space exploration and from a CCF analysis, both likely obstacles for an accurate abundance estimation.

            Comparing with low-resolution data analysis, our retrieved H$_2$O abundance is fully consistent with the value found by \citet{Tsiaras2018} (H$_2$O $\log_{10}($VMR$)$ $= -2.51 \pm 0.9$), and the value found by \citet{Zhang2020} at $10^3$ Pa (10 mbar, H$_2$O $\log_{10}($VMR$)$ $= -2.5 \pm 0.3$). In contrast, our result is not consistent with the values found by \citet{Madhusudhan2014}, \citet{Pinhas2019}, and \citet{Welbanks2019} (respectively H$_2$O $\log_{10}($VMR$)$ of $-5.13^{+1.68}_{-0.18}$\footnote{Assuming a H$_2$ VMR of 0.85.}, $-5.04^{+0.46}_{-0.30}$, $-4.66^{+0.35}_{-0.33}$). For \citet{Pinhas2019}, \citet{Zhang2020} suggested that the high number of retrieved parameters they used may have lead them to obtain an overfitting solution. A similar explanation may apply to \citet{Welbanks2019}, as they use the same retrieval setup than \citet{Pinhas2019}, and to \citet{Madhusudhan2014}, as they retrieved 13 parameters on $\approx$ 30 data points. In addition, we note that all of the above works use the same Hubble Space Telescope's Wide Field Camera 3 (HST WFC3) data reduced by \citet{McCullough2014}. In contrast, the previously mentioned low-resolution studies that show an agreement with our value either use an independent reduction method \citep{Tsiaras2018}, or both a different data set and an independent reduction method \citep{Zhang2020}.
            
            Finally, we also note that studies analysing the emission spectrum of HD~189733~b, that is, the dayside of the planet, generally find H$_2$O $\log_{10}($VMR$)$ ranging from $\approx -3$ to $< -5$ \citep[e.g.][]{birkby2013detection, Barstow2014, Line2014, brogi2018exoplanet, Cabot2018, Brogi2019, Finnerty2024}, but making comparison with these studies is not necessarily straightforward as they probe a different region of the planet with different sensitivities to some parameters.
            
            We also note that there is an anti-correlation between the temperature and the H$_2$O MMR. In \autoref{subsubsec:isothermal_temperature} we estimated that our retrieved temperature is likely underestimated. Consequently, our retrieved H$_2$O MMR should be regarded as an upper estimation. However, it is difficult to quantify this bias. In addition to this anti-correlation, as discussed above, 3-D models predict an underestimation of the retrieved H$_2$O abundance in case of difference in chemical composition between the two terminators \citep{MacDonald2020}. The 3-D-induced bias and the parameter correlation can add or compensate each other, hence estimating the overall effect would require a more detailed study. 

        \subsubsection{Other species abundances}
            With our P-12 setup we can derive $95\%$ upper limits for the abundance of other species, summarised in \autoref{tab:summary_retrieved_values}. 

            These constraints are compatible with a slightly sub-solar metallicity atmosphere simulated by our Exo-REM model (see \autoref{tab:exorem_z10_vmr}). For CH$_4$, CO and HCN, those are also compatible with an atmospheric C/O ratio of 0.1, which would be consistent with the H$_2$O VMR we retrieve. We note that a sub-solar C/O ratio ($0.3 \pm 0.1$)  and super-solar O/H ratio ($6_{-4}^{+9}$ times solar) were derived by \citet{Finnerty2024}. The relatively low upper limits for H$_2$S and NH$_3$ could also be interpreted as a sulfur and nitrogen depletion, or as a hint for unknown chemical processes. These constraints should be taken cautiously though, as our retrieval also favors an atmosphere consisting only of H$_2$, He and H$_2$O at $2.4 \sigma$ (from the $\Delta\ln(\mathcal{Z})$ between our P-12 and P-14 setups), which is unlikely for a Jupiter-like planet, but confirms that adding those species does not significantly improves the data explanation. The low temperature we retrieve and the limits of our models may also play a role in this results. Data with better S/N and covering stronger molecular bands would be necessary to be more confident.

            Regarding our CO estimation, it should be highlighted that HD~189733 present CO lines in its spectrum. This is known to perturbate CO detection in CCF analysis if the Rossiter-McLaughlin effect is not accurately corrected \citep[e.g.,][]{Brogi2016, Chiavassa2019}. It can be noted that, in the CARMENES spectral range, the stellar CO lines are much weaker than in the K band probed by these studies. However, CO stellar lines may remain in our prepared data, potentially affecting our result. In conclusion, we strongly advice caution regarding our CO abundance estimation, even if it seems compatible with other studies.
            
            We can also compare some of these constraints with other studies:
            \begin{itemize}
                \item Our 99$\%$ upper limit for the CH$_4$ $\log_{10}($MMR$)$ ($< -5.9$) is significantly lower than the one reported by \citet{Finnerty2024} ($< -4.6$), and our 95$\%$ CH$_4$ $\log_{10}($VMR$)$ upper limit is not consistent with the value retrieved by \citet{Line2014} ($-4.70^{+0.08}_{-0.30}$). However, we note that all previous estimations of the CH$_4$ abundances \citep{Madhusudhan2009, Swain2009, Lee2012, Lee2014} found only upper limits for CH$_4$, ranging from $\lessapprox -2$ to $\lessapprox -7$. In addition, all of these estimations were obtained using emission spectra. 
                \item Our upper limit for the CO $\log_{10}($VMR$)$ is on the lower end of the abundance found in emission spectroscopy by \citet{Madhusudhan2009, Lee2012, Line2014, Brogi2019, Finnerty2024} (respectively $\lessapprox -1.7$, $-1.9 \pm 0.5$, $-1.7^{+0.2}_{-2.8}$, $\gtrapprox -3$, $-3.3 \pm 0.5$).
                \item The abundance of H$_2$S is unconstrained in \citet{Finnerty2024}. That is, it is at least $< -3$ $\log_{10}($MMR$)$, their upper prior boundary. Our upper limit is significantly lower.
                \item Our upper limit for the HCN $\log_{10}($VMR$)$ is consistent with the estimation found by \citet{Cabot2018} ($\approx -6$), although this value was estimated using a CCF analysis. We note that \citet{Finnerty2024} reported for this species that their posteriors shown a "weak preference" for a HCN $\log_{10}($VMR$)$ of $\approx 4.8$, which would be inconsistent with our results if confirmed, and $\approx$ 3 orders of magnitude larger than what is expected by our Exo-REM model.
                \item Our 99$\%$ upper limit for the NH$_3$ $\log_{10}($VMR$)$ ($< -6.6$) is significantly lower than the one reported by \citet{Finnerty2024} ($< -4.5$).
            \end{itemize}

        \subsubsection{Clouds and hazes}
            We find no evidence for clouds at the pressures probed by our data. With our P-15 setup, we constrain an opaque cloud layer to be located at $\gtrapprox 10^5$ Pa (at $1 \sigma$). This is consistent with the constraints established by \citet{McCullough2014}, \citet{Pinhas2019} ($\gtrapprox 10^4$ Pa), \citet{Boucher2021} ($\gtrapprox 2.0\times10^4$ Pa), and \citet{Finnerty2024} ($5.479_{-1.06}^{+1.02}$ $\log_{10}$(Pa)), as well as with our 1-D self-consistent temperature profile and estimated condensation curves (\autoref{fig:exorem_tpr}). We also find no evidence of hazes having an effect at our observed wavelengths. We note that, based on low-resolution data, this planet has been inferred to be strongly hazy \citep[e.g.][]{Sing2016, Zhang2020}. This is not surprising as \citet{alonso2019multiple} (based on the work by \citet{Pino2018}) already established that haze effects (or equivalent star spots effects) affecting water vapor bands in the CARMENES data near-infrared wavelength range should be negligible. Our results are thus consistent with a clear atmosphere in the near-infrared, and not inconsistent with a strong haze-like spectral feature at bluer wavelengths.

    \subsection{Retrieved error bars}
        \label{subsec:retrieved_error_bars}
        We note that our retrieved error bars for $T$, $T_0$ and the H$_2$O $\log_{10}$(VMR) are respectively 10, 8 and 5 times tighter than expected from our retrieval on simulated data, consistently with the high $\Delta\log(\mathcal{Z})$ we obtained (see \autoref{subsec:retrievals}). This might be caused by the simplicity of our transit light curve model, or by 3-D effects, for example. Another cause could be hidden residuals in our prepared data, although none of the median value we retrieve for other parameters are of particular concern if we compare them to other published values -- sometimes obtained from different techniques or kind of data --, as discussed in this section. We thus recommend caution when considering our error bars. Analysis of other datasets and the enhancement of our models may be necessary to better understand this behaviour.

    \section{Summary and conclusion}
        We have introduced a robust retrieval framework for high-resolution, telluric contaminated data described by \autoref{eq:log_likelihood} and demonstrated analytically that it produced unbiased parameter estimations, with any preparing pipeline, provided that:
        \begin{itemize}
            \item the pipeline effect can be written as in \autoref{eq:pipeline},
            \item the model accurately describes the data,
            \item the data uncertainties are accurate,
            \item there is no degeneracy between the retrieved parameters,
            \item the BPM (\autoref{eq:bpm}) is $\approx 0$.
        \end{itemize}
        We confirmed our analytical demonstration with a simulated retrieval. We introduced the aforementioned BPM (Bias Pipeline Metric) that can be used to compare the risk of bias of different preparing pipelines.
        
        We presented the result of a retrieval using a 1-D model on 3-D simulated data at high-resolution, enabled by the pRT-Orange model (Mollière et al., in prep.). The results show that several biases can be expected, according to this model:
        \begin{itemize}
            \item the retrieved resolving power is lower than the truth due to line-broadening by the planet rotation and its winds,
            \item the surface gravity is biased toward higher values, for the same reasons,
            \item the posteriors are overall slightly wider than expected probably because the 1-D model is struggling to fit 3-D-induced spectral features,
            \item no significant line shift is retrieved \citep[in contradiction with e.g.][]{Flowers2019},
            \item no significant temperature bias is retrieved (in contradiction with the observations).
        \end{itemize}
        The contradictions reported above may be partially caused by the pRT-Orange model assumption that the atmospheric properties are constant with latitude, although the latitude-dependence of the GCM velocity field is fully taken into account. This will be investigated further in a future work.

        We re-analysed CARMENES data of a HD~189733~b transit, covering wavelengths from 0.96 to 1.71 $\mu$m at a resolving power of $\approx 80\,400$ -- studied by \citet{alonso2019multiple} and \citet{sanchez2019water} -- using our "Polyfit" preparing pipeline, both with a classical CCF analysis, and our retrieval framework
        . Our CCF analysis detects a signal consistent with a transit of HD~189733~b at $\approx 12.4 \sigma$ (based on modeling the observation, we predict a detection significance of $\approx 10 \sigma$). With our retrieval, a simple model with H$_2$O spectral features is favoured at $\approx 20 \sigma$ using Polyfit compared to a featureless model. 
        
        We obtain significantly tighter constraints on $T$ and the H$_2$O abundance compared to what we could have expected from our simulated retrievals (i.e., respectively a $\approx 10$ K standard deviation instead of an uncertainty of $\pm 100$ K, and an uncertainty of $\pm 0.15$ instead of $\pm 0.75$). Finding the cause of this discrepancy may require improved atmospheric forward models. 
        
        Our results on the kinematics parameters are consistent with the previous analysis, both with a new CCF analysis and with the retrievals. Our retrieval analysis is able to reduce the error bars on $K_p$ and $V_\mathrm{rest}$ by a factor 2 for the former and a factor 4 for the latter compared to our CCF analysis. We observe the blueshift of the spectral absorption reported by several other studies, including \citet{alonso2019multiple} and \citet{sanchez2019water}. Our fiducial value for $V_\mathrm{rest}$ is $-5.51^{+0.66}_{-0.53}$ km$\cdot$s$^{-1}$, which is similar to what was observed by, for example, \citet{Boucher2021}, but is significantly more than the prediction of models including planetary rotation and winds, such as our pRT-Orange model as well as the one described by \citet{Flowers2019}. Additional modelling effort \citep[e.g., by including the formalism of][Appendix A]{Klein2024} and observations may be required to understand this discrepancy.

        Our retrieval analysis gave us an estimation of some atmospheric properties of HD~189733~b: 
        \begin{itemize}
            \item The temperature we retrieve ($\approx 650$ K) is significantly lower than expected by 3-D models ($T \approx 1200$ K), but consistent with some of the other studies. This is likely a bias, possibly caused by terminator asymmetry \citep{MacDonald2020}. A wider spectral range and the use of equilibrium chemistry seem to decrease this effect \citep{Zhang2020}.
            \item The H$_2$O $\log_{10}$(VMR) we retrieve ($\approx -2.4$), is consistent with a 10 times solar metallicity atmosphere or a 3 times solar metallicity atmosphere with a significantly sub-solar C/O ratio ($\approx 0.1$). However, the anti-correlation between the temperature and the H$_2$O abundance, as well as the expected 3-D-induced bias \citep{MacDonald2020} limit the accuracy of this result.
            \item We derived $95 \%$ upper limits for the abundances of CH$_4$, CO, and HCN, consistent with a slightly sub-solar atmospheric metallicity, or with a 3 times solar metallicity atmosphere with a C/O ratio $\approx 0.1$. However, we regard these constraints with caution as the low temperature we retrieve and the limits of our models may impact these results. Our CO upper limit may also be impacted by the Rossiter-McLaughlin effect and should be taken with even further caution.
            \item We derive upper limits for the abundance of H$_2$S and NH$_3$, consistent with a sub-solar atmospheric metallicity. We also consider these results with caution for the reason pointed above.
            \item Our results are consistent with no clouds at the pressures probed by our data. Other studies obtained the same result.
            \item We do not detect hazes at the wavelengths of our data, as expected by \citet{alonso2019multiple}. We note that a haze-like spectral feature has been observed in the optical and near-UV \citep[e.g.][]{Sing2016, Zhang2020}, which is not inconsistent with our near-infrared results.
        \end{itemize}

        Comparing our results with other works, we noted that, for $V_\mathrm{rest}$, studies are split between two inconsistent values: either $\approx -1.5 \pm 1.0$ km$\cdot$s$^{-1}$, or a value in agreement with our findings of $\approx -4.5 \pm 1.0$ km$\cdot$s$^{-1}$. Similarly, the H$_2$O $\log_{10}$(VMR) abundance is retrieved to be $\approx -2.5 \pm 0.5$, close to what we retrieve, or $\approx -5.0 \pm 0.5$, depending on the study.
        Another similar discrepancy can be found with CH$_4$, although this molecule has overall been less studied than H$_2$O. In all cases, the cause for these discrepancies has, to our knowledge, yet to be identified. For the molecular abundances, these discrepancies might be explained by overfitting retrievals, the use of a biased retrieval framework, or to the data reduction method used. In all cases, we suggest that including the complex 3-D-induced line shapes into models, or a more formal-based approach for the choice the log-likelihood equation and preparing pipelines used, could help solving this issue. In addition, we highlight the correlation between the mid-transit time and the rest velocity, and argue in favour of also retrieving the former -- even if it is well-known -- in order to ensure more accurate estimations of $V_\mathrm{rest}$.
        
        In this work we focused on the presentation of our retrieval framework and its comparison with a CCF analysis. 
        The model we chose for the analysis of these HD~189733~b transit spectra is a relatively simple one, but we consider it is enough to demonstrate the capabilities of our framework. Using more complex models taking into account terminator asymmetry and the evolution of the transmission lines shape with time for a more accurate analysis of transit spectra are obvious next steps. This added complexity might actually be necessary to accurately retrieve atmospheric properties in high-resolution data where species' lines are abundant and overlapping. Indeed, while HD~189733~b is a well-studied object, it seems that there are important discrepancies between different works regarding most notably the observed blueshift of the planet spectral absorption as well as the H$_2$O abundance. These two parameters are crucial in our understanding of hot jupiters atmospheric dynamics and formation process. Finding the origin of these discrepancies and reconciling these results will be important for obtaining a more accurate view of this planet.

\begin{acknowledgements}
    This research has made use of the Spanish Virtual Observatory (http://svo.cab.inta-csic.es) supported by the MINECO/FEDER through grant AyA2017-84089.7. This research has made use of the NASA Exoplanet Archive, which is operated by the California Institute of Technology, under contract with the National Aeronautics and Space Administration under the Exoplanet Exploration Program. This research has made use of the Exoplanet Follow-up Observation Program (ExoFOP; DOI: 10.26134/ExoFOP5) website, which is operated by the California Institute of Technology, under contract with the National Aeronautics and Space Administration under the Exoplanet Exploration Program. The authors thank M. Zechmeister and L. Nortmann for their insights on the CARACAL reduction pipeline and the CARMENES time stamps. D.B. thanks L. Pino for informing him about the correct term to use for the Hadamard product. D.B. also thanks C. Bousardo for her additional proofreading of the article. A.S.L. acknowledges funding from the European Research Council under the European Union's Horizon 2020 research and innovation program under grant agreement No 694513 and acknowledges financial support from the Severo Ochoa grant CEX2021-001131-S funded by MCIN/AEI/ 10.13039/501100011033. P.M. thanks J. Wardenier for running comparison calculations with gCMCRT to benchmark pRT-Orange. P.M. also thanks F. Debras for extracting GCM atmospheric structures from public tables for use in pRT-Orange.
\end{acknowledgements}

\bibliography{bibliography}{}
\bibliographystyle{aasjournal}

\begin{appendix}

\section{UTC to TDB timestamps conversion code}
    \label{anx:utc_to_tdb_code}
    We display here the Python code snippet that we used to convert the CARACAL timestamps from UTC to TDB. The variable \lstinline|times_tdb| is the one we used as our TDB timestamps. This code is also implemented in the petitRADTRANS package. This was inspired by the SERVAL\footnote{\url{https://github.com/mzechmeister/serval}} code \citep{Zechmeister2018}.

\begin{lstlisting}[belowskip=-1\medskipamount]
import astropy.units as u
from astropy.coordinates import (EarthLocation, 
                                 SkyCoord)
from astropy.time import Time

site_name = "CAHA"  # Calar Alto astropy site name
ra = 300.1821223 * u.deg  # (degree) for HD 189733
dec = 22.7097759 * u.deg  # (degree) for HD 189733

times_utc = ...  # placeholder to load MJD_UTC times

observer_location = EarthLocation.of_site(site_name)

target_coordinates = SkyCoord(
    ra=ra,
    dec=dec
)

times_utc = Time(times_utc, format="jd", scale="utc")
times_tdb = (
        times_utc.tdb
        + times_utc.light_travel_time(
            target_coordinates,
            location=observer_location
        )
)
\end{lstlisting}

\section{SysRem pipeline effect}
    \label{anx:sysrem_pipeline_effect}
    \subsection{SysRem preparing pipeline implementation}
        \label{subsec:sysrem_pipeline}
        We have implemented SysRem in our framework for comparison with Polyfit (\autoref{subsec:preparing_pipeline}). SysRem \citep{Tamuz2005} is a preparing pipeline developed to remove systematics of light curves obtained from photometric observations, but is also widely used to prepare high resolution data \citep[e.g.,][]{alonso2019multiple, Pelletier2021, Gibson2022}. The algorithm iterates to find the set of vectors "$a(t)$" and "$c(\lambda)$", such that $a(t) \circ c(\lambda)$ best fit the data. It can be applied repeatedly ("passes") to find several "hidden" linear trends in the data. Our implementation follows the steps described below:

        \paragraph{Step 1:} This step is almost identical to step 1 of Polyfit (\autoref{subsec:preparing_pipeline}). We obtain the same "normalised" spectrum $\mathbf{F}_{\overline{\mathbf{X}}}(t, \lambda)$ and correct the uncertainties in the same way to obtain $\mathbf{U}_{\mathbf{N},\overline{\mathbf{X}}}(t,\lambda)$. In addition, we follow \citet{alonso2019multiple} and mask $\mathbf{F}_{\overline{\mathbf{X}}}(t, \lambda)$ where its values are below 0.8, corresponding to the core of the strongest telluric lines. We note that for this "normalisation" step, various techniques are used in other works, that involve for example mean division, polynomial fitting, Gaussian filtering, combined in a variety of ways. The efficiency of these different techniques to remove $\mathbf{X}$ has to our knowledge never been formally discussed, but such discussion is beyond the scope of this work.

        \paragraph{Step 2:} We subtract the average over wavelengths\footnote{This is done using the \lstinline|numpy| 1.24.3 function \lstinline|numpy.ma.average| with parameters $a = \mathbf{F}_{\overline{\mathbf{X}}}(t, \lambda)$, $\mathrm{weights} = \mathbf{1}$, and $\mathrm{axis} = -1$.} of $\mathbf{F}_{\overline{\mathbf{X}}}(t, \lambda)$ following:
        \begin{eqnarray}
            \label{eq:sysrem_step2}
        		\mathbf{F}_a(t, \lambda) &=& \mathbf{F}_{\overline{\mathbf{X}}}(t, \lambda) - \langle \mathbf{F}_{\overline{\mathbf{X}}}(t, \lambda) \rangle(t).
        \end{eqnarray}
        At this step, $\mathbf{F}_a(t, \lambda)$ is on average $\approx 0$. This is crucial for the core of the SysRem algorithm to work properly. Because we only subtract a value from the data, this step has no impact on the uncertainties.

\begin{figure*}
       \centering
       \includegraphics[width=\hsize]{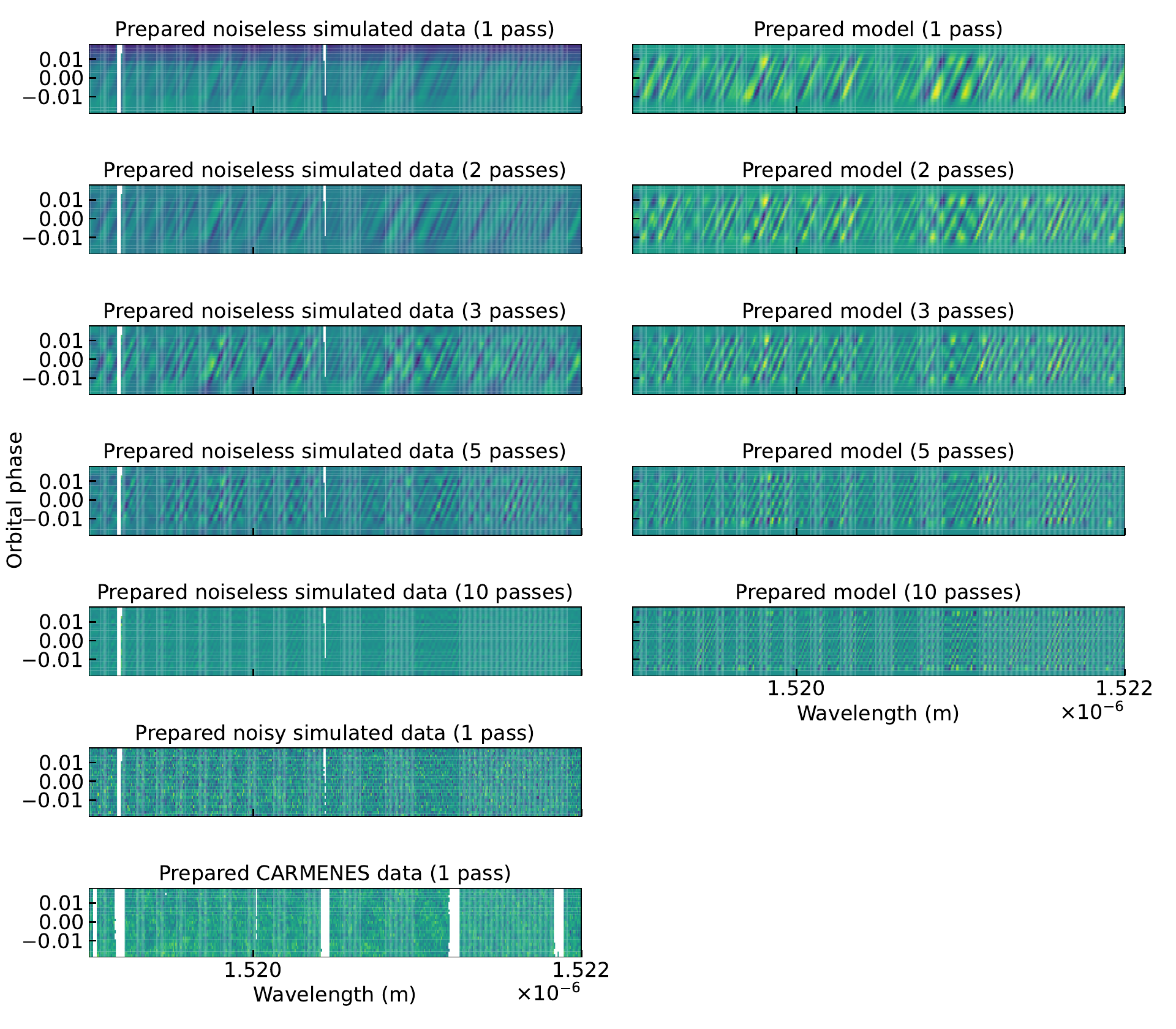}
          \caption{
            Illustration of the effect of our implementation of the SysRem preparing pipeline on the order 46 of the studied data, for a selection of number of passes. The left column represents prepared data, the right column prepared models. Row 1 to 5: prepared noiseless simulated data and model with respectively, 1, 2, 3, 5 and 10 SysRem passes. The simulated data used are the same as in step 9 of \autoref{fig:simulated_data_steps}. Sixth row: same as first row but with added noise. The lines are no longer visible to the naked eye. Bottom row: prepared CARMENES data. The white vertical stripes are masked telluric lines.
        }
        \label{fig:preparing_pipeline_sysrem}
    \end{figure*}

        \paragraph{Step 3:} We apply the SysRem algorithm as described in \citet{Tamuz2005}. The algorithm requires to start with an estimation of either the parameter "$a(t)$", which can be seen as similar to the airmass, or the parameter "$c(\lambda)$", which can be seen as similar to the extinction coefficients. However, as long as the algorithm is able to converge, the starting parameters and their initial values have no importance. We chose to start with $c(\lambda) = \mathbf{1}$. Following \citet{alonso2019multiple}, we run one SysRem pass with 15 iterations. We obtain an estimation of the systematics $a(t) \circ c(\lambda)$, that we subtract\footnote{In this setup, we found that skipping step 2 and dividing the systematics instead of subtracting them, in order to obtain prepared data akin to \autoref{eq:pipeline}, lead to inaccurate retrievals on simulated data. As a division is not considered to remove the systematics in \citet{Tamuz2005}, we did not investigate this further.} to the data to obtain the prepared spectrum after 1 pass ${}^{(1)}\mathbf{F}_S(t, \lambda)$ with:
        \begin{eqnarray}
            \label{eq:sysrem_step3}
        		{}^{(1)}\mathbf{F}_S(t, \lambda) &=& \mathbf{F}_a(t, \lambda) - a(t) \circ c(\lambda),
        \end{eqnarray}
        where the superscript $(1)$ represents the number SysRem passes. This last step can be repeated any number of time $n$ to remove more linear trends. In that case, $\mathbf{F}_a$ in the above equation is replaced by ${}^{(n-1)}\mathbf{F}_S$, that is, the spectrum obtained from the previous pass. Because we again only subtract a value from the data, this step has no impact on the uncertainties. Note that because of the subtractions used, this preparing pipeline cannot be represented with \autoref{eq:pipeline}, and has no preparing matrix as we defined it. We will use the notation $P_S$ to represent all of the SysRem pipeline steps, with $P_S(\mathbf{F}) = {}^{(n)}\mathbf{F}_S(t, \lambda)$. Note that as with $P_\mathbf{R}$, the result of $P_S$ will change depending on the input. The effect of the SysRem preparing pipeline is displayed in \autoref{fig:preparing_pipeline_sysrem}.
    
    \subsection{"Noiseless BPM"}
    \label{subanx:bpm}
    As previously stated, and as it can be seen in \autoref{eq:sysrem_step3}, SysRem does not follow \autoref{eq:pipeline}, so the BPM that we derived cannot be calculated for this preparing pipeline. A workaround to this issue is to remove the noise from \autoref{eq:bpm}, so that $\mathbf{R}_\mathbf{F}$ disappears from the equation, and a value for SysRem can be obtained. Note that this is not a strictly correct way to evaluate the BPM: the impact of the noise is not measured, and we stress that the equation has been established assuming the pipeline respects \autoref{eq:pipeline}. Doing so and adding $\mathbf{1}$ to the results of our SysRem implementation, we obtain a "noiseless BPM" of $5\times10^{-10}$ for "Polyfit", while we find $9\times10^{-6}$ and $2\times10^{-9}$ for SysRem with respectively 1 and 15 passes. We are giving these "noiseless BPM" values as indications, but their meaningfulness is questionable.

    \subsection{Retrieval results with SysRem}
    \begin{figure*}
       \centering
       \includegraphics[width=\hsize]{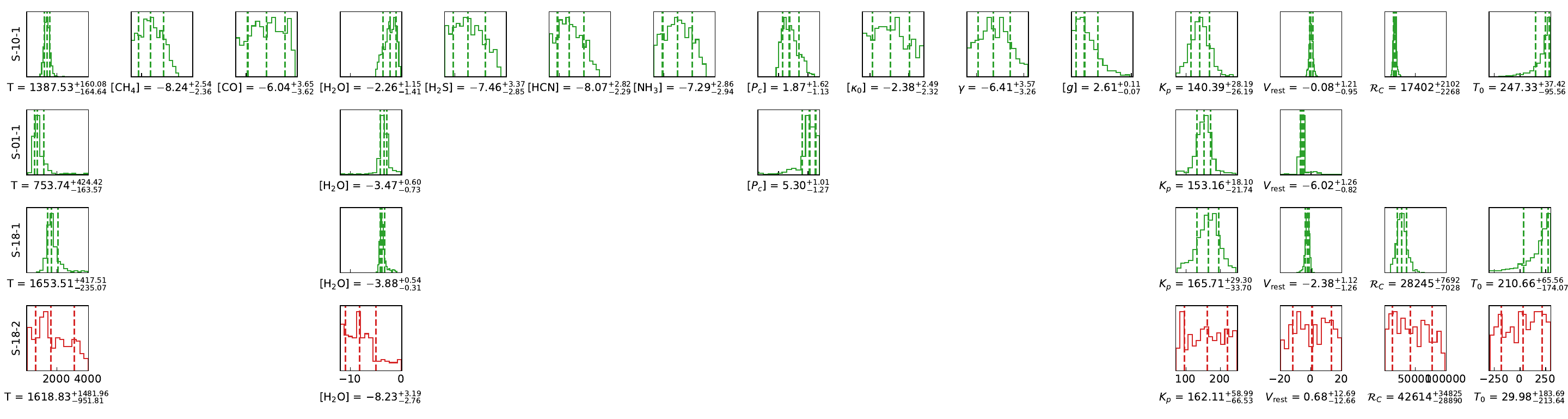}
          \caption{
            Posterior probability distributions for a selection of SysRem setups. The dashed vertical lines represents the 0.16, 0.50 and 0.84 quantiles (i.e. median and 1-$\sigma$ error bar) of the distributions obtained. Below each panel is indicated the median and the 1-$\sigma$ error bar of the retrieved parameters using the setup. The units are the same as in \autoref{fig:expected_retrieval}.
        }
        \label{fig:retrievals_posteriors_sysrem}
    \end{figure*}
    
    When using SysRem, we use the exact same setup as for Polyfit, only replacing $P_\mathbf{R}$ with $P_S$ and $\mathbf{U}_\mathbf{R}$ with $\mathbf{U}_{\mathbf{N},\overline{\mathbf{X}}}$ in \autoref{eq:log_likelihood}. The setups are denoted "S-xx-y", where "xx" corresponds to the same setup as in \autoref{tab:retrievals_comparison}, and "y" corresponds to the number of SysRem passes used. For example, S-01-1 has the same retrieved parameters and priors than P-01, and 1 SysRem pass was applied. 
    
    In \autoref{fig:retrievals_posteriors_sysrem}, we show the posteriors of a selection of SysRem retrievals.
    With 1 SysRem pass, the results obtained with SysRem are quite different than those obtained with Polyfit. Most notably, for the S-01-1 and S-18-1 setups, the H$_2$O MMR is 2 orders of magnitude lower compared with the latter and compatible at $\approx 2 \sigma$. The mid transit time posterior is also clearly truncated, and the constraints obtained are much softer. With 2 SysRem passes, we obtain essentially flat posteriors. We also observe flat posteriors with 3, 5 and 10 passes (not shown here). This behaviour might seems surprising as it is common in other works to use more than one SysRem pass \citep[e.g.][used 10 SysRem passes]{alonso2019multiple}. However, in our framework we apply the preparing pipeline on both the data and the models, which is not necessarily what is done in other works. In \autoref{fig:preparing_pipeline_sysrem}, it can be seen that the models and the noiseless simulated data are not deformed in a similar way, which might lead to the biased results we observe with our framework. In addition, it seems that the simulated noiseless data is quickly removed by SysRem, and that the planet's atmospheric lines in the simulated noiseless data are well preserved with at most 2 passes (although we did not quantify this). This "line-preserving" number of passes may depend on the spectral window, the kinematic properties of the observed target, and of the instrument used. This behaviour may also be different in a noisy case.
    
    This highlights that not all preparing pipelines are suitable with all frameworks. We stress that the results presented in this appendix are only valid with our framework, and that in other published work using PCA-based methods, the forward model is prepared for retrieval in a different manner. The apparently successful use of SysRem and related PCA-based approaches in previous retrieval studies \citep[e.g.][]{Pelletier2021, Gibson2022} suggests that formally justified frameworks using SysRem-like preparing pipelines may exist, but investigating them is beyond the scope of this work.

\section{Retrieval with all orders}
    \label{anx:retrieval_with_all_orders}
\begin{figure*}
        \centering
        \includegraphics[width=\hsize]{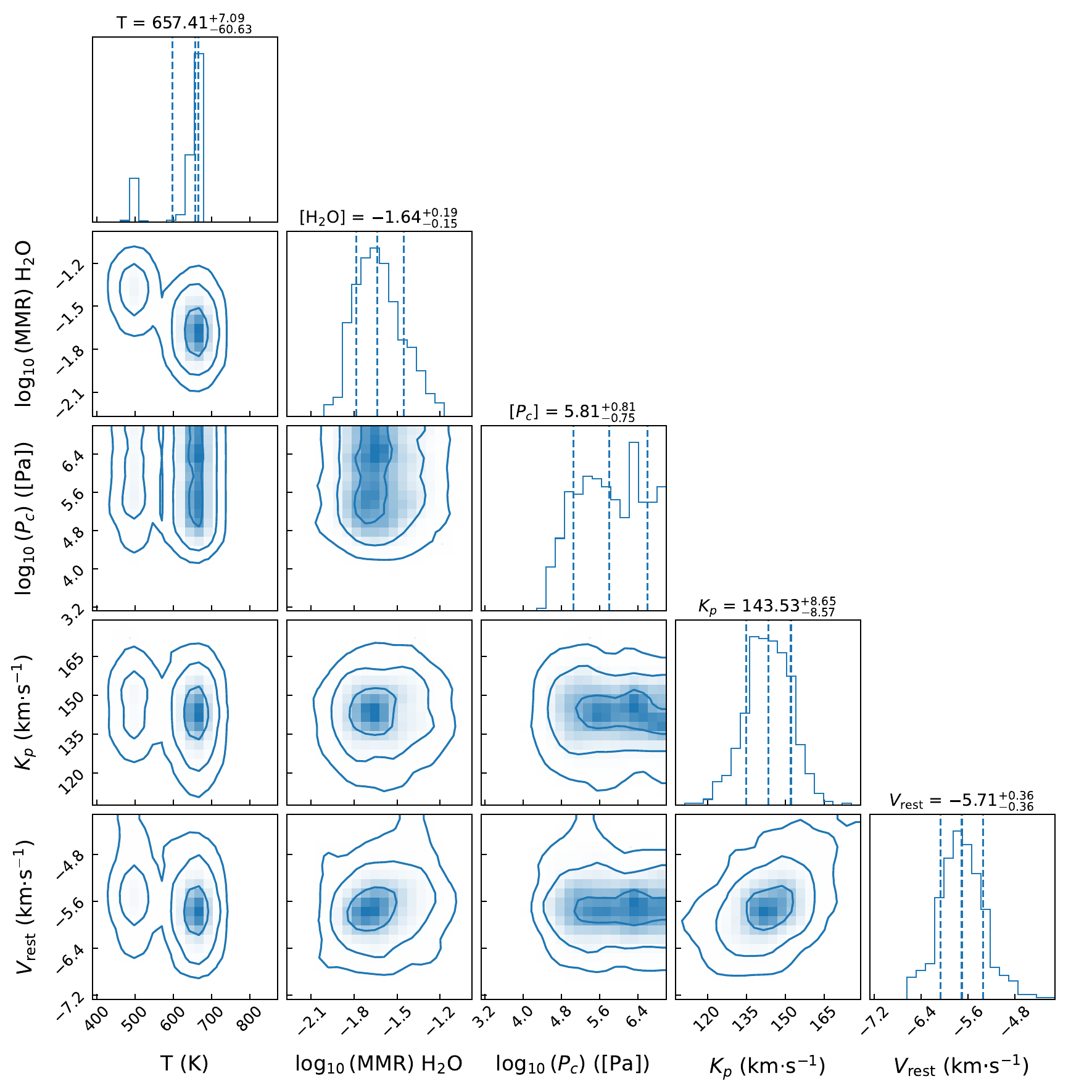}
            \caption{Posterior probability distributions for a Polyfit setup, with all orders, using 100 live points. Blue: results using the Polyfit preparing pipeline. The dashed vertical blue lines represents the 0.16, 0.50 and 0.84 quantiles (i.e. median and 1-$\sigma$ error bar) of the distributions obtained with Polyfit. On top of each column is indicated the median and the 1-$\sigma$ error bar of the retrieved parameters using Polyfit. The contours in the 2D histograms corresponds to the 1, 2 and 3-$\sigma$ contours.}
            \label{fig:simple_retrieval_all_orders}
    \end{figure*}

    We display the result of a retrieval with all orders in \autoref{fig:simple_retrieval_all_orders}. The values retrieved with the Polyfit setup are consistent with those retrieved with the same setup but with our order selection.

\section{Risking biased results: automatic order selection algorithms}
    \label{anx:order_selection_algorithm}
    
    \begin{figure}
        \centering
        \includegraphics[width=\hsize]{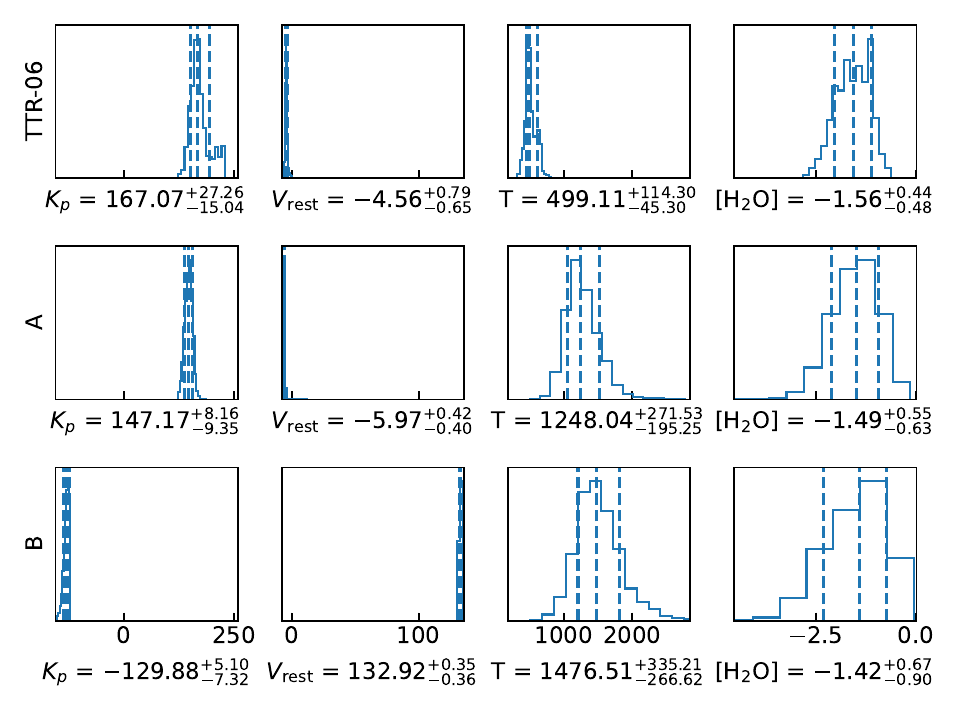}
          \caption{
            Posterior probability distributions for several order selections and the exposures corresponding to $T_{\mathrm{transit}}$. Each row represent a series of posteriors obtained with a different order selection. TTR-06: order selection used in body of the article, for comparison. A: order selection obtained when running our algorithm nominally. B: order selection obtained when running our algorithm, starting with orders that places the co-added CCF maximum at $K_p$ and $V_{\mathrm{rest}}$ far from the expected ones. The dashed vertical blue lines represents the 0.16, 0.50 and 0.84 quantiles (i.e. median and 1-$\sigma$ error bar) of the distributions obtained. Below each panel is indicated the median and the 1-$\sigma$ error bar of the retrieved parameters using the setup.
        }
        \label{fig:retrievals_posteriors_auto_order}
    \end{figure}

    The results presented in this appendix are derived from models following \autoref{subsec:spectrum_model}, except that step 6 is not performed. The time array used here are also in BJD$_\textrm{UTC}$ instead of the BJD$_\textrm{TDB}$ used in the body of the article. As stated in the conclusion of this appendix, we recommend against using the technique described below, and discarded it ourselves. Nevertheless, we chose to report the following results as they show that using a similar technique can dangerously bias the data analysis.
    
    The order selection algorithm follows the steps below:
    
    \paragraph{Step 1:} based on the Exo-REM model generated in \autoref{subsec:self-consistent_model} and using the fact that most of the transmission signal comes from a pressure of $\approx 10^4$ Pa, we fix the absorbers MMR at the values listed in \autoref{tab:prt_base_model_mmr}. We don't include any cloud nor haze effect: indeed, we don't expect the MnS cloud to have a significant contribution to the spectrum. We use the parameters listed in \autoref{tab:general_parameters}, the CARACAL uncertainties ($\mathbf{U}_\mathbf{N}$), and the 19 exposures centered around $T_0$.
    
    \paragraph{Step 2:} we cross-correlate this model with the prepared data using the setup described in \autoref{subsec:ccf_setup}. No co-addition of CCF is performed yet. We choose a limited set of orders, that we estimate are the least impacted by telluric lines and instrumental effects, have strong H$_2$O absorption or minor contributions from other species, and cover most of the CARMENES wavelength range. For these data, we started with orders 3, 7, 9, 13, 25, 28, 29, 46, 54.
    
    \paragraph{Step 3:} we calculate the co-added CCF of the selected orders data, and compute the corresponding S/N as described in \autoref{subsec:ccf_setup}. In addition, we determine:
    \begin{itemize}
    	\item the S/N at the co-added CCF maximum,
    	\item the number of elements of our co-added CCF matrix that is greater or equal to 0.68 times the function maximum (we will call this quantity $\mathcal{A}$),
    	\item the position of this maximum in the $K_p$--$V_{\mathrm{rest}}$ space.
    \end{itemize}
    
    \paragraph{Step 4:} we add one of the remaining orders to the set of included orders, and perform step 3 with this new set. We then test for the following:
    \begin{itemize}
    	\item the S/N at the co-added CCF maximum calculated with the new set is higher than the one calculated with the previous set,
    	\item the new $\mathcal{A}$ is lower than the previous one,
    	\item the position of the new co-added CCF maximum is between 145 and 160 km$\cdot$s$^{-1}$ along the $K_p$ axis,
    	\item the position of the new co-added CCF maximum is between -10 and 10 km$\cdot$s$^{-1}$ along the $V_{\mathrm{rest}}$ axis.
    \end{itemize}
    The order is kept within the set if all these conditions are met. Otherwise, the order is removed from the set. This step is repeated with the new set, until each remaining order has been tested.
    
    \paragraph{Step 5:} this step is similar to step 4, but instead we remove a order from the set of orders obtained at the end of step 4. The same tests are performed. The order is removed from the set if all of the conditions are met. Otherwise, the order is kept within the set. This step is repeated until each order in the set has been tested.
    
    We repeat the whole procedure, starting with the obtained order selection from the last iteration, until the set does not change. At the end, we obtain the following stable order selection: 1, 3, 7, 8, 9, 14, 25, 28, 29, 30, 46, 54. We will call this set our "cleaned-up" order selection. 
    
    The procedure assumes that:
    \begin{itemize}
    	\item The model used for the cross-correlation is a good representation of the data.
    	\item There is enough planet signal to prevent a false-positive CCF peak.
    	\item The planet signal was emitted within the procedure $K_p$ and $V_{\mathrm{rest}}$ test ranges and does not vary strongly from order to order. 
    	\item An order containing valuable information will necessarily increase the S/N of the CCF maximum, and narrow the CCF peak in $K_p$--$V_{\mathrm{rest}}$ space. This narrowing can be estimated with $\mathcal{A}$.
    	\item An order not containing valuable information will necessarily decrease the S/N of the CCF maximum or expand the CCF peak in $K_p$--$V_{\mathrm{rest}}$ space. 
    \end{itemize}
    If some of these assumptions are not true, there is a risk to reject orders containing valuable information, and a risk to select worthless orders biasing our results toward our favoured atmospheric model. This risk has already been noted by e.g. \cite{Boucher2021} and \citet{Cheverall2023}.

    We used the orders selected by this procedure for several retrievals. We show a result example in \autoref{fig:retrievals_posteriors_auto_order}. When using the orders selection mentioned above, we obtain the posteriors in row "A". When instead we start with orders 11, 18, 19, 25, 26, 32, 33, 34 and remove the constraints on the co-added CCF maximum location in $K_p$--$V_{\mathrm{rest}}$ space, we obtain the following order selection:  1,  9, 10, 11, 13, 18, 19, 24, 26, 27, 29, 31, 32, 33, 34. In the latter case, the co-added CCF maximum is located at unrealistic values: $K_p = -142$ km$\cdot$s$^{-1}$ and $V_{\mathrm{rest}} = 133$ km$\cdot$s$^{-1}$. We also adapted our $K_p$ and $V_{\mathrm{rest}}$ priors to respectively $\mathcal{U}(-192, -122)$ km$\cdot$s$^{-1}$ and $\mathcal{U}(100, 160)$ km$\cdot$s$^{-1}$. This gives the posteriors in row "B". Despite the difference in $K_p$ and $V_{\mathrm{rest}}$ values, the retrieved temperature and H$_2$O MMR are similar in rows "A" and "B", and close to the values of the model used to calculate the CCF (\autoref{tab:general_parameters}, \autoref{tab:prt_base_model_mmr}). This most probably indicates that the algorithm is biased towards the model used to calculate the CCF. The bias may propagate into the retrieval, despite retrievals being much more flexible than a CCF analysis. Due to the strong risk of bias, we estimate that it is ill-advised to use this algorithm.

\section{Retrieval on noisy simulated data}
    \label{anx:retrieval_on_simulated_data_with_simulated_noise}
\begin{figure*}
        \centering
        \includegraphics[width=\hsize]{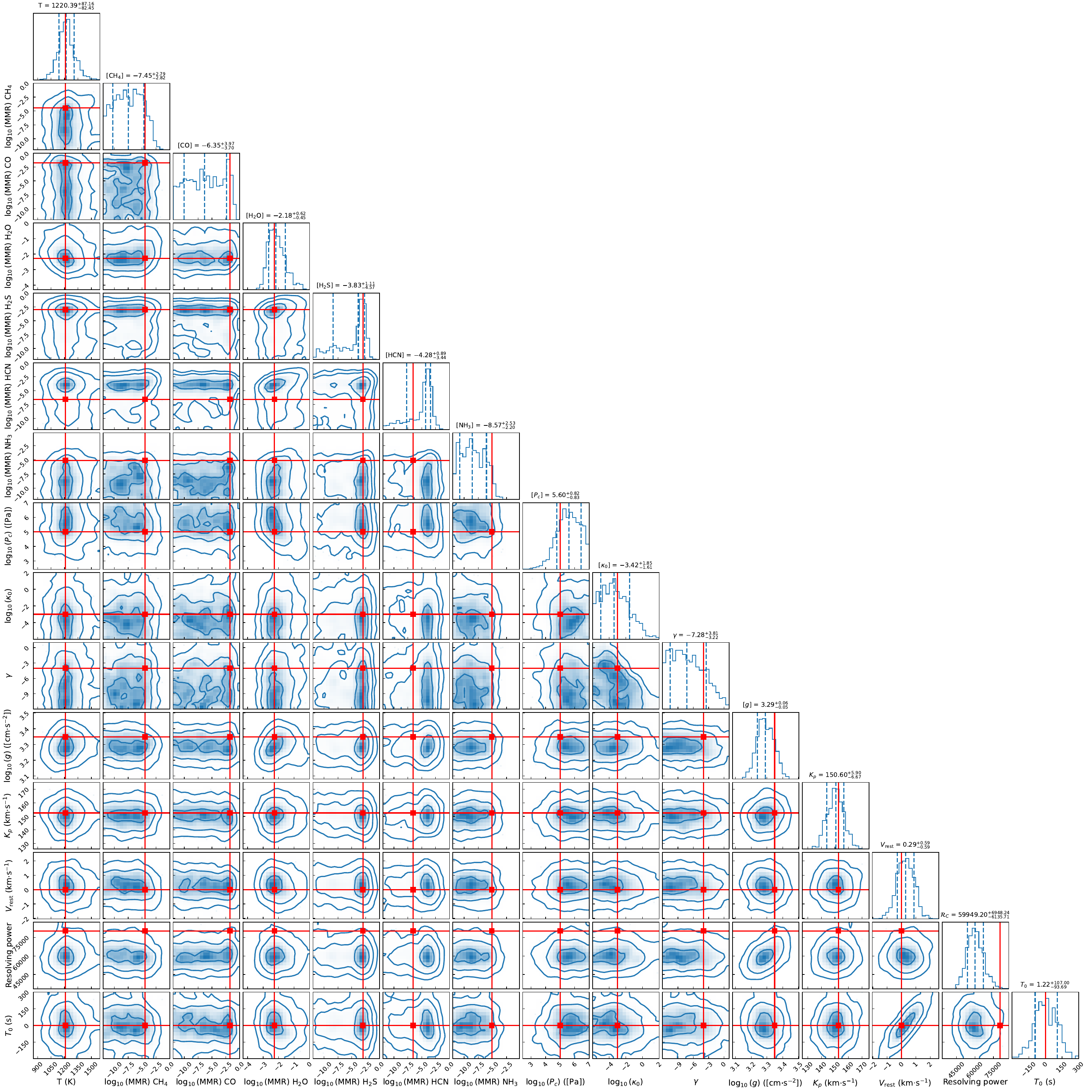}
            \caption{Posterior probability distributions of our noisy simulations. The noise matrix generation is detailed in \autoref{subsec:simulated_data}, the seed used was $12345$. Blue: pRT 1-D simulation using Polyfit. The solid red lines corresponds to the model input values for the 1-D model. The dashed vertical blue lines represents the 0.16, 0.50 and 0.84 quantiles (i.e. median and 1-$\sigma$ error bar) of the distributions for the 1-D simulation. On top of each column is indicated the median and the 1-$\sigma$ error bar of the retrieved parameters for the 1-D simulation. The contours in the 2D histograms corresponds to the 1, 2 and 3-$\sigma$ contours.}
            \label{fig:expected_retrieval_noisy}
    \end{figure*}

    We can evaluate the performance of our framework against noisy simulated data (see \autoref{subsubsec:validation_simulated_retrievals}) using the Polyfit pipeline. The result is shown in \autoref{fig:expected_retrieval_noisy}. With Polyfit, we obtained a best fit $\chi^2_\nu = 0.999$ (not corrected by $k_\sigma$). Note that this is true only for this specific noise realisation and no general conclusion can be driven from this result, in contrast with our noiseless simulated retrievals results (\autoref{subsubsec:validation_simulated_retrievals}).
    
\section{Retrieval on noiseless simulated data with fewer free parameters}
    \label{anx:retrieval_on_noisless_simulated_data_with_fewer_parameters}

\begin{figure*}
        \centering
        \includegraphics[width=\hsize]{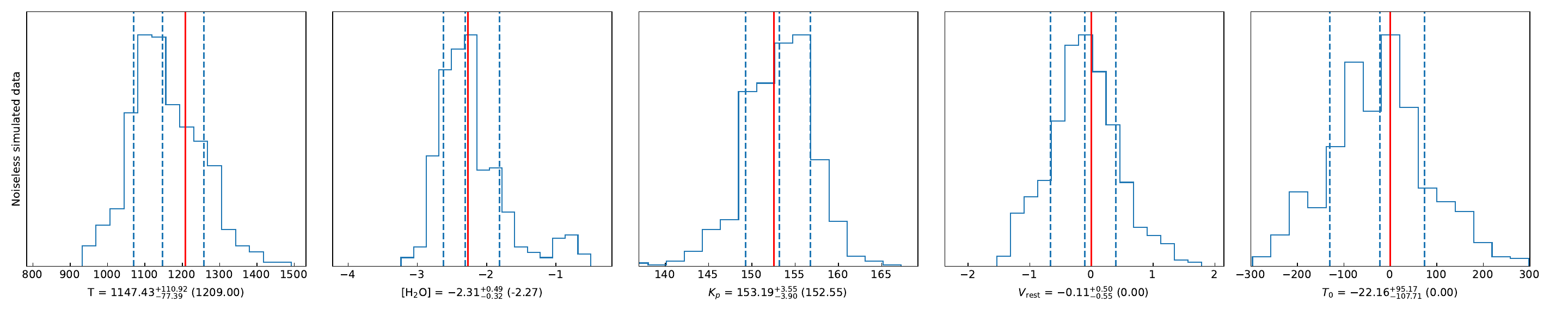}
            \caption{Posterior probability distributions for of our noiseless simulation, with few retrieved parameters. The solid red lines corresponds to the model input values for the 1-D model. The dashed vertical lines represents the 0.16, 0.50 and 0.84 quantiles (i.e. median and 1-$\sigma$ error bar) of the distributions obtained. Below each panel is indicated the median and the 1-$\sigma$ error bar of the retrieved parameters using the setup, as well as the true value between parenthesis. The units are the same as in \autoref{fig:expected_retrieval}
        }
            \label{fig:retrievals_noiseless_simulated_data_few_parameter}
    \end{figure*}

    In \autoref{subsubsec:validation_simulated_retrievals}, we attribute the slight bias of the H$_2$O abundance retrieved from our noiseless simulation mainly to the retrieval of other parameters. This can be verified by retrieving the H$_2$O abundance on the same simulated data, but retrieving with parameters that cannot compete to reproduce the H$_2$O lines. We display the result of such retrieval in \autoref{fig:retrievals_noiseless_simulated_data_few_parameter}. The peak of the H$_2$O abundance posterior is arguably on the true value. The temperature posterior peaks at temperatures slightly lower than the truth but well within the error bar. This is again caused by our pipeline's imperfections.

\section{Marginalising over the uncertainties}
    \label{anx:marginalizing_the_uncertainties}

    \begin{figure*}
       \centering
       \includegraphics[width=\hsize]{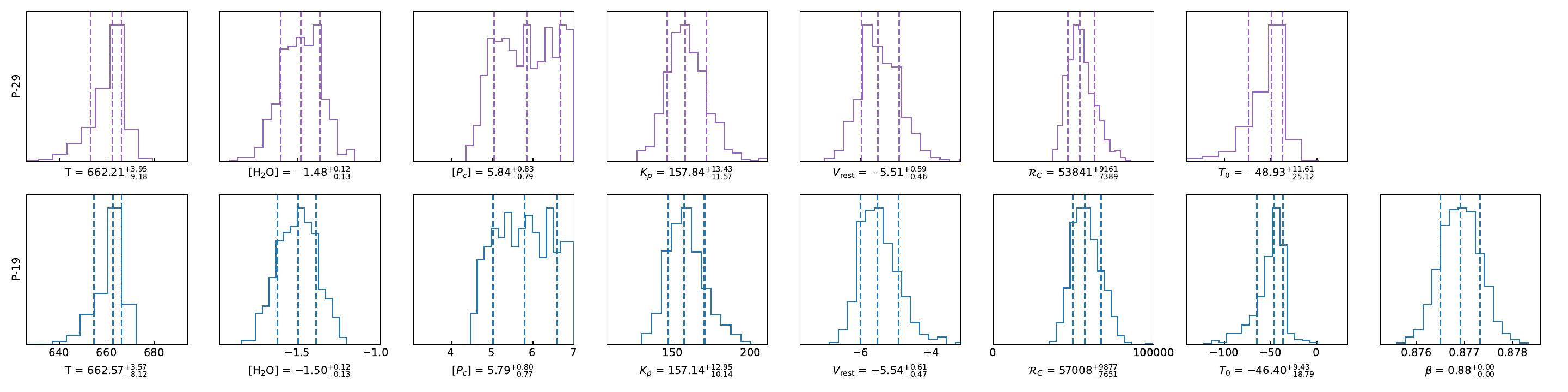}
          \caption{
            Posterior probability distributions for a selection of Polyfit setups. The dashed vertical lines represents the 0.16, 0.50 and 0.84 quantiles (i.e. median and 1-$\sigma$ error bar) of the distributions obtained. Below each panel is indicated the median and the 1-$\sigma$ error bar of the retrieved parameters using the setup. The units are the same as in \autoref{fig:expected_retrieval}.
        }
        \label{fig:retrievals_posteriors_beta}
    \end{figure*}
    
    Our framework is also able to retrieve an uncertainty scaling parameter ($\beta$) or to use the uncertainty "nulling" procedure described in \citet{Brogi2019} and \citet{Gibson2020}. As in \citet{Gibson2020}, the uncertainties are scaled in the likelihood function following:
    \begin{eqnarray}
        \label{eq:log_likelihood_beta}
            \ln(\mathcal{L}) &=& - \frac{1}{2} \sum \left( \frac{P_\mathbf{R}(\mathbf{F}) - P_\mathbf{R}(\mathbf{M}_\theta)}{\beta\mathbf{U}_\mathbf{R}} \right)^2,
    \end{eqnarray}
    with the scalar $\beta$ being retrieved. For the "nulling" procedure, we followed \citet{Gibson2020}, giving the following log-likelihood function:
    \begin{eqnarray}
        \label{eq:log_likelihood_nulling}
            \ln(\mathcal{L}) &=& - \frac{N}{2} \log \left( \frac{1}{N} \sum \left( \frac{P_\mathbf{R}(\mathbf{F}) - P_\mathbf{R}(\mathbf{M}_\theta)}{\mathbf{U}_\mathbf{R}} \right)^2 \right),
    \end{eqnarray}
    where $N$ is the number of non-masked data points. For the reason exposed in \autoref{subsubsec:log_l}, we did not used this feature to obtain our fiducial results. 
    
    Nevertheless, we did run retrievals with these features enabled, that we label P-19 (which is similar to P-14, but with $\beta$ as an additional retrieved parameter) and P-29 (which is similar to P-14, but using the nulling procedure). For $\beta$, we used a uniform prior $\mathcal{U}(0.5, 100)$. We display the results in \autoref{fig:retrievals_posteriors_beta}. In both cases, the results are almost identical with P-14, only with slightly tighter posteriors. The scaling parameter $\beta$ corresponds well to $1/k_\sigma$. For the P-19 setup, since we essentially "force" $\chi^2_\nu = 1$, we unsurprisingly obtain a larger log-evidence than with the other setups, with $\log(\mathcal{Z}) = -807\,500$. With P-29, the log-likelihood equation is changed and we obtain this time a positive log-evidence, with $\log(\mathcal{Z}) = 307\,300$.

\end{appendix}

\end{document}